\documentclass[showpacs,aps,prd,reprint,superscriptaddress,nofootinbib,longbibliography,preprintnumbers]{revtex4-1}
\usepackage[colorlinks=true, pdfstartview=FitV, linkcolor=magenta,citecolor=blue, urlcolor=magenta,
bookmarks=true, bookmarksnumbered=true, breaklinks]{hyperref}
\usepackage[dvipdfmx]{graphicx}
\usepackage{here}
\usepackage{url}
\usepackage{amsmath,amssymb,bm,color,longtable,mathrsfs,amsfonts,slashed}
\usepackage{ulem}

\newcommand{\Slash}[1]{{\ooalign{\hfil/\hfil\crcr$#1$}}}

\begin{document}

\preprint{YITP-23-172, RIKEN-iTHEMS-Report-23}

\title{Mass spectrum of spin-one hadrons in dense two-color QCD:\\
Novel predictions by extended linear sigma model}

\author{Daiki~Suenaga}
\email[]{daiki.suenaga@riken.jp}
\affiliation{Few-body Systems in Physics Laboratory, RIKEN Nishina Center, Wako 351-0198, Japan}
\affiliation{Research Center for Nuclear Physics,
Osaka University, Ibaraki 567-0048, Japan }

\author{Kotaro~Murakami}
\email[]{kotaro.murakami@yukawa.kyoto-u.ac.jp}
\affiliation{Department of Physics, Tokyo Institute of Technology, 2-12-1 Ookayama, Megro, Tokyo 152-8551, Japan}
\affiliation{Interdisciplinary Theoretical and Mathematical Sciences Program (iTHEMS), RIKEN, Wako 351-0198,
Japan}

\author{Etsuko~Itou}
\email[]{itou@yukawa.kyoto-u.ac.jp}
\affiliation{Interdisciplinary Theoretical and Mathematical Sciences Program (iTHEMS), RIKEN, Wako 351-0198,
Japan}
\affiliation{Yukawa Institute for Theoretical Physics, Kyoto University, Kyoto 606-8502, Japan}

\author{Kei~Iida}
\email[]{iida@kochi-u.ac.jp,}
\affiliation{Department of Mathematics and Physics, Kochi University, 2-5-1 Akebono-cho, Kochi 780-8520, Japan}

\date{\today}

\begin{abstract}
We construct an extended version of the linear sigma model in such a way as to describe spin-$1$ hadrons as well as spin-$0$ hadrons in two-color QCD (QC$_2$D) by respecting the Pauli-G\"{u}rsey $SU(4)$ symmetry. Within a mean-field approximation, we therefrom examine a mass spectrum of the spin-$1$ hadrons at finite quark chemical potential ($\mu_q$) and zero temperature. Not only mean fields of scalar mesons and scalar-diquark baryons but also of vector mesons and vector-diquark baryons are incorporated.
As a result, we find that, unless all of those four types of mean fields are taken into account, neither lattice result for the critical $\mu_q$ that corresponds to the onset of baryon superfluidity nor for $\mu_q$ dependence of the pion mass can be reproduced. We also find that a slight suppression of the $\rho$ meson mass in the superfluid phase, which was suggested by the lattice simulation, is reproduced by subtle mixing effects between spin-$0$ and spin-$1$ hadrons. Moreover, we demonstrate the emergence of an axialvector condensed phase and possibly of a vector condensed phase by identifying the values of $\mu_q$ at which the corresponding hadron masses vanish. The possible presence of iso-triplet $1^-$ diquarks that may be denoted by a tensor-type quark bilinear field is also discussed.
 \end{abstract}

\pacs{}

\maketitle

\section{Introduction}
\label{sec:Introduction}

Revealing the characteristics of hadrons in cold dense matter stands as a significant pursuit in quantum chromodynamics (QCD), since these particles serve as good probes to explore medium modifications of QCD symmetry properties, such as chiral symmetry restoration. For this reason, thus far, tremendous theoretical and experimental effort has been devoted to shed light on the hadronic properties in dense nuclear matter~\cite{Holt:2014hma,Metag:2017yuh}.   First-principles lattice Monte Carlo simulations, however, are difficult to apply to such a cold and dense regime due to the so-called sign problem~\cite{Aarts:2015tyj, Nagata:2021ugx}. Hence, our understanding of how the hadronic properties are modified in cold dense matter is limited as compared to the case of hot QCD matter.

Although lattice simulations remain to be effective in three-color QCD at finite quark chemical potential ($\mu_q$), exceptionally for two-color QCD (QC$_2$D) with even numbers of quark flavors, the cumbersome sign problem disappears and indeed the simulations turn out to be applicable at nonzero $\mu_q$~\cite{Muroya:2003qs}. So far, many lattice simulations have been performed toward the delineation of hadron modifications as well as of the phase structure in cold dense QC$_2$D~\cite{Hands:1999md,Kogut:2001na,Hands:2001ee,Muroya:2002ry,Chandrasekharan:2006tz,Hands:2006ve,Hands:2007uc,Hands:2010gd,Cotter:2012mb,Hands:2012yy,Boz:2013rca,Braguta:2016cpw,Puhr:2016kzp,Boz:2018crd,Astrakhantsev:2018uzd,Iida:2019rah,Wilhelm:2019fvp,Buividovich:2020gnl,Iida:2020emi,Astrakhantsev:2020tdl,Bornyakov:2020kyz,Buividovich:2020dks,Buividovich:2021fsa,Iida:2022hyy,Murakami:2023ejc} (see Ref.~\cite{Braguta:2023yhd} and references therein). In concert with those {\it numerical experiments}, theoretical examinations have been made for qualitative understanding by using hadronic and microscopic models~\cite{Kogut:1999iv,Kogut:2000ek,Lenaghan:2001sd,Splittorff:2001fy,Ratti:2004ra,Sun:2007fc,Fukushima:2007bj,Brauner:2009gu,Kanazawa:2009ks,Harada:2010vy,Andersen:2010vu,Zhang:2010kn,He:2010nb,Strodthoff:2011tz,Imai:2012hr,Strodthoff:2013cua,Khan:2015puu,Duarte:2015ppa,Chao:2018czo,Adhikari:2018kzh,Contant:2019lwf,Suenaga:2019jjv,Khunjua:2020xws,Khunjua:2021oxf,Kojo:2021knn,Suenaga:2021bjz,Kojo:2021hqh,Suenaga:2022uqn,Kawaguchi:2023olk}.

In QC$_2$D, diquarks, i.e., bound states of two quarks, emerge as color-singlet hadrons thanks to the pseudoreality of $SU(2)_c$ color group, unlike in three-color QCD. As a consequence, diquarks and mesons can be embedded into single multiplets and described collectively. Moreover, the pseudoreality allows us to extend $SU(2)_L\times SU(2)_R$ chiral symmetry to the so-called Pauli-G\"{u}rsey $SU(4)$ symmetry~\cite{Kogut:1999iv,Kogut:2000ek}, and then, the chiral condensate induces the symmetry breaking of $SU(4)\to Sp(4)$. Thus, chiral models in QC$_2$D are constructed based on this symmetry-breaking pattern and, additionally, small violation of the Pauli-G\"{u}rsey $SU(4)$ symmetry to account for a finite pion mass.

Another noteworthy feature of QC$_2$D is emergence of the {\it diquark condensed phase}: Diquarks are bosonic hadrons carrying the quark number in QC$_2$D so that they start to form a Bose-Einstein condensate (BEC) at certain $\mu_q$~\cite{Kogut:1999iv,Kogut:2000ek}. This distinctive phase violates $U(1)_B$ baryon-number symmetry spontaneously, and hence, the diquark condensed phase is also referred to as the {\it baryon superfluid phase}. Meanwhile, the stable phase at smaller $\mu_q$, which no longer contains BECs, is simply called the {\it hadronic phase}. In the latter phase all thermodynamic quantities show no $\mu_q$ dependence at zero temperature, and such a salient property is called the {\it Silver-Blaze property}.

Recently, a mass spectrum of the low-lying spin-$0$ hadrons carrying negative and positive parities was simulated on lattice at finite $\mu_q$~\cite{Murakami:2022lmq,Murakami2022}. The simulation result indicates that $\eta$ mesons (iso-singlet $0^-$ mesons) are lighter than pions in the superfluid phase, which is in contrast to our naive expectation that $\eta$ mesons are heavier than pions due to the $U(1)_A$ anomaly effects. Motivated by this characteristic mass inversion, in Ref.~\cite{Suenaga:2022uqn} we constructed the linear sigma model (LSM) based on the (approximate) Pauli-G\"{u}rsey $SU(4)$ symmetry, which is capable of describing not only $0^-$ mesons and $0^+$ (anti)diquark baryons but also $0^+$ mesons and $0^-$ (anti)diquark baryons. Based on the LSM, indeed, we succeeded in explaining the mass inversion by showing that the $\eta$ mass is sufficiently suppressed in the superfluid phase owing to mixing with $0^-$ (anti)diquark baryons, which is triggered by the $U(1)_B$ baryon-number violation.

In this paper, we extend the LSM by newly incorporating spin-$1$ hadrons, i.e., $1^\pm$ mesons and (anti)diquark baryons, but still respecting the Pauli-G\"{u}rsey $SU(4)$ symmetry. Then, we demonstrate the importance of mixing effects between spin-$0$ and spin-$1$ hadrons for the $\mu_q$ dependence of physical quantities such as the diquark condensate and the quark-number density. Besides, we present the predicted masses of spin-$1$ hadrons in cold matter and possible novel phases triggered by mass-vanishing spin-$1$ hadrons such as axialvector and vector condensed phases.

The predictions given by the present study on how the spin-$1$ hadrons have their masses modified in cold matter are expected to be checked by future lattice QC$_2$D simulations. Ultimately, our comprehensive model, which allows us to simultaneously describe the spin-$0$ and spin-$1$ hadrons without incorporating the quark degrees of freedom explicitly even at high densities, could serve as a guideline on how to use a hadronic model in cold matter. In addition, as diquarks themselves are observable, QC$_2$D possesses an advantage over three-color QCD where diquark dynamics can be solely seen through, e.g., singly heavy baryons (SHBs) made of one heavy quark and one diquark~\cite{Kawakami:2018olq,Kawakami:2019hpp,Harada:2019udr,Kim:2020imk,Kawakami:2020sxd,Suenaga:2021qri,Suenaga:2022ajn,Suenaga:2023tcy,Takada:2023evq}. The SHBs are now under intensive investigation in accordance with the recent development of experimental techniques. In this regard, our findings on the diquarks in QC$_2$D would also serve as good references for understanding the SHBs dynamics from chiral symmetry.

This article is organized as follows. In Sec.~\ref{sec:Model} we introduce quark-bilinear fields for spin-$0$ and spin-$1$ hadrons
and construct an effective model describing these hadrons based on the Pauli-G\"{u}rsey $SU(4)$ symmetry. After explaining, in Sec.~\ref{sec:Inputs}, our procedure to fix model parameters for later numerical calculations, we show, in Sec.~\ref{sec:NumericalMF}, $\mu_q$ dependence of the mean fields within the present model and, in Sec.~\ref{sec:MassSpectrum}, our main results, namely, the hadron mass spectrum at finite $\mu_q$. Besides, the chiral partner structure for the spin-$1$ hadrons is demonstrated in Sec.~\ref{sec:ChiralPartner}. Sections~\ref{sec:Discussions} and \ref{sec:Conclusions} are devoted to discussions and conclusions, respectively.

\section{Model construction}
\label{sec:Model}

In this section, we construct our effective model describing both the spin-$0$ and spin-$1$ hadrons based on the linear realization of the Pauli-G\"{u}rsey $SU(4)$ symmetry.

\subsection{Spin-0 hadron fields}
\label{sec:Spin0}

For the purpose of constructing the effective Lagrangian, we first introduce a useful building block of the spin-$0$ hadrons whose $SU(4)$ symmetry properties are manifest, following the previous work~\cite{Suenaga:2022uqn}.

In two-flavor QC$_2$D, $SU(2)_L\times SU(2)_R$ chiral symmetry is extended to the Pauli-G\"{u}rsey $SU(4)$ symmetry due to the pseudoreal property of $SU(2)_c$ gauge group, as shown in Appendix~\ref{sec:SU2Nf}. The extended symmetry enables us to treat mesons and diquark baryons in a unified way. Then, as shown in Ref.~\cite{Suenaga:2022uqn}, it is useful to introduce a $4\times4$ matrix $\Sigma$ corresponding to both the spin-$0$ mesons and diquarks baryons as defined by
\begin{eqnarray}
\Sigma = \sum_{a=0}^5 (\mathscr{S}^a+i \mathscr{P}^a)X^aE \ . \label{SigmaSum}
\end{eqnarray}
In this equation, $X^{a=0}=\frac{1}{2\sqrt{2}}{\bm 1}_{4\times4}$ and $X^{a=1-5}$ are generators belonging to the Lie algebra of $SU(4)/Sp(4)$, the expressions for which are given by Eq.~(\ref{XDef}) in Appendix~\ref{sec:Generators}. Besides, $E$ is the $4\times4$ symplectic matrix defined by
\begin{eqnarray}
E = \left(
\begin{array}{cc}
0 & {\bm 1}_f  \\
-{\bm 1}_f & 0 \\
\end{array}
\right) \ . \label{SympDef}
\end{eqnarray}
In Eq.~(\ref{SigmaSum}) $\mathscr{S}^a$ and $\mathscr{P}^a$ represent a set of the spin-$0$ hadron fields as
\begin{eqnarray}
&& \sigma = \mathscr{S}^0\ , \ \ a_0^0 = \mathscr{S}^3\ , \ \ a_0^\pm = \frac{\mathscr{S}^1\mp i \mathscr{S}^2}{\sqrt{2}}\ , \nonumber\\
&& \eta = \mathscr{P}^0\ , \ \ \pi^0= \mathscr{P}^3 \ , \ \ \pi^\pm=\frac{\mathscr{P}^1\mp i\mathscr{P}^2}{\sqrt{2}}\ ,  \nonumber\\
&& B = \frac{\mathscr{S}^5-i\mathscr{S}^4}{\sqrt{2}}\ , \ \ \bar{B} = \frac{\mathscr{S}^5+i\mathscr{S}^4}{\sqrt{2}} \ , \nonumber\\
&&  B' = \frac{\mathscr{P}^5-i\mathscr{P}^4}{\sqrt{2}}\ , \ \ \bar{B}' =\frac{\mathscr{P}^5+i\mathscr{P}^4}{\sqrt{2}} \ , \label{Spin0SP}
\end{eqnarray}
where $\sigma$, $a_0$, $\eta$, and $\pi$ are mesons while $B$ and $B'$ ($\bar{B}$ and $\bar{B}'$) are (anti)diquark baryons. The quantum numbers carried by those hadrons are summarized in Table~\ref{tab:Spin0}. With the correspondence~(\ref{Spin0SP}), the $4\times4$ matrix $\Sigma$ reads
\begin{widetext}
\begin{eqnarray}
\Sigma =  \frac{1}{2}\left(
\begin{array}{cccc}
0 & -B'+iB &\frac{\sigma-i\eta+a^0-i\pi^0}{\sqrt{2}} & a^+-i\pi^+ \\
 B'-iB & 0 & a^--i\pi^- & \frac{\sigma-i\eta-a^0+i\pi^0}{\sqrt{2}} \\
-\frac{\sigma-i\eta+a^0-i\pi^0}{\sqrt{2}} & -a^-+i\pi^- & 0 & - \bar{B}' + i\bar{B} \\
-a^++i\pi^+&-\frac{\sigma-i\eta-a^0+i\pi^0}{\sqrt{2}} & \bar{B}'-i\bar{B}& 0 \\
\end{array}
\right) \ . \label{SigmaDef}
\end{eqnarray}
\end{widetext}

In terms of the quark doublet operator $\psi=(u,d)^T$, the hadrons are denoted by
\begin{eqnarray} 
&& \sigma \sim \bar{\psi}\psi\ ,  \ \ a_0^\pm \sim\frac{1}{\sqrt{2}} \bar{\psi}\tau_f^\mp \psi  \ ,\ \ a_0^0 \sim \bar{\psi}\tau_f^3\psi \ , \nonumber\\
&& \eta \sim \bar{\psi}i\gamma_5\psi \ , \ \ \pi^{\pm} \sim \frac{1}{\sqrt{2}}\bar{\psi}i\gamma_5\tau_f^\mp\psi  \ , \ \ \pi^0 \sim \bar{\psi}i\gamma_5\tau_f^3\psi \ , \nonumber\\
&& B \sim -\frac{i}{\sqrt{2}} \psi^T{\cal C}\gamma_5\tau_c^2\tau_f^2\psi \ , \ \  B' \sim- \frac{1}{\sqrt{2}} \psi^T{\cal C}\tau_c^2\tau_f^2\psi  \ , \nonumber\\
&&  \bar{B} \sim -\frac{i}{\sqrt{2}}\psi^\dagger {\cal C}\gamma_5\tau_c^2\tau_f^2\psi^* \ , \ \ \bar{B}' \sim \frac{1}{\sqrt{2}}\psi^\dagger {\cal C}\tau_c^2\tau_f^2\psi^* \ , \label{Spin0Psi}
\end{eqnarray}
where ${\cal C}=i\gamma^2\gamma^0$ is the charge-conjugation Dirac matrix and $\tau_f^\pm=\tau_f^1\pm i\tau_f^2$. Using Eqs.~(\ref{Spin0SP}) and~(\ref{Spin0Psi}), one can see that, in terms of the four-component quark field $\Psi$ defined by Eq.~(\ref{PsiFour}), the $4\times4$ matrix $\Sigma$, Eq.~(\ref{SigmaDef}), reads
\begin{eqnarray}
\Sigma_{ij} \sim \Psi_j^T\sigma^2\tau_c^2\Psi_i\ . \label{SigmaInt}
\end{eqnarray}
Thus, under the $SU(4)$ transformation, $\Sigma$ transforms as
\begin{eqnarray}
\Sigma \to U \Sigma U^T \label{SigmaSU4}
\end{eqnarray}
with $U\in SU(4)$. This symmetry property plays a significant role in constructing the effective Lagrangian describing the spin-$0$ hadrons.

\begin{table}[t]
\begin{center}
  \begin{tabular}{c|ccc}  \hline\hline
Hadron & $J^P$ & Quark number & Isospin \\ \hline
$\sigma$ & $0^+$ & $0$ & $0$ \\ 
$a_0$ & $0^+$ & $0$ & $1$ \\
$\eta$ & $0^-$ & $0$ & $0$ \\
$\pi$ & $0^-$ & $0$ & $1$ \\
$B$ ($\bar{B}$) & $0^+$ & $+2$ ($-2$) & $0$ \\
$B'$ ($\bar{B}'$) & $0^-$ & $+2$ ($-2$) & $0$ \\
\hline \hline
 \end{tabular}
\caption{Quantum numbers of the spin-$0$ hadrons.}
\label{tab:Spin0}
\end{center}
\end{table}

Here, it is well known that the sigma field ($\sigma$) can acquire its mean-field value:  $\sigma_0\equiv\langle\sigma\rangle$, to mimic the chiral condensate which results in the chiral symmetry breaking. Within such a mean-field level, the matrix~(\ref{SigmaDef}) is reduced to
\begin{eqnarray}
\Sigma \to \Sigma_0 \equiv \frac{\sigma_0}{2\sqrt{2}}E\ ,
\end{eqnarray}
where $E$ is the symplectic matrix, Eq.~(\ref{SympDef}). In general, $\Sigma_0$ is not invariant under $SU(4)$: $\Sigma_0\to U\Sigma_0 U^T$. 
Only when the element $U$ is generated by $h$ satisfying
\begin{eqnarray}
hEh^T=E\ , \label{SP4}
\end{eqnarray}
however, $\Sigma_0$ turns out to be invariant. Equation~(\ref{SP4}) shows that the spontaneous symmetry-breaking pattern triggered by the chiral condensate is $SU(4)\to Sp(4)$ in QC$_2$D.

Before moving on to the spin-$1$ hadrons, we comment on the properties of $\mathscr{S}^a$ and $\mathscr{P}^a$. With the correspondence~(\ref{Spin0SP}), for instance, one can see that the pseudoscalar mesons ($0^-$) and scalar (anti)diquaks ($0^+$), parities of which are opposite, are collectively denoted by $\mathscr{P}^a$, i.e., these hadrons belong to the same multiplet of the $SU(4)$ algebra. The difference of parities are understandable from their intrinsic parities; the pseudoscalar mesons contain one quark and one antiquark while the scalar (anti)diquarks contain two (anti)quarks. Also, from this consideration it can be understood that those hadrons are the ground states in the hadronic phase because of their $S$ wave nature in a quark-model description. Likewise, the scalar mesons ($0^+$) and pseudoscalar (anti)diquarks ($0^-$) are collectively represented by $\mathscr{S}^a$, while they belong to the same multiplet. Those hadrons can be identified as $P$-wave excited states.

\begin{table}[t]
\begin{center}
  \begin{tabular}{c|ccc}  \hline\hline
Hadron & $J^P$ & Quark number & Isospin \\ \hline
$\omega$ & $1^-$ & $0$ & $0$ \\ 
$\rho$ & $1^-$ & $0$ & $1$ \\
$f_1$ & $1^+$ & $0$ & $0$ \\
$a_1$ & $1^+$ & $0$ & $1$ \\
$B_S$ ($\bar{B}_S$) & $1^+$ & $+2$ ($-2$) & $1$ \\
$B_{AS}$ ($\bar{B}_{AS}$) & $1^-$ & $+2$ ($-2$) & $0$ \\
\hline \hline
 \end{tabular}
\caption{Quantum numbers of the spin-$1$ hadrons.}
\label{tab:Spin1}
\end{center}
\end{table}

\subsection{Spin-1 hadron fields}
\label{sec:Spin1}

In Sec.~\ref{sec:Spin0}, we have introduced the $4\times4$ matrix $\Sigma$ corresponding to the spin-$0$ mesons and diquark baryons. Next, in this subsection we consider another $4\times4$ matrix $\Phi^\mu$, which is essential to describe the spin-$1$ hadrons toward construction of our effective Lagrangian.

In two-flavor QC$_2$D, the relevant low-lying spin-$1$ mesons and diquark baryons are
\begin{eqnarray}
&& \omega^\mu \sim \bar{\psi}\gamma^\mu\psi \ , \ \ f_1^\mu \sim \bar{\psi}\gamma_5\gamma^\mu\psi \ , \nonumber\\
&& \rho^{0,\mu} \sim \bar{\psi}\tau_f^3\gamma^\mu\psi \ , \ \  \rho^{\pm,\mu} \sim \frac{1}{\sqrt{2}}\bar{\psi}\tau_f^\mp\gamma^\mu\psi  \ , \nonumber\\
&& a_1^{0,\mu} \sim \bar{\psi}\tau_f^3\gamma_5\gamma^\mu\psi \ , \ \ a_1^{\pm,\mu} \sim \frac{1}{\sqrt{2}}\bar{\psi}\tau_f^\mp\gamma_5\gamma^\mu\psi \ ,  \label{Spin1Meson}
\end{eqnarray}
and 
\begin{eqnarray}
&& B_{S}^{I_z=0,\mu} \sim-\frac{i}{\sqrt{2}}\psi^T{\cal C}\gamma^\mu\tau_c^2\tau_f^1\psi  \nonumber\\
&& B_{S}^{I_z=\pm1,\mu} \sim -\frac{i}{2}\psi^T{\cal C}\gamma^\mu\tau_c^2({\bm 1}_f\pm \tau_f^3)\psi \ , \nonumber\\
&& B_{AS}^\mu \sim -\frac{1}{\sqrt{2}}\psi^T{\cal C}\gamma_5\gamma^\mu\tau_c^2\tau_f^2\psi \nonumber\\
&& \bar{B}_{S}^{I_z=0,\mu} = (B_{S}^{I_z=0,\mu})^\dagger \ , \ \ \bar{B}_{S}^{I_z=\pm1,\mu} = (B_{S}^{I_z=\mp1,\mu})^\dagger \ , \nonumber\\
&& \bar{B}_{AS}^\mu = (B_{AS}^\mu)^\dagger \ , \label{Spin1Badyon}
\end{eqnarray}
respectively, in terms of the quark operator. The quantum numbers for those states are summarized in Table~\ref{tab:Spin1}. Here, for the spin-$1$ diquark baryons, the subscripts $S$ and $AS$ denote the ``symmetric'' and ``anti-symmetric'' structure of the flavor contents, respectively, that is, the former is iso-triplet while the latter is iso-singlet. Besides, the superscript $I_z=0,\pm1$ for $B_S$ ($\bar{B}_S$) stands for the eigenvalues of the isospin ``$z$ components.'' We note that the axialvector and vector (anti)diquark baryons are iso-triplet and iso-singlet, respectively, as dictated by the Pauli principle.\footnote{In three-color QCD, the diquarks are translated into SHBs. Indeed, the axialvector and vector diquarks, $B_S$ and $B_{AS}$, would correspond to $\Sigma_c(2455)$ [and its heavy-quark spin partner $\Sigma_c(2520)$] and $\Lambda_c(2595)$ [and $\Lambda_c(2625)$], respectively.}

For spin-$1$ hadrons, it would be convenient to introduce a $4\times4$ matrix $\Phi^\mu$ by the following assignment:
\begin{widetext}
\begin{eqnarray}
\Phi^\mu &=& \frac{1}{2} \left(
\begin{array}{cccc}
\frac{\omega+\rho^{0}-(f_1+a_1^0)}{\sqrt{2}}  & \rho^+-a_1^+& \sqrt{2}  B_S^{I_z=+1} & B_S^{I_z=0}-B_{AS} \\
\rho^--a_1^- & \frac{\omega-\rho^{0}-(f_1-a_1^0)}{\sqrt{2}}  & B_S^{I_z=0} + B_{AS} & \sqrt{2}B_S^{I_z=-1} \\
\sqrt{2}\bar{B}_S^{I_z=-1} & \bar{B}_S^{I_z=0}+ \bar{B}_{AS} & -\frac{\omega+\rho^{0}+f_1+a_1^0}{\sqrt{2}} & -(\rho^-+a_1^-) \\
\bar{B}_S^{I_z=0}-\bar{B}_{AS} & \sqrt{2}\bar{B}_S^{I_z=+1} & -(\rho^++a_1^+) & -\frac{\omega-\rho^{0}+f_1-a_1^0}{\sqrt{2}} \\
\end{array}
\right)^\mu \ . \label{PhiDef}
\end{eqnarray}
\end{widetext}
In fact, in terms of the four-component quark field $\Psi$, Eq.~(\ref{PsiFour}), and the interpolating fields, Eqs.~(\ref{Spin1Meson}) and~(\ref{Spin1Badyon}), $\Phi^\mu$ can be simply rewritten in the form of
 \begin{eqnarray}
\Phi_{ij} \sim \Psi_j^\dagger\sigma^\mu\Psi_i\ , \label{PhiInt}
\end{eqnarray}
where $\sigma^\mu = ({\bm 1},\sigma^i)$ ($\sigma^i$ is the Pauli matrix in the two-component spinor space). Thus, the matrix~(\ref{PhiDef}) fulfills the following homogeneous $SU(4)$ transformation law:
\begin{eqnarray}
\Phi^\mu \to U\Phi^\mu U^\dagger\ , \label{PhiSU4}
\end{eqnarray}
which allows us to construct an effective Lagrangian straightforwardly. For this reason, in what follows we employ $\Phi^\mu$ as a building block of the spin-$1$ hadrons. We note that, with the help of $U(4)$ generators $X^a$ and $S^i$, Eqs.~(\ref{XDef}) and~(\ref{SDef}), the spin-$1$ hadron matrix $\Phi^\mu$ can be expressed as 
\begin{eqnarray}
\Phi^\mu = \left(\sum_{i=1}^{10}V^i S^i -\sum_{a=0}^5V'^aX^a \right)^\mu \ ,\label{PhiHadron}
\end{eqnarray}
where
\begin{eqnarray}
&& \omega = V^0\ , \ \  \rho^\pm = \frac{V^1\mp iV^2}{\sqrt{2}}\ , \ \ \rho^0 = V^3\ , \nonumber\\
&& f_1=V'^0 \ , \ \ a_1^\pm = \frac{V'^1\mp iV'^2}{\sqrt{2}}\ , \ \ a_1^0 = V'^3\ , \nonumber\\
&& B_S^{I_z=0} = \frac{V^9 + iV^{10}}{\sqrt{2}}\ , \ \  \bar{B}_S^{I_z=0} = \frac{V^9 - iV^{10}}{\sqrt{2}}  \ , \nonumber\\
&& B_S^{I_z=\pm1} = \frac{(V^5+iV^6) \pm (V^7+iV^8)}{2} \ , \nonumber\\
&& \bar{B}_S^{I_z=\pm1} = \frac{(V^5-iV^6) \mp (V^7-iV^8)}{2} \ , \nonumber\\
&& B_{AS} = \frac{V'^5-iV'^4}{\sqrt{2}}\ , \ \ \bar{B}_{AS} = \frac{V'^5+iV'^4}{\sqrt{2}}\ .
\end{eqnarray}

The reduced form~(\ref{PhiHadron}) is useful to see symmetry properties of the spin-$1$ hadrons. For instance, from Eq.~(\ref{PhiHadron}) one can see that the vector mesons ($1^-$) and axialvextor (anti)diquarks ($1^+$), parities of which are opposite, belong to the $Sp(4)$ algebra proportional to $S^i$ and hence to the same multiplet. Likewise, the axialvector mesons ($1^+$) and vector (anti)diquarks ($1^-$) are the elements of the remaining algebra. The difference of parities between the mesons and (anti)diquarks in a single multiplet can be understood from their intrinsic parities as in the case of the spin-$0$ hadrons. We note that $V^i$ and $V'^a$ are the ground and excited states, respectively, in the hadronic phase, since the former and latter are identifiable as $S$-wave and $P$-wave states, respectively.

\subsection{Extended linear sigma model}
\label{sec:ELSM}

In Sec.~\ref{sec:Spin0} and Sec.~\ref{sec:Spin1}, the $4\times4$ matrices corresponding to the spin-$0$ and spin-$1$ hadrons in two-flavor QC$_2$D, $\Sigma$ and $\Phi^\mu$, have been introduced. In this subsection, by making the most of these building blocks, we construct an effective Lagrangian to describe interactions among those hadrons.

The $SU(4)$ transformation laws for $\Sigma$ and $\Phi^\mu$ are given by Eqs.~(\ref{SigmaSU4}) and~(\ref{PhiSU4}). Toward construction of the effective Lagrangian, in addition to the $SU(4)$ properties it is necessary to examine the discrete symmetries: parity and charge conjugation invariance. Those discrete transformation laws of $\Sigma$ and $\Phi^\mu$ can be read off from the interpolating fields~(\ref{SigmaInt}) and~(\ref{PhiInt}). From Eq.~(\ref{PsiFour}) the four-component quark field $\Psi$ transforms as
\begin{eqnarray}
\Psi(x) \overset{\cal P}{\to} \ \Omega\tau_c^2\sigma^2{\Psi}^*(x_P)\ , \ \ \Psi \overset{\cal C}{\to}  iE^T\tau^2\Psi\ , \label{PsiPC}
\end{eqnarray}
under parity and charge conjugation with $x_P=(x^0,-{\bm x})$, where $E$ is the symplectic matrix~(\ref{SympDef}) and $\Omega$ is defined by
\begin{eqnarray}
\Omega = \left(
\begin{array}{cc}
0 & {\bm 1}_f  \\
{\bm 1}_f & 0 \\
\end{array}
\right)\ . 
\end{eqnarray}
 Thus, the resultant transformation laws of $\Sigma$ and $\Phi^\mu$ read
\begin{eqnarray}
\Sigma(x) \overset{\cal P}{\to} \Omega\Sigma^\dagger(x_P)\Omega &,&  \Sigma \overset{\cal C}{\to} E^T\Sigma E \ , \nonumber\\
\Phi^\mu(x) \overset{\cal P}{\to} -\Omega\Phi^T_\mu(x_P)\Omega &,& \Phi^\mu \overset{\cal C}{\to} E^T\Phi^\mu E\ . \label{SigmaPhiPC}
\end{eqnarray}

Using the transformation laws given by Eqs.~(\ref{SigmaSU4}),~(\ref{PhiSU4}) and~(\ref{SigmaPhiPC}), one can construct the following effective Lagrangian in such a way as to preserve the Pauli-G\"{u}rsey $SU(4)$ symmetry as well as parity and charge-conjugation invariance:
\begin{widetext}
\begin{eqnarray}
{\cal L}_{\rm eLSM} &=& {\rm tr}[D_\mu \Sigma^\dagger D^\mu\Sigma]-m_0^2{\rm tr}[\Sigma^\dagger\Sigma]-\lambda_1\big({\rm tr}[\Sigma^\dagger\Sigma]\big)^2-\lambda_2{\rm tr}[(\Sigma^\dagger\Sigma)^2]+{\rm tr}[H^\dagger\Sigma+\Sigma^\dagger H] +c( {\rm det}\Sigma + {\rm det}\Sigma^\dagger) \nonumber\\
&-& \frac{1}{2}{\rm tr}[\Phi_{\mu\nu}\Phi^{\mu\nu}] + m_1^2{\rm tr}[\Phi_\mu\Phi^\mu] + ig_3{\rm tr}\big[\Phi_{\mu\nu}[\Phi^\mu,\Phi^\nu]\big] + h_1{\rm tr}[\Sigma^\dagger\Sigma]{\rm tr}[\Phi_\mu\Phi^\mu] + h_2{\rm tr}[\Sigma\Sigma^\dagger\Phi_\mu\Phi^\mu]\nonumber\\
&+&h_3{\rm tr}[\Phi_\mu^T\Sigma^\dagger\Phi^\mu\Sigma] + g_4{\rm tr}[\Phi_\mu\Phi_\nu\Phi^\mu\Phi^\nu] + g_5{\rm tr}[\Phi_\mu\Phi^\mu\Phi_\nu\Phi^\nu] + g_6{\rm tr}[\Phi_\mu\Phi^\mu]{\rm tr}[\Phi_\nu\Phi^\nu] + g_7{\rm tr}[\Phi_\mu\Phi_\nu]{\rm tr}[\Phi^\mu\Phi^\nu] \ . \nonumber\\
\label{ELSMLag}
\end{eqnarray}
\end{widetext}
In this Lagrangian,
\begin{eqnarray}
\Phi_{\mu\nu} &\equiv& D_\mu\Phi_\nu-D_\nu\Phi_\mu 
\end{eqnarray}
is the field strength of $\Phi^\mu$, and the covariant derivatives read
\begin{eqnarray}
&& D_\mu\Sigma \equiv \partial_\mu\Sigma-iG_\mu\Sigma-i\Sigma G^T_\mu-ig_1\Phi_\mu\Sigma-ig_2\Sigma\Phi_\mu^T\ , \nonumber\\
&& D_\mu \Phi_\nu \equiv \partial_\mu\Phi_\nu-i[G_\mu,\Phi_\nu]\ ,
\end{eqnarray}
where $G_\mu$ is an external field, the transformation law of which is $G_\mu\to UG_\mu U^\dagger-i\partial_\mu UU^\dagger$. In the present analysis, $G_\mu$ is replaced by the quark chemical potential $\mu_q$ to access a finite-density system in a systematic way. The replacement can be performed by 
\begin{eqnarray}
G_\mu \to \mu_q\delta_{\mu0}J
\end{eqnarray}
with
\begin{eqnarray}
{J} \equiv \left(
\begin{array}{cc}
{\bm 1}_f & 0 \\
0 & -{\bm 1}_f \\
\end{array}
\right)\ ,
\end{eqnarray}
[see Eq.~(\ref{BaryonPsi})]. Besides, in constructing the model, we have included contributions allowed by the relevant symmetries up to fourth order in $\Sigma^{(\dagger)}$ and $\Phi^\mu$. We note that the $4\times4$ matrix $H$ in Eq.~(\ref{ELSMLag}) is responsible for explicit breaking of the Pauli-G\"{u}rsey $SU(4)$ symmetry to yield the finite pion mass, which takes the form of
\begin{eqnarray}
H = h_qE\ .
\end{eqnarray}

The Lagrangian~(\ref{ELSMLag}) can be understood as an extension of the previous LSM established in Ref.~\cite{Suenaga:2022uqn} where only spin-$0$ hadrons are treated. For this reason we call this model the extended linear sigma model (eLSM). We note that this eLSM is the QC$_2$D version of the one invented for three-color QCD by the Frankfurt group~\cite{Parganlija:2012fy} and applied to the case of finite density by one of the present authors~\cite{Suenaga:2019urn}.

In the eLSM Lagrangian~(\ref{ELSMLag}), with the help of the antisymmetric property: $\Sigma^T = -\Sigma$, the kinetic term for $\Sigma$ can be expanded as
\begin{eqnarray}
{\rm tr}[D_\mu\Sigma^\dagger D^\mu\Sigma] &=& {\rm tr}[\partial_\mu\Sigma^\dagger\partial^\mu\Sigma] + (g_1+g_2){\rm tr}[\Sigma\Sigma^\dagger(\Phi_\mu G^\mu  \nonumber\\
&+& G_\mu\Phi^\mu) ]+ 2(g_1+g_2){\rm tr}[\Phi_\mu^T\Sigma^\dagger G^\mu\Sigma] \nonumber\\
&+& i(g_1+g_2){\rm tr}[\Phi_\mu(\partial^\mu\Sigma\Sigma^\dagger-\Sigma\partial^\mu\Sigma^\dagger)]  \nonumber\\
&+& (g_1^2+g_2^2){\rm tr}[\Sigma\Sigma^\dagger\Phi_\mu\Phi^\mu] \nonumber\\
&+& 2g_1g_2{\rm tr}[\Phi_\mu^T\Sigma^\dagger\Phi^\mu\Sigma]\ . \label{DerivativeExp}
\end{eqnarray}
The trace structure of the last two pieces in the right-hand side (RHS) is equivalent to the $h_2$ and $h_3$ terms in Eq.~(\ref{ELSMLag}), while the remaining interactions are proportionally dependent on the combination of $g_1+g_2$ alone. Hence, the four parameters $g_1$, $g_2$, $h_2$, and $h_3$ can be absorbed into three new combinations
\begin{eqnarray}
C_1 &\equiv& g_1+g_2\ , \nonumber\\
C_2 &\equiv& g_1^2+g_2^2 + h_2\ , \nonumber\\
C_3 &\equiv& 2g_1g_2+h_3\ . \label{C123Def}
\end{eqnarray}

When the spectrum includes spin-$1$ hadrons, it is well known that the Zweig rule, i.e., the large $N_c$ suppression of the interactions, works phenomenologically. In other words, diagrams that are not linked by a single quark line are not expected to play significant roles as far as the spin-$1$ hadrons are concerned. In the following analysis,  therefore, we will leave only terms including a single trace for $\Phi^\mu$'s, which allows us to work with the following reduced eLSM:
\begin{eqnarray}
{\cal L}^{\rm red.}_{\rm eLSM} &=& {\rm tr}[D_\mu \Sigma^\dagger D^\mu\Sigma]-m_0^2{\rm tr}[\Sigma^\dagger\Sigma]-\lambda_1\big({\rm tr}[\Sigma^\dagger\Sigma]\big)^2 \nonumber\\
&-& \lambda_2{\rm tr}[(\Sigma^\dagger\Sigma)^2]+{\rm tr}[H^\dagger\Sigma+\Sigma^\dagger H] \nonumber\\
&+& c( {\rm det}\Sigma + {\rm det}\Sigma^\dagger) -\frac{1}{2}{\rm tr}[\Phi_{\mu\nu}\Phi^{\mu\nu}]  \nonumber\\
&+& m_1^2{\rm tr}[\Phi_\mu\Phi^\mu] +ig_3{\rm tr}\big[\Phi_{\mu\nu}[\Phi^\mu,\Phi^\nu]\big] \nonumber\\
&+& h_2{\rm tr}[\Sigma\Sigma^\dagger\Phi_\mu\Phi^\mu] + h_3{\rm tr}[\Phi_\mu^T\Sigma^\dagger\Phi^\mu\Sigma] \nonumber\\
&+& g_4{\rm tr}[\Phi_\mu\Phi_\nu\Phi^\mu\Phi^\nu] + g_5{\rm tr}[\Phi_\mu\Phi^\mu\Phi_\nu\Phi^\nu] \ . 
\label{ELSMRed}
\end{eqnarray}
As we will see, the masses of the hadrons can be read off from quadratic terms of the corresponding fields on top of the appropriate mean fields.

\section{Inputs}
\label{sec:Inputs}

In Sec.~\ref{sec:ELSM} we have constructed the eLSM to describe both the spin-$0$ and spin-$1$ hadrons at arbitrary $\mu_q$. In this section, before numerical investigation of $\mu_q$ dependence of the mean fields and the hadron masses, we explain our procedure to determine various parameters of the reduced eLSM.

The reduced eLSM~(\ref{ELSMRed}) includes $12$ parameters: $m_0^2$, $m_1^2$, $\lambda_1$, $\lambda_2$, $h_q$, $c$, $g_3$, $g_4$, $g_5$, $C_1$, $C_2$, and $C_3$. In this exploratory study, we try to reduce the number of the parameters as much as possible to avoid unnecessary complexities in the following numerical analysis. First, as discussed in Ref.~\cite{Suenaga:2022uqn}, contributions from the $\lambda_1$ and $c$ terms, which only affect the mass spectrum of the spin-$0$ hadrons, are expected to be small from the $N_c$ counting. Thus, we take $\lambda_1=c=0$. The main aim of the present paper is to delineate the behavior of the spin-$1$ hadrons at finite $\mu_q$, so that this simplification does not affect the following arguments considerably. Next, as for the couplings among the spin-$1$ hadrons, there is no {\it a priori} way to determine all of the parameters due to the currently limited lattice data. When we derive the present eLSM from the ${\cal O}(p^2)$ hidden-local-symmetry (HLS) Lagrangian~\cite{Harada:2010vy}, however, it is expected that the interactions among the spin-$1$ hadrons satisfy the following ``gauge-principle parametrization:''
\begin{eqnarray}
g_3 = g_\Phi\ , \ \ g_4=-g_5=g_\Phi^2\ . \label{GPhiDef}
\end{eqnarray}
From this consideration, we assume Eq.~(\ref{GPhiDef}) to employ only a single parameter $g_\Phi$ instead of $g_3$, $g_4$, and $g_5$. In addition, one can see from Eq.~(\ref{C123Def}) that the roles of $C_1$, $C_2$, and $C_3$ are essentially the same; these three parameters control the interaction strength between the spin-$0$ and spin-$1$ hadrons. For this reason we assume
\begin{eqnarray}
C_1= C_2 \equiv C \label{CDef}
\end{eqnarray} 
for simplicity. In Eq.~(\ref{CDef}) we have not included $C_3$ to define the common coupling, since $C_3$ is uniquely fixed by inputs as will be explained below.

Now the number of the model parameters {\bf is} reduced to seven: $m_0^2$, $m_1^2$, $\lambda_2$, $h_q$, $g_\Phi$, $C$, and $C_3$. The present study is devoted to unveiling properties of the spin-$1$ hadrons, and hence, we take $C$ and $g_\Phi$, which control couplings related to the spin-$1$ hadrons, as free parameters. In order to determine the remaining five parameters, as the first four inputs, we use the masses of $\pi$, $B'(\bar{B}')$, $\rho$, and $a_1$ at vanishing $\mu_q$ simulated in Refs.~\cite{Murakami:2022lmq,Murakami2022}. The input mass values read\footnote{The physical scale is fixed such that the pseudo-critical temperature of the chiral phase transition reads $T_c=200$ MeV at $\mu_q=0$~\cite{Iida:2020emi}.}
\begin{eqnarray}
&& m_\pi^{(\rm H)} = 738\, {\rm MeV}\ , \ \ m_{B'(\bar{B}')}^{(\rm H)}\big|_{\mu_q=0} = 1611\, {\rm MeV}\ , \nonumber\\
&& m^{(\rm H)}_\rho = 908\, {\rm MeV} \ , \ \ m_{a_1}^{\rm (H)} = 1614\, {\rm MeV} \ . \label{MassInput}
\end{eqnarray}
As for the last input, following Ref.~\cite{Suenaga:2022uqn} we take the mean field value of $\sigma$ as
\begin{eqnarray}
\sigma_0^{\rm (H)}=250\,  {\rm MeV}\ , \label{Sigma0Input}
\end{eqnarray}
which gives a typical strength of the chiral symmetry breaking. Here, in Eqs.~(\ref{MassInput}) and~(\ref{Sigma0Input}), the superscript $({\rm H})$ is attached to emphasize that the quantities are defined in the hadronic phase where no diquark condensates emerge. Using analytic expressions for the hadron masses derived in Appendix~\ref{sec:MassHadronic}, together with Eqs.~(\ref{MassInput}) and~(\ref{Sigma0Input}), the remaining parameters can finally be determined as
\begin{eqnarray}
m_0^2 &=& -\frac{1}{2}\left[\big(m_{a_0}^{\rm (H)}\big)^2-3Z_\pi^{-2}\big(m_{\pi}^{\rm (H)}\big)^2\right]\ , \nonumber\\
m_1^2 &=& \frac{1}{2}\left[\big(m_{\rho}^{\rm (H)}\big)^2 + \big(m_{a_1}^{\rm (H)}\big)^2\right] - \frac{C}{8}\big(\sigma_0^{(\rm H)}\big)^2\ , \nonumber\\
\lambda_2 &=& \frac{2}{\big(\sigma_0^{\rm (H)}\big)^2}\left[\big(m_{a_0}^{\rm (H)}\big)^2-Z_\pi^{-2}\big(m_{\pi}^{\rm (H)}\big)^2\right]\ , \nonumber\\
h_q &=& \frac{\sigma_0^{(\rm H)}}{2\sqrt{2}}Z_\pi^{-2}\big(m_\pi^{(\rm H)}\big)^2 \ , \nonumber\\
C_3 &=& \frac{4}{\big(\sigma_0^{(\rm H)}\big)^2}\left[\big(m_{a_1}^{\rm (H)}\big)^2-\big(m_{\rho}^{\rm (H)}\big)^2\right]
\end{eqnarray}
for a given $C$, where the renormalization factor $Z_\pi$ is of the form
\begin{eqnarray}
Z_\pi = \left(1-\frac{C^2\big(\sigma_0^{(\rm H)}\big)^2}{8\big(m_{a_1}^{(\rm H)}\big)^2}\right)^{-1/2}\ .
\end{eqnarray}
We note that the analytic expression for $h_q$, which links the magnitude of explicit breaking of chiral symmetry to the pion mass, can be derived from a stationary condition of the effective potential with respect to $\sigma_0^{(\rm H)}$. 

In the following analysis, we will regard $g_\Phi$ and $C$ as free parameters to explore how the mean fields as well as hadron masses behave at finite $\mu_q$. Recall that $g_\Phi$ is the coupling constant that controls the interaction strength among only spin-$1$ hadrons. Meanwhile, $C$ is particularly responsible for the transition between the spin-$0$ and spin-$1$ hadrons through derivative couplings, as indicated by Eq.~(\ref{DerivativeExp}). Thus, $C$ can be regarded as a parameter that measures the magnitude of the spin-$0$ and spin-$1$ mixing effect. For instance, in the hadronic phase, nonzero $C$ induces $\pi$ - $a_1$ mixing and $\eta$ - $f_1$ mixing, which can be captured by the renormalization constants, $Z_\pi$ and $Z_\eta$, as derived in Appendix~\ref{sec:MassHadronic}.

\begin{figure}[t]
\centering
\hspace*{-0.5cm} 
\includegraphics*[scale=0.1]{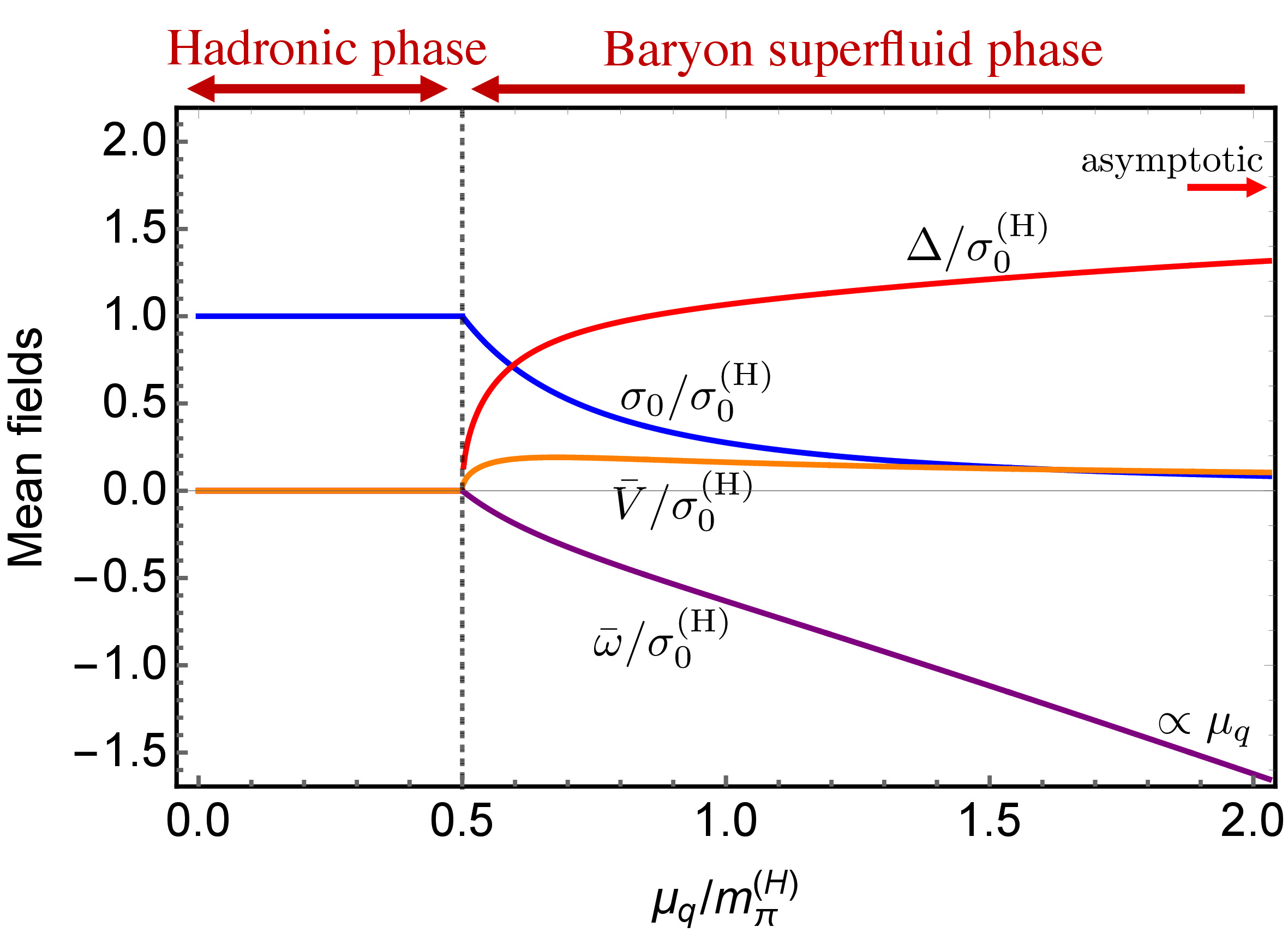}
\caption{$\mu_q$ dependence of the mean fields: $\sigma_0$, $\Delta$, $\bar{\omega}$, and $\bar{V}$, with $C=12$. The value indicated by the arrow represents the asymptotic constant value of $\Delta$: $\Delta\sim 0.34m_1=1.74\sigma_0^{(\rm H)}$. }
\label{fig:MeanFields}
\end{figure}

\begin{figure}[t]
\centering
\hspace*{-0.5cm} 
\includegraphics*[scale=0.1]{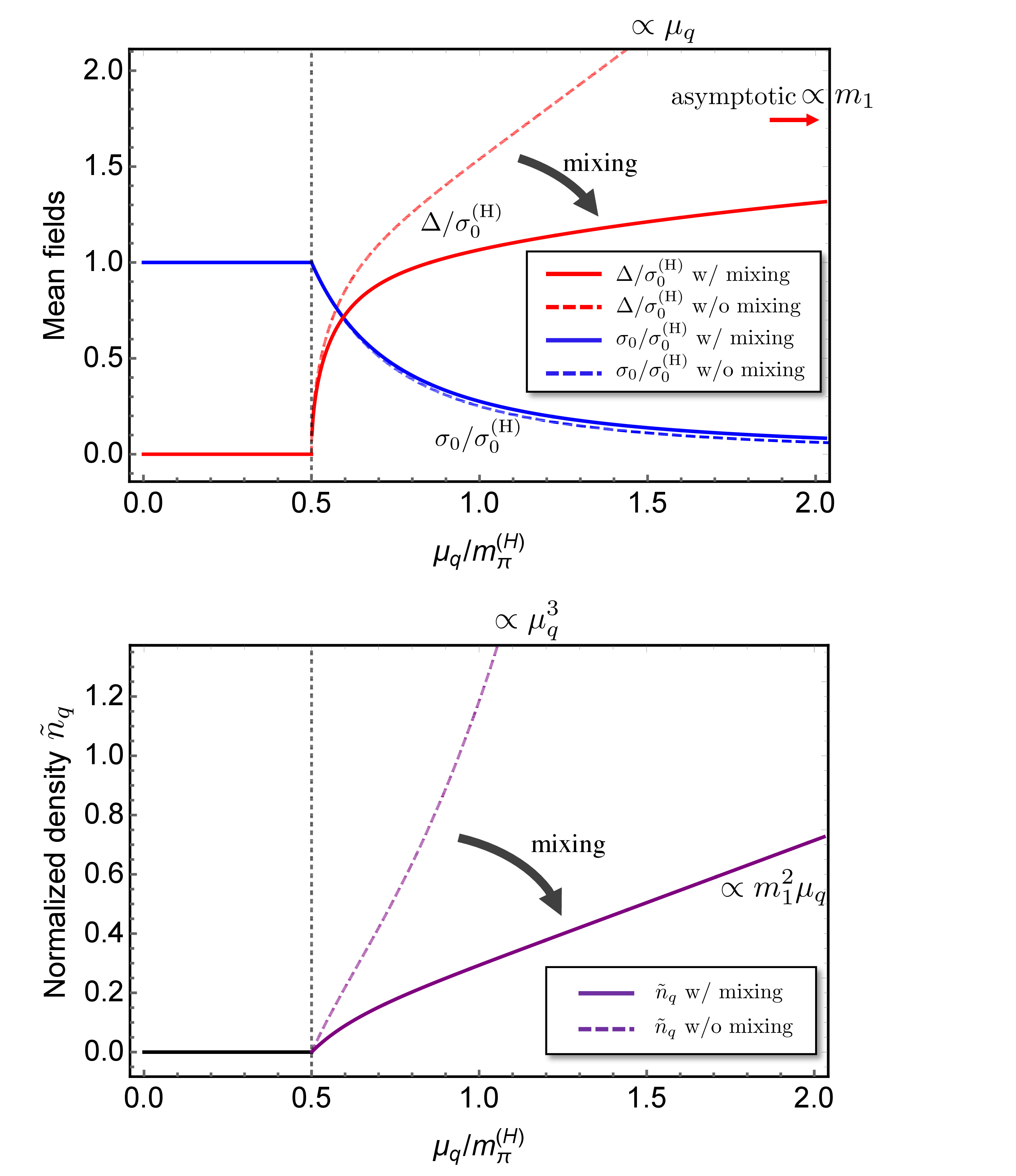}
\caption{$\mu_q$ dependence of $\sigma_0$ and $\Delta$ (top) and of the quark-number density $\tilde{n}_q$ (bottom) with and without spin-$0$ and spin-$1$ mixing, $C=12$ and $C=0$.}
\label{fig:MixMF}
\end{figure}

\section{Mean fields}
\label{sec:NumericalMF}

In this section, employing a mean-field approximation, we numerically explore $\mu_q$ dependence of the mean fields at zero temperature from the reduced eLSM~(\ref{ELSMRed}) with the inputs presented in Sec.~\ref{sec:Inputs}.

At finite $\mu_q$, not only the sigma meson $\sigma$ but also the (anti)diquark baryon $B$ ($\bar{B}$) can acquire a nonzero mean field value, resulting in the appearance of the baryon superfluid phase~\cite{Kogut:1999iv,Kogut:2000ek}. Following Ref.~\cite{Suenaga:2022uqn} we take $\Delta \equiv \langle B^5\rangle$ to express the mean field of the (anti)diquark. In addition to those spin-$0$ hadrons, violation of the Lorentz invariance yields a mean field of the $\omega$ meson~\cite{Walecka:1974qa}. Assuming the parity invariance, only the time component of $\omega$ can have a nonzero mean-field value: $\bar{\omega} \equiv \langle\omega_{\mu=0}\rangle$. Furthermore, in the baryon superfluid phase, one can expect that the $\omega$ meson mixes with the vector (anti)diquark $B_{AS}$ ($\bar{B}_{AS}$) due to the baryon-number violation. Thus, these diquarks are also capable of acquiring nonzero mean-field values. When the phase of $\Delta$ is chosen according to $\Delta=\langle B^5\rangle$, only $\bar{V} \equiv\langle V'^4_{\mu=0}\rangle$ becomes nonzero.\footnote{In this phase choice, one can indeed prove that $\langle V'^5_{\mu=0}\rangle$ must be always zero by solving the stationary conditions explicitly. Inversely, if we choose the phase of $\Delta$ such that $\langle B^4\rangle\neq0$ but $\langle B^5\rangle=0$, then $\langle V'^4_{\mu=0}\rangle=0$ but $\langle V_{\mu=0}'^5\rangle\neq0$ is obtained for $\bar{V}$.} To summarize, in the present analysis we take into account the following four mean fields without loss of generality:
\begin{eqnarray}
&& \sigma_0 = \langle \sigma\rangle \ , \ \ \Delta = \langle B^5\rangle\ , \nonumber\\
&& \bar{\omega} = \langle\omega_{\mu=0}\rangle\ , \ \  \bar{V} \equiv\langle V'^4_{\mu=0}\rangle\ . \label{MeanFields}
\end{eqnarray}
The $\mu_q$ dependence of these mean fields can be determined by solving the corresponding stationary conditions.

Depicted in Fig.~\ref{fig:MeanFields} is the resultant $\mu_q$ dependence of the mean fields: $\sigma_0$, $\Delta$, $\bar{\omega}$, and $\bar{V}$, normalized by $\sigma_0^{(\rm H)}$. In obtaining this plot, the strength of the spin-$0$ and spin-$1$ mixing effect is chosen as $C=12$. As shown in Sec.~\ref{sec:Numerical}, this value leads to a slight reduction of the $\rho$ mass in the superfluid phase, a feature consistent with the lattice data~\cite{Murakami:2022lmq,Murakami2022}. We note that the stationary conditions hold independently of $g_\Phi$, so that the resultant $\mu_q$ dependence of the mean fields is not affected by $g_\Phi$. Figure~\ref{fig:MeanFields} implies that the baryon superfluid phase is triggered by the onset of nonzero $\Delta$ at a critical chemical potential $\mu_q=\mu_q^{\rm cr}$ with
\begin{eqnarray}
\mu_q^{\rm cr} \equiv m_\pi^{(\rm H)}/2\ , \label{MuCritical}
\end{eqnarray}
which is irrespective of whether or not the spin-$1$ hadrons are present. This critical value is universal in the sense that it is also derived from chiral effective models involving only spin-$0$ hadrons~\cite{Kogut:1999iv,Kogut:2000ek,Ratti:2004ra,Suenaga:2022uqn}. At the same time as the onset, both $\bar{V}$ and $\bar{\omega}$ begin to acquire nonzero values. Detailed analyses in the vicinity of the phase transition will be provided in Sec.~\ref{sec:PhaseTransition}. At asymptotically high $\mu_q$, $\sigma_0$ and $\bar{V}$ vanish. Meanwhile, $\Delta$ converges to a constant (see the arrow in Fig.~\ref{fig:MeanFields}) whose value can be evaluated as $\Delta/\sigma_0^{(\rm H)} \approx 0.34\big(m_1/\sigma_0^{(\rm H)}\big)=1.74$ for the present values of $C$ and $C_3$, whereas $\bar{\omega}$ grows in the negative direction with a power of $\mu_q$.
 
Here, we note that the onsets of nonzero $\bar{\omega}$ and $\bar{V}$ do not necessarily coincide with the critical chemical potential~(\ref{MuCritical}), since those mean fields are unphysical, in other words, they are gauge-dependent quantities within the gauge-field description. In fact, the mean field of $\omega$ was found to be proportional to $\mu_q$ in the hadronic phase within the HLS formalism~\cite{Harada:2010vy}.

\begin{figure*}[t]
\centering
\hspace*{-0.5cm} 
\includegraphics*[scale=0.09]{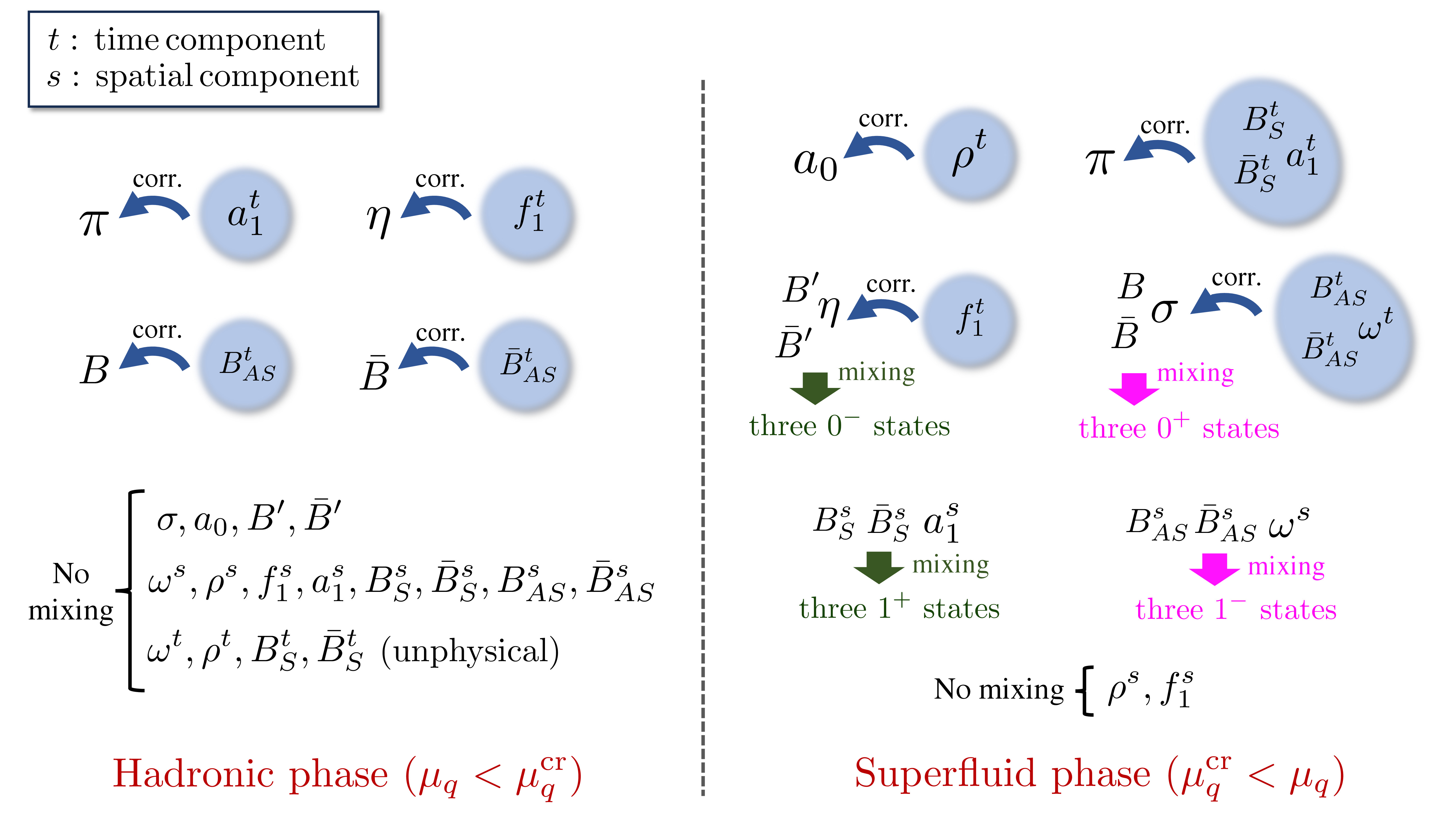}
\caption{Schematic picture of corrections to the hadron masses from various mixings. }
\label{fig:Schematic}
\end{figure*}

The converging behavior of $\Delta$ is distinct from the result within the conventional LSM in the absence of spin-$1$ hadrons~\cite{Suenaga:2022uqn}. To take a closer look at this difference, we depict the $\mu_q$ dependence of $\sigma_0$ and $\Delta$ with and without the spin-$0$ and spin-$1$ mixing effect, $C=12$ and $C=0$, in the top panel of Fig.~\ref{fig:MixMF}. The figure indicates that $\sigma_0$ does not have its $\mu_q$ dependence corrected considerably by the mixing while $\Delta$ has its $\mu_q$ dependence significantly modified by the mixing; the asymptotically converging behavior of $\Delta$ is induced only when the spin-$0$ and spin-$1$ mixing takes effect. In fact, the asymptotic constant value of $\Delta$ with the mixing effect is essentially determined by the bare mass of the spin-$1$ hadrons, $m_1$, together with a factor stemming from the mixing strength $C$ and $C_3$. The diverging behavior of $\Delta$ in the absence of the mixing is lost, but instead the negatively diverging growth of $\bar{\omega}$ appears as indicated in Fig.~\ref{fig:MeanFields}. We note that spin-$1$ mean fields $\bar{\omega}$ and $\bar{V}$ are always zero when $C=0$:
\begin{eqnarray}
\bar{\omega} = 0\ \ {\rm and}\ \ \bar{V}=0 \  \ \ ({\rm at\ any}\ \mu_q\ {\rm for}\ C=0)\ ,
\end{eqnarray}
as will be argued in Sec.~\ref{sec:PhaseTransition}. We also note that, when we take $C$ to be negative, the signs of the induced $\bar{\omega}$ and $\bar{V}$ in the superfluid phase become positive and negative, respectively.

The significant corrections due to the spin-$0$ and spin-$1$ mixing effect are also reflected by $\mu_q$ dependence of the quark-number density as displayed in the bottom panel of Fig.~\ref{fig:MixMF}. In this figure we have plotted the normalized density
\begin{eqnarray}
\tilde{n}_q = \frac{n_q}{16f_\pi^2 m_\pi^{(\rm H)}}\ ,
\end{eqnarray}
where $n_q$ is the ordinary quark-number density given by
\begin{eqnarray}
n_q &=& \frac{\partial{\cal L}^{\rm red.}_{\rm eLSM}}{\partial\mu_q}\Big|_{\rm mean\, field} \nonumber\\
&=& 4\Delta^2\mu_q + \frac{C\Delta}{\sqrt{2}}(\Delta\bar{\omega}-\bar{V}\sigma_0)
\end{eqnarray}
with the mean fields~(\ref{MeanFields}), and $f_\pi=\sigma_0^{\rm (H)}/\sqrt{2}$ is the pion decay constant. The figure shows that the asymptotic growth that is proportional to $\mu_q^3$ as can be derived in the absence of the mixing is changed into the $m_1^2\mu_q$ dependence due to the mixing. As a result, increment in the density gets milder. It should be noted that the Silver-Blaze property for the quark-number density is obvious since $n_q$ is proportional to $\Delta$.\footnote{Although the present eLSM predicts $n_q\propto \mu_q$ for larger $\mu_q$, at some point, such a hadronic description would be violated and quark matter would appear, which leads eventually to $n_q\propto\mu_q^3$.}

Before moving on to evaluation of the hadron masses at finite $\mu_q$, we give comments on roles of the spin-$1$ mean fields $\bar{\omega}$ and $\bar{V}$ in controlling the onset of the baryon superfluid phase. As indicated in Fig.~\ref{fig:MeanFields}, the superfluid phase emerges once $\mu_q$ reaches the critical chemical potential~(\ref{MuCritical}), which is exactly what lattice simulations suggest~\cite{Hands:2001ee,Braguta:2016cpw,Iida:2019rah}. Note, however, that this is only when we include all the four mean fields: $\sigma_0$, $\Delta$, $\bar{\omega}$, and $\bar{V}$. If any of the mean fields were dropped, the critical chemical potential would not coincide with Eq.~(\ref{MuCritical}) in the presence of the spin-$0$ and spin-$1$ mixing effect. We have checked this property by choosing various parameter sets.\footnote{In Sec.~\ref{sec:PhaseTransition}, we, indeed, analytically prove that Eq.~(\ref{MuCritical}) holds as long as all of $\sigma_0$, $\Delta$, $\bar{\omega}$, and $\bar{V}$ are included based on a certain assumption on their critical behaviors.} In fact, if $\bar{\omega}$ or $\bar{V}$ is neglected while keeping $C=12$, then the critical chemical potential is found to change to approximately $0.379m_\pi^{(\rm H)}$, which is lower than $\mu_q^{\rm cr} = m_\pi^{\rm (H)}/2$ suggested by lattice simulations. Accordingly, one can show that the pion mass in the superfluid phase is not given by $m_\pi=2\mu_q$ that other chiral effective models commonly predict~\cite{Kogut:1999iv,Kogut:2000ek,Ratti:2004ra,Suenaga:2022uqn}. Furthermore, the Nambu-Goldstone (NG) boson associated with the breakdown of $U(1)_B$ symmetry does not emerge in this case. From these observations, we conclude that $\bar{\omega}$ and $\bar{V}$ as well as $\sigma_0$ and $\Delta$ play important roles in cold and dense QC$_2$D matter when the spin-$0$ and spin-$1$ mixing effect is present.

\section{Mass spectrum}
\label{sec:MassSpectrum}

In this section we examine a mass spectrum of the spin-$1$ hadrons at finite $\mu_q$ by expanding the Lagrangian~(\ref{ELSMRed}) {on top of the mean fields obtained in the previous section}.

\subsection{General properties}
\label{sec:Analytic}

Before showing numerical results for the $\mu_q$ dependence of the spin-$1$ hadron masses, we start with general properties of such hadrons in a medium. In what follows, we consider the rest frame of the medium.

The quantum numbers carried by (axial)vector mesons are identical to those of (pseudo)scalar mesons with a derivative, e.g., $a_1^{a,\mu}$ and $\partial^{\mu}\pi^a$, so that such two kinds of mesons can mix with each other even in the hadronic phase with nonzero $C$. Similarly, (anti)baryons can mix with the corresponding spin-$1$ (anti)baryons through a derivative. In the rest frame of the hadronic medium, therefore, the following four mixings appear:
\begin{eqnarray}
(\partial_0\pi, a_1^t)\,  ,\    (\partial_0\eta, f_1^t) \, , \  (\partial_0B, B_{AS}^t)\,  ,\  (\partial_0\bar{B}, \bar{B}_{AS}^t) \ , \label{MixingHadronic}
\end{eqnarray}
where each bracket represents the mixing partners and the isospin indices are suppressed for simplicity. Here, the superscript ``$t$'' stands for the time component ($\mu=0$) of the respective spin-$1$ hadrons. The time components of the spin-$1$ hadrons are unphysical, so the mixings in Eq.~(\ref{MixingHadronic}) only lead to modifications of $\pi$, $\eta$, $B$, and $\bar{B}$, while the physical components, i.e., the spatial components of the spin-$1$ hadrons remain unaffected. We note that the remaining spin-$0$ hadrons: $\sigma$, $a_0$, $B'$, and $\bar{B}'$, are not contaminated by any mixing.

Meanwhile, in the baryon superfluid phase the mixings get more complicated due to the $U(1)_B$ baryon-number violation as well as the creation of a baryonic medium. Taking into account the unbroken $SU(2)_I$ isospin symmetry, one can see that the following six mixings are possible:
\begin{eqnarray}
&& (\partial_0 a_0, \rho^t)\ , \nonumber\\
&& (\partial_0 B', \partial_0 \bar{B}' , \partial_0\eta, f_1^t)\ , \nonumber\\
&&  (\partial_0 \pi,B_S^t,\bar{B}_S^t, a_1^t)\, , \ (B_S^s,\bar{B}_S^s, a_1^s) \nonumber\\
&& (\partial_0 B, \partial_0 \bar{B}, \partial_0\sigma, B_{AS}^t, \bar{B}^t_{AS}, \omega^t)\, , \ (B_{AS}^s,\bar{B}_{AS}^s, \omega^s)\ , \nonumber\\
\label{MixingSuper}
\end{eqnarray}
where each bracket again represents the mixing partners. That is, all of the spin-$0$ hadronic states are corrected by the corresponding unphysical spin-$1$ states. Besides, the physical components of $B_S, \bar{B}_S, a_1$ and of $B_{AS}, \bar{B}_{AS}, \omega$ are also corrected by the mixings.
A schematic picture of the mixings among the hadrons both in the hadronic and superfluid phases are shown in Fig.~\ref{fig:Schematic}.

In the present paper, the hadron masses are evaluated at tree level in the presence of the mean fields~(\ref{MeanFields}). The analysis is straightforward but is complicated and lengthy since, particularly in the superfluid phase, we need to find pole positions of the respective propagator matrix for the mixed states in Eq.~(\ref{MixingSuper}). For this reason, we leave the detailed procedure to compute the hadron masses to Appendix~\ref{sec:MassFormulas}.

\begin{figure*}[t]
  \begin{center}
    \begin{tabular}{cc}

      \begin{minipage}[c]{0.5\hsize}
       \centering
       \hspace*{-2.5cm} 
         \includegraphics*[scale=0.08]{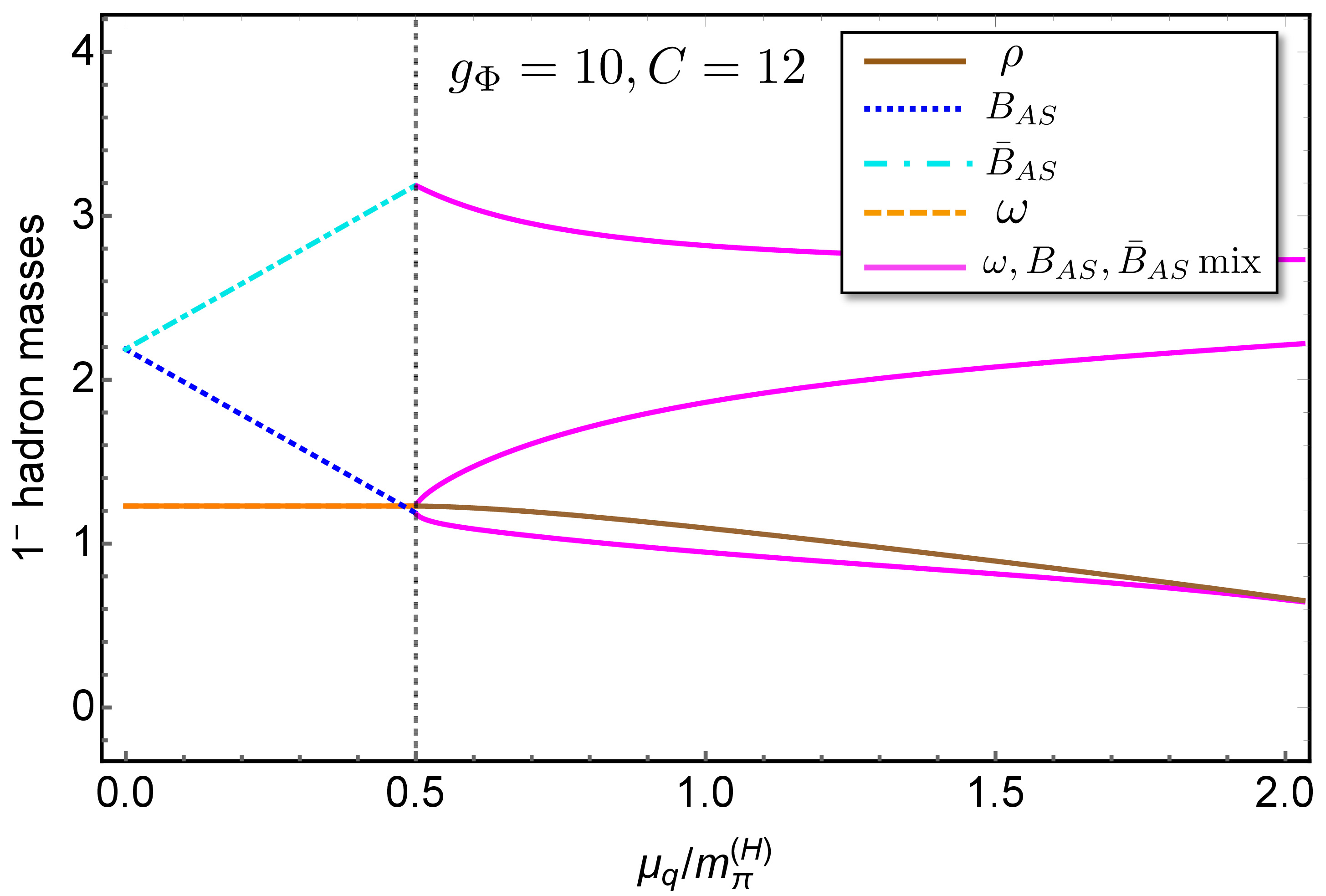}\\
         \end{minipage}

      \begin{minipage}[c]{0.4\hsize}
       \centering
        \hspace*{-1.1cm} 
          \includegraphics*[scale=0.08]{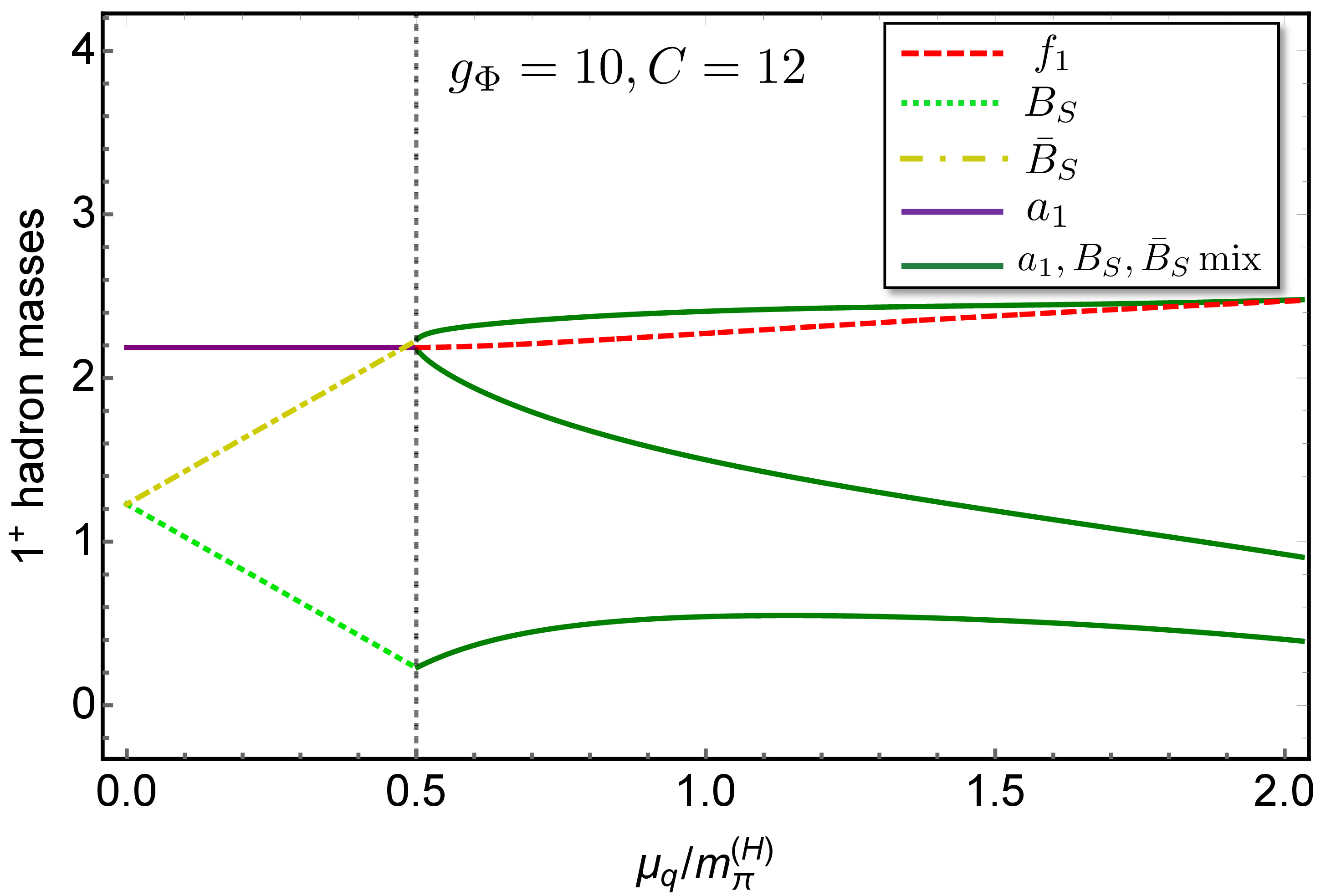}\\
      \end{minipage}

      \end{tabular}
 \caption{$\mu_q$ dependence of the masses of negative-parity (left) and positive-parity (right) spin-$1$ hadrons for $(g_\Phi,C)=(10,12)$. In this figure the masses are normalized by $m_\pi^{(\rm H)}$.} 
\label{fig:Spin1_G10_C12}
  \end{center}
\end{figure*}

\begin{figure*}[t]
  \begin{center}
    \begin{tabular}{cc}

      \begin{minipage}[c]{0.5\hsize}
       \centering
       \hspace*{-2.5cm} 
         \includegraphics*[scale=0.08]{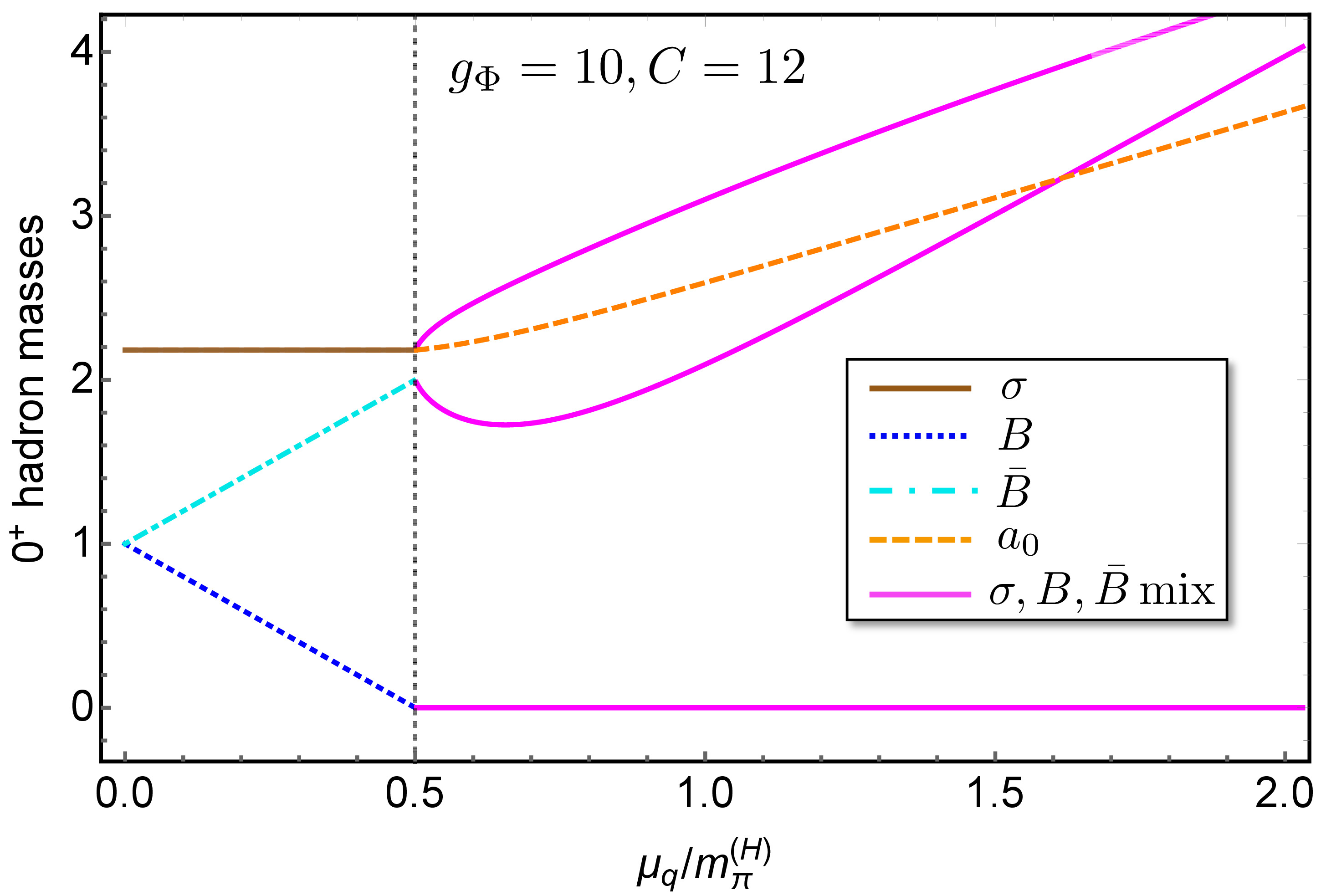}\\
         \end{minipage}

      \begin{minipage}[c]{0.4\hsize}
       \centering
        \hspace*{-1.1cm} 
          \includegraphics*[scale=0.08]{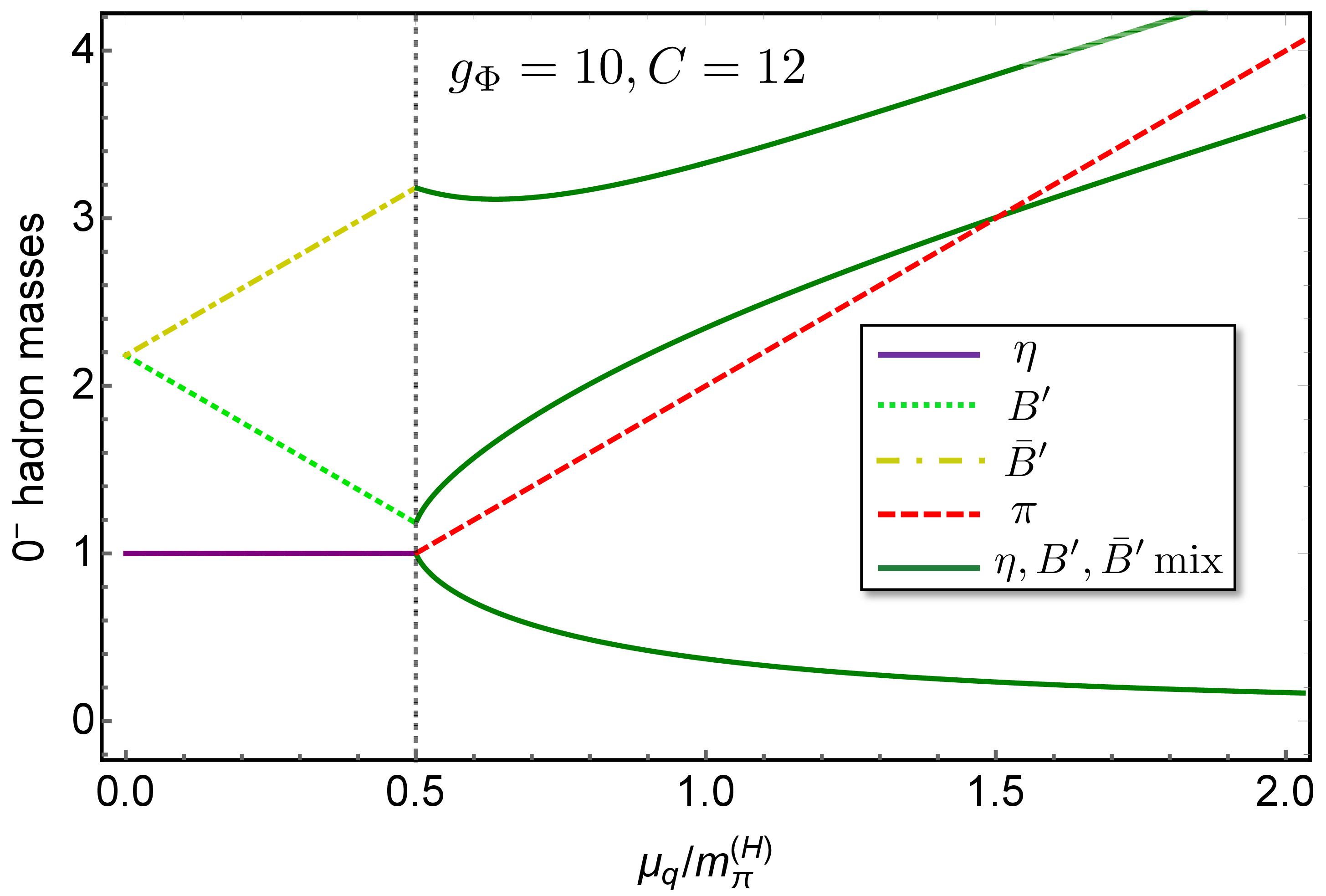}\\
      \end{minipage}

      \end{tabular}
 \caption{$\mu_q$ dependence of the masses of positive-parity (left) and negative-parity (right) spin-$0$ hadrons for $(g_\Phi,C)=(10,12)$. In this figure the masses are normalized by $m_\pi^{(\rm H)}$.} 
\label{fig:Spin0_G10_C12}
  \end{center}
\end{figure*}

\begin{figure*}[t]
  \begin{center}
    \begin{tabular}{cc}

      \begin{minipage}[c]{0.5\hsize}
       \centering
       \hspace*{-2.5cm} 
         \includegraphics*[scale=0.08]{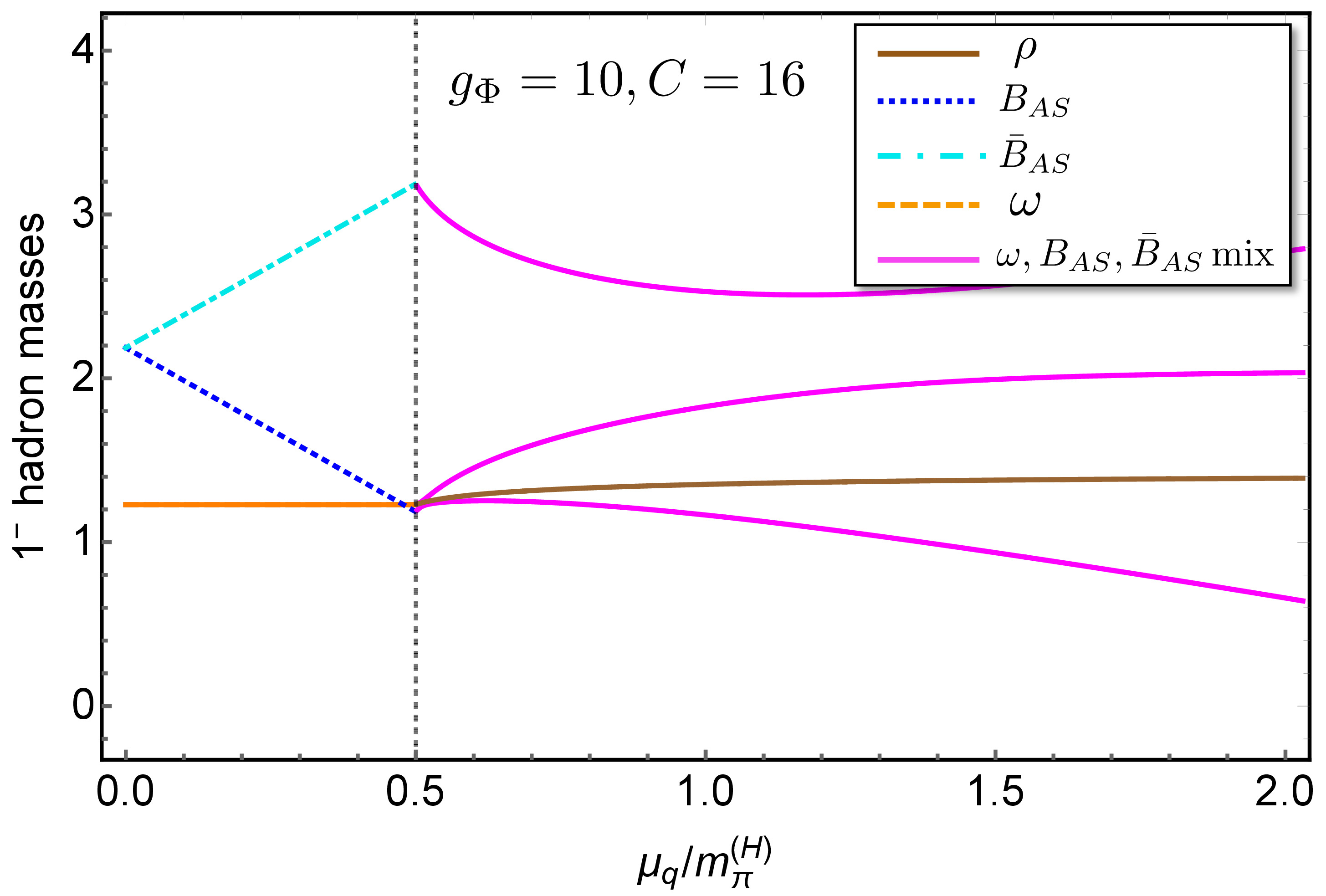}\\
         \end{minipage}

      \begin{minipage}[c]{0.4\hsize}
       \centering
        \hspace*{-1.1cm} 
          \includegraphics*[scale=0.08]{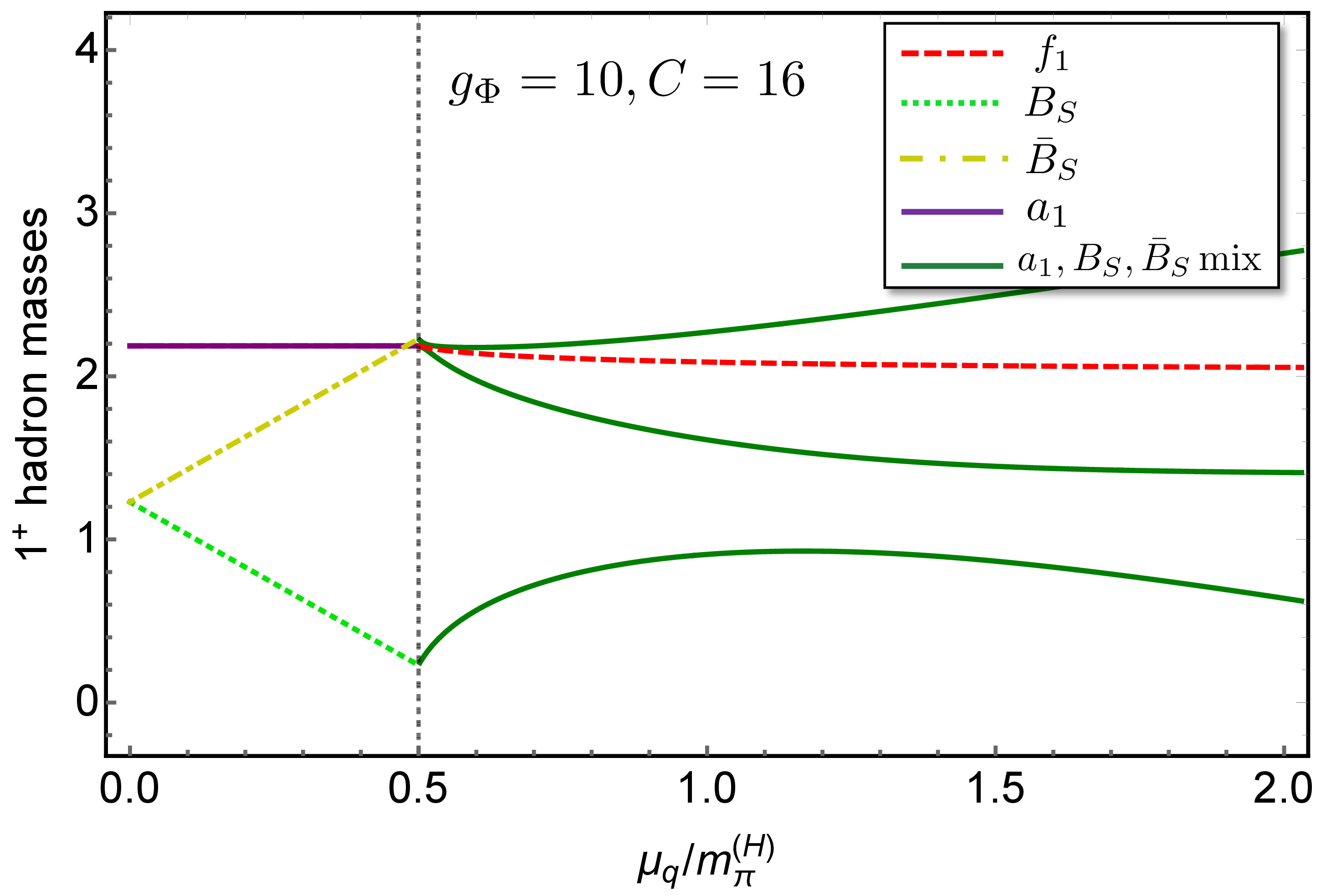}\\
      \end{minipage}

      \end{tabular}
 \caption{Same as Fig.~\ref{fig:Spin1_G10_C12} but for $(g_\Phi,C)=(10,16)$. } 
\label{fig:Spin1_G10_C16}
  \end{center}
\end{figure*}

\begin{figure*}[t]
  \begin{center}
    \begin{tabular}{cc}

      \begin{minipage}[c]{0.5\hsize}
       \centering
       \hspace*{-2.5cm} 
         \includegraphics*[scale=0.08]{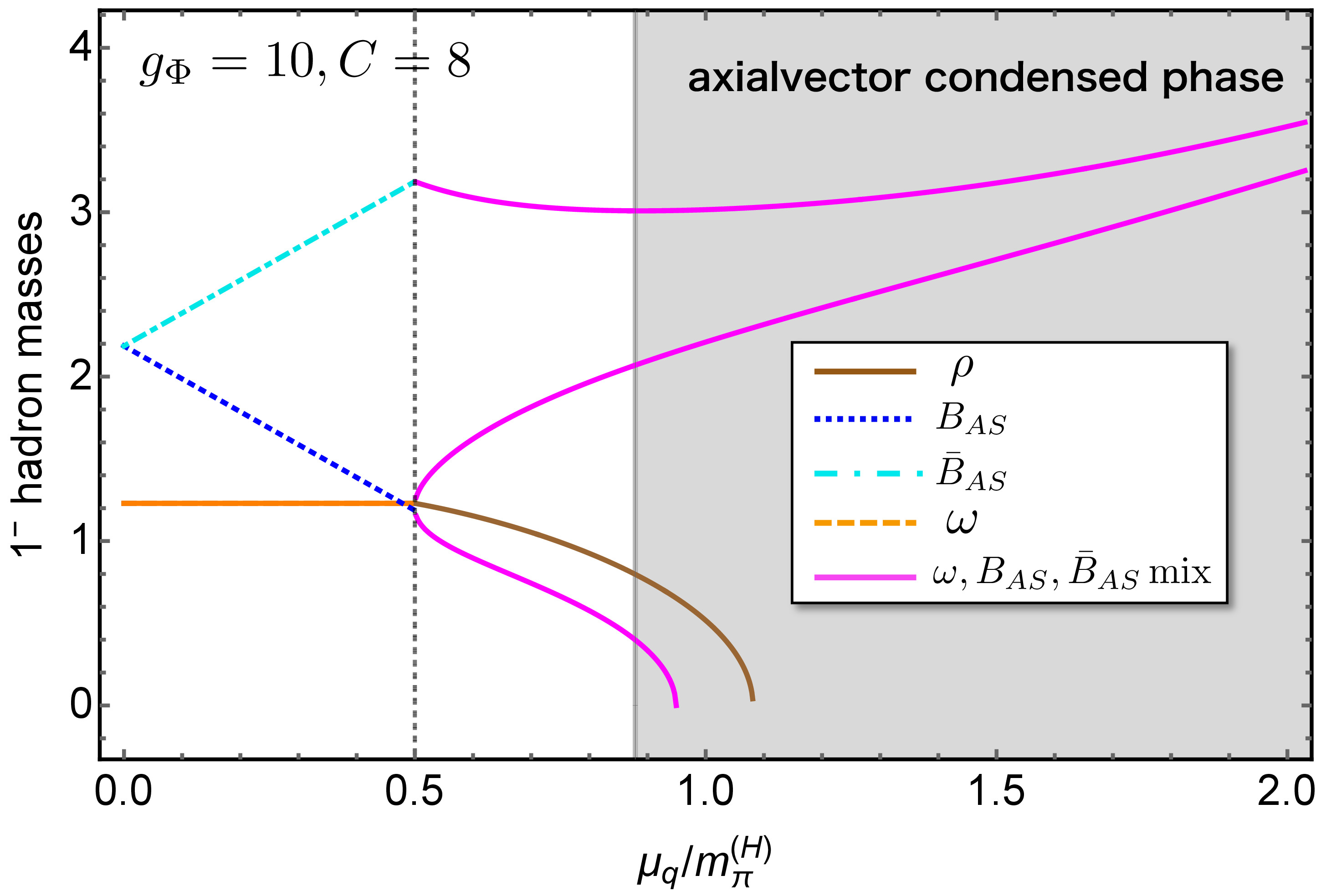}\\
         \end{minipage}

      \begin{minipage}[c]{0.4\hsize}
       \centering
        \hspace*{-1.1cm} 
          \includegraphics*[scale=0.08]{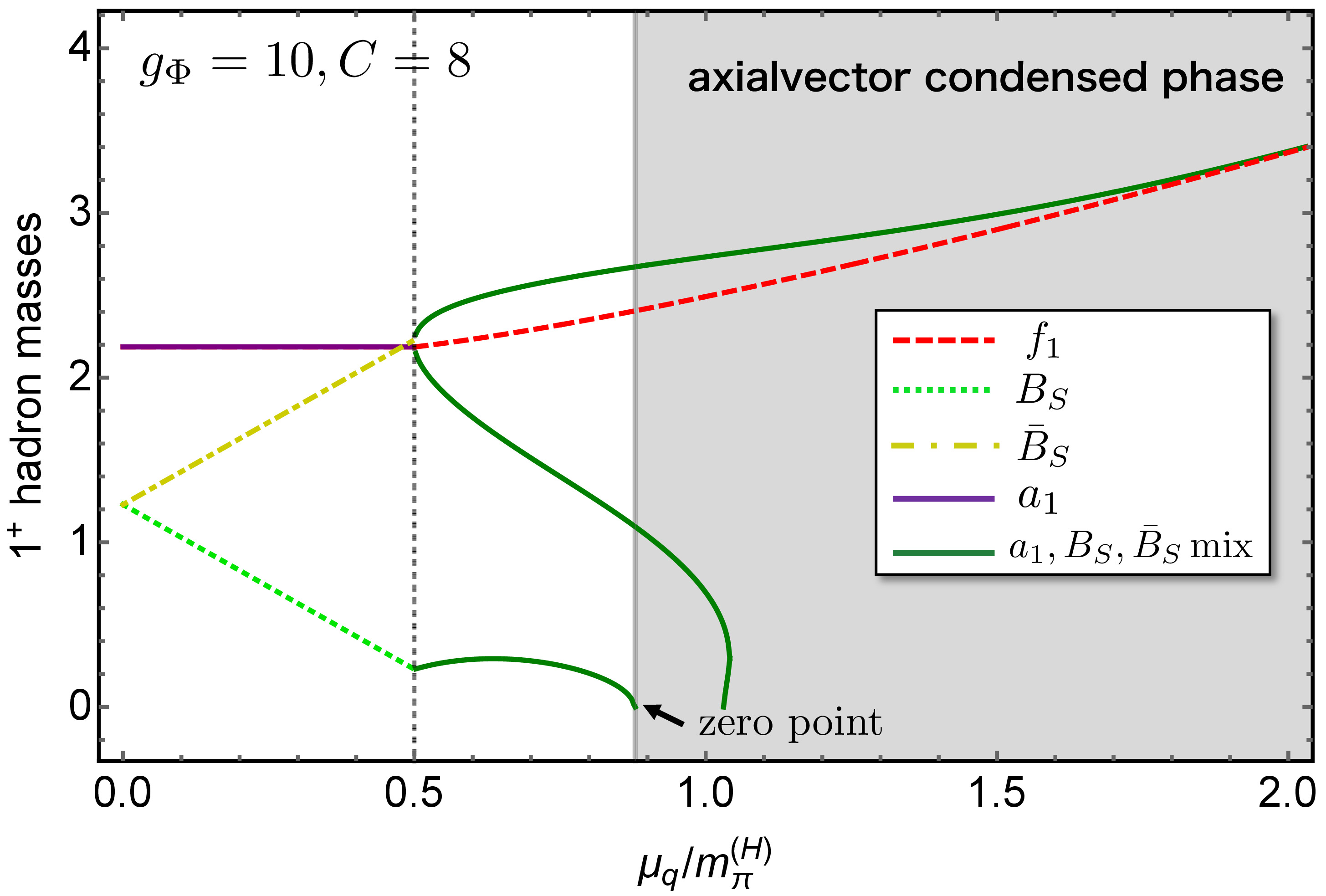}\\
      \end{minipage}

      \end{tabular}
 \caption{Same as Fig.~\ref{fig:Spin1_G10_C12} but for $(g_\Phi,C)=(10,8)$. } 
\label{fig:Spin1_G10_C8}
  \end{center}
\end{figure*}

\subsection{Numerical results {\bf for} the hadron masses at {\bf nonzero} $\mu_q$}
\label{sec:Numerical}

In this subsection, based on the expressions derived in Appendix~\ref{sec:MassFormulas} we numerically elucidate $\mu_q$ dependence of the masses of spin-$1$ hadrons. As for the free parameters $g_\Phi$ and $C$, we first take $(g_\Phi,C)=(10,12)$ to draw a typical prediction of the mass spectrum of the spin-$1$ hadrons. Next, we vary the value of $C$ {with $g_\Phi$ kept fixed to unveil how the spin-$0$ and spin-$1$ mixing affects the mass spectrum.  More concretely we take $(g_\Phi,C)=(10,16)$ and $(g_\Phi,C)=(10,8)$. Finally, we also choose $(g_\Phi,C)=(5,12)$ to see influence of the coupling $g_\Phi$ on the mass spectrum.

Depicted in Fig.~\ref{fig:Spin1_G10_C12} is the resultant $\mu_q$ dependence of the spin-$1$ hadron masses at $(g_\Phi,C)=(10,12)$. As can be seen from this figure, at vanishing chemical potential, the masses of $\omega$, $\rho$, $B_{S}$, and $\bar{B}_S$ degenerate and so do those of $f_1$, $a_1$, $B_{AS}$, and $\bar{B}_{AS}$. 
In the hadronic phase, as $\mu_q$ increases, all the meson masses do not change and the (anti)baryon masses simply modified linearly. These stable behaviors are reminiscent of the Silver-Blaze property. Here, the slopes of the linear decrement (increment) of diquark (antidiquark) baryons can be understood by their baryon numbers. More explicitly, their mass formulas are given by Eqs.~(\ref{MassBSHad}) and~(\ref{MassBASHad}). In the baryon superfluid phase, on the other hand, due to the $U(1)_B$ baryon-number violation, $\omega$ - $B_{AS}$ - $\bar{B}_{AS}$ mixing and $a_1$ - $B_S$ - $\bar{B}_S$ mixing take place, leading to non-monotonic $\mu_q$ dependence of the masses of the resulting mixed states. Most remarkably, the $a_1$ - $B_S$ - $\bar{B}_S$ mixing has been observed by the recent lattice simulation~\cite{Murakami2022}. Meanwhile, the $\rho$ and $f_1$ mesons are not contaminated by any mixing so that those masses depend on $\mu_q$ fairly monotonically. We note that the present parameter set yields a slight reduction of the $\rho$ meson mass in the superfluid phase, which is consistent with the lattice data~\cite{Murakami:2022lmq,Murakami2022}.

In the presence of the spin-$0$ and spin-$1$ mixing effect, it is worth examining the mass spectrum of the spin-$0$ hadrons in addition to that of the spin-$1$ hadrons as depicted in Fig.~\ref{fig:Spin1_G10_C12}. In Fig.~\ref{fig:Spin0_G10_C12}, we thus draw the resultant $\mu_q$ dependence of the spin-$0$ hadron masses at $(g_\Phi,C)=(10,12)$. In this figure, the hadron masses are again shown to depend on $\mu_q$ monotonically in the hadronic phase, which is consistent with the Silver-Blaze property.
Besides, in the superfluid phase, the left panel indicates that a massless mode emerges in the $\sigma$ - $B$ - $\bar{B}$ mixed state. This mode corresponds to a NG boson associated with the spontaneous breakdown of $U(1)_B$ baryon-number symmetry. Moreover, from the right panel, the pion mass is found to increase linearly in the superfluid phase; this numerical result can be reproduced by a simple formula,
\begin{eqnarray}
m_\pi = 2\mu_q\ ,
\end{eqnarray}
which is consistent with other chiral models~\cite{Kogut:1999iv,Kogut:2000ek,Ratti:2004ra,Suenaga:2022uqn}. We again emphasize that all these reasonable results stem from the present correct treatment of the four mean fields~(\ref{MeanFields}) within the eLSM.

\begin{figure}[t]
\centering
\hspace*{-0.5cm} 
\includegraphics*[scale=0.07]{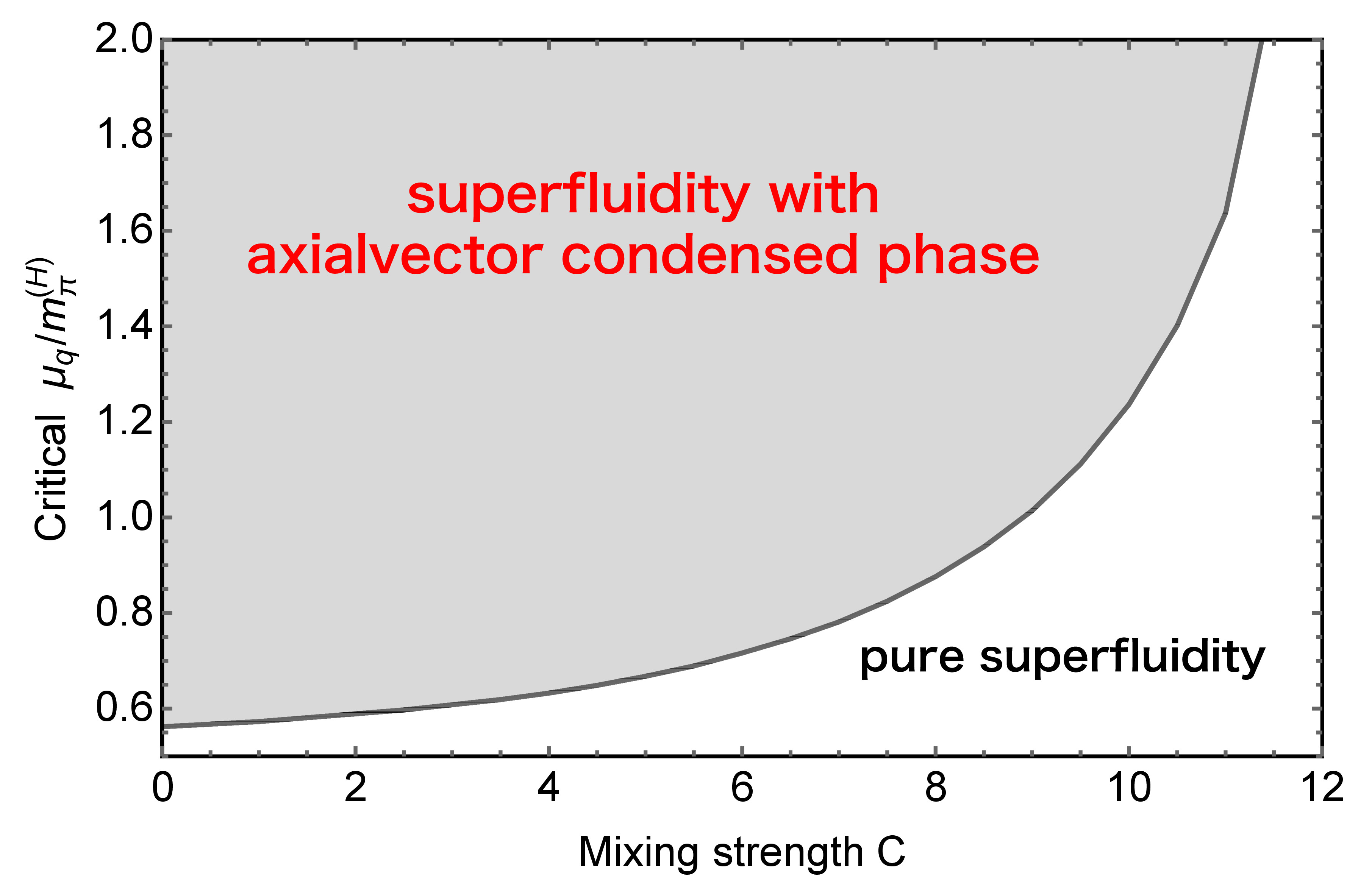}
\caption{Critical chemical potential for the axialvector condensation as a function of mixing strength $C$. We take $g_\Phi=10$.}
\label{fig:AVCritical}
\end{figure}

\begin{figure*}[t]
  \begin{center}
    \begin{tabular}{cc}

      \begin{minipage}[c]{0.5\hsize}
       \centering
       \hspace*{-2.5cm} 
         \includegraphics*[scale=0.08]{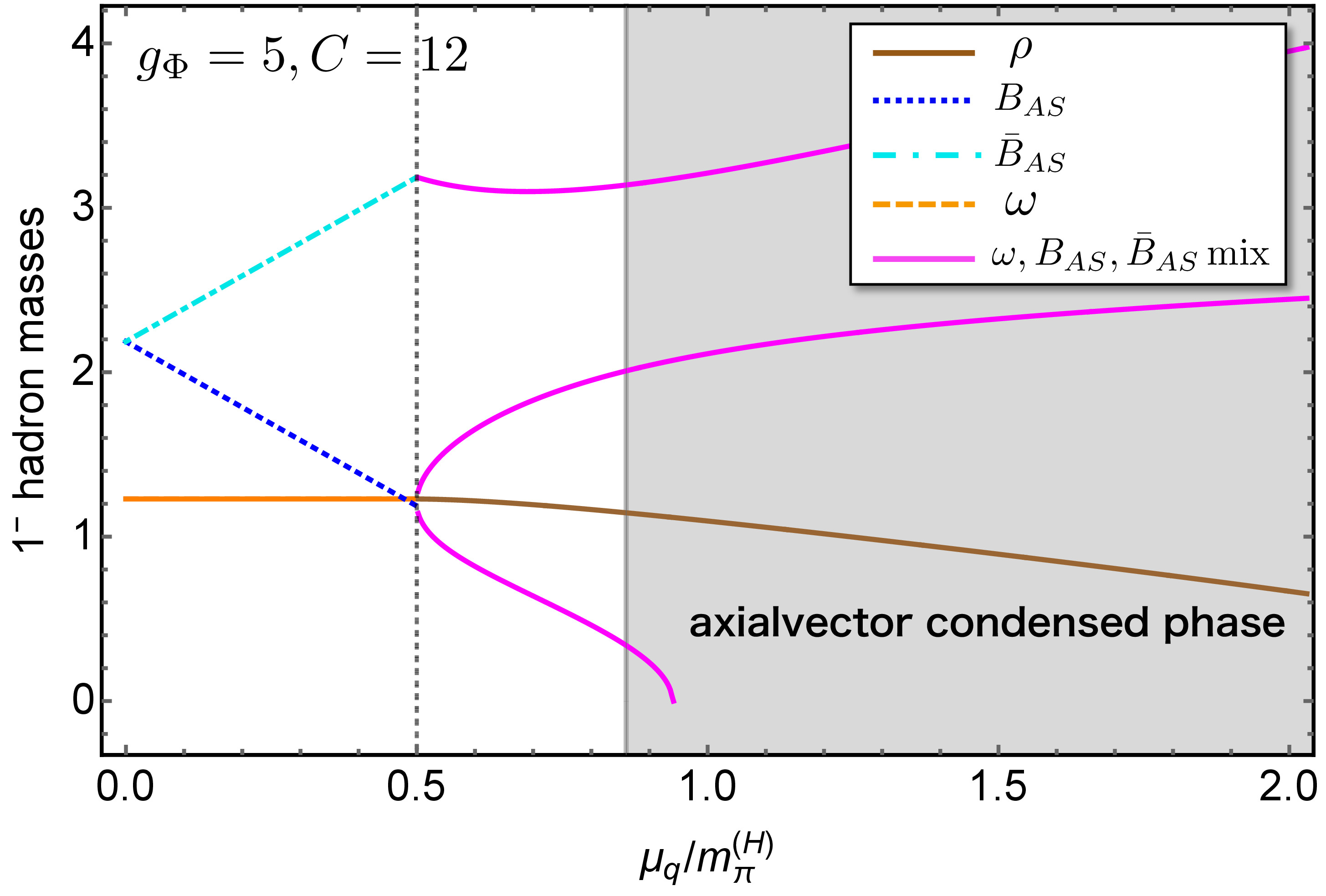}\\
         \end{minipage}

      \begin{minipage}[c]{0.4\hsize}
       \centering
        \hspace*{-1.1cm} 
          \includegraphics*[scale=0.08]{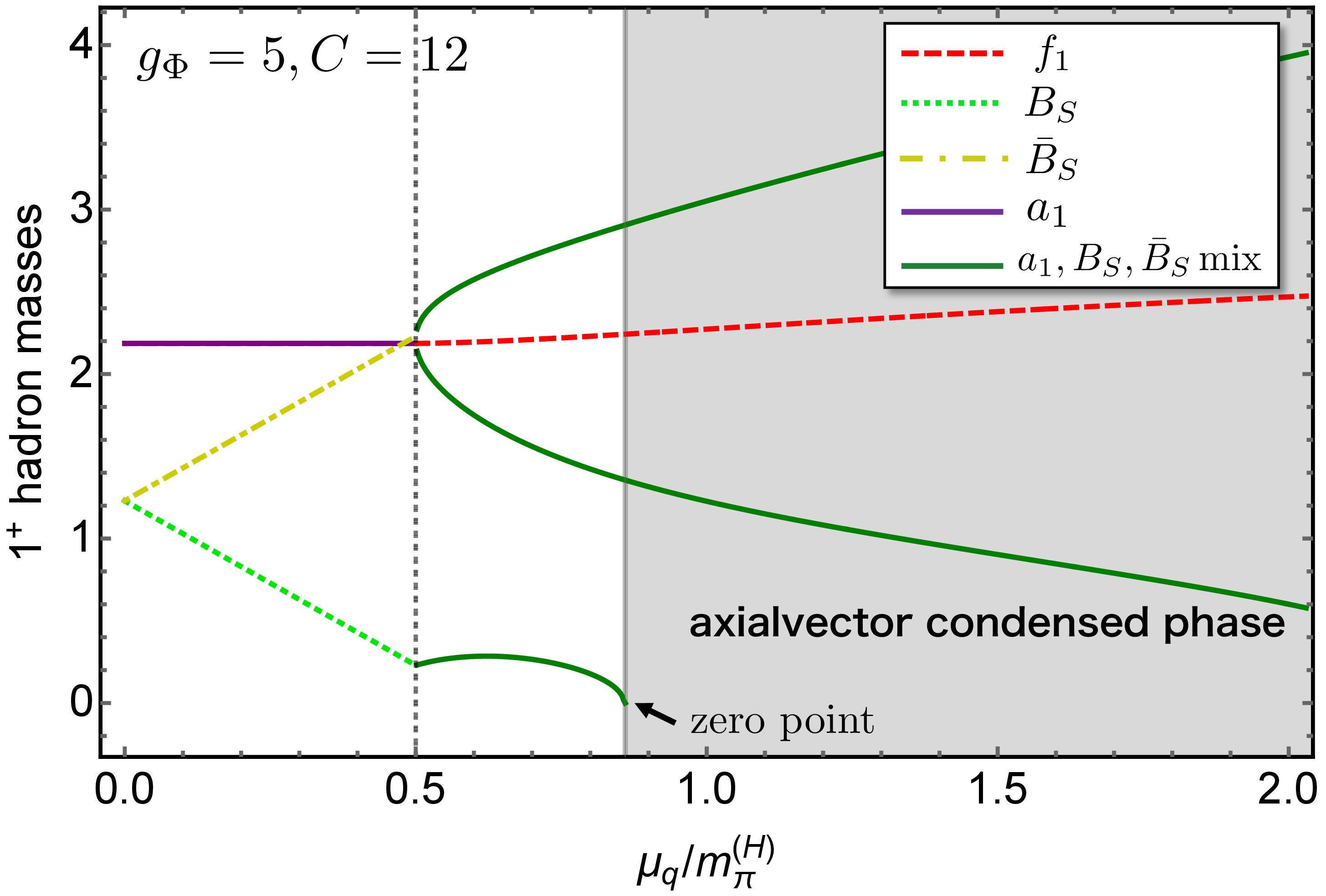}\\
      \end{minipage}
      \end{tabular}
 \caption{Same as Fig.~\ref{fig:Spin1_G10_C12} but for $(g_\Phi,C)=(5,12)$. } 
\label{fig:Spin1_G5_C12}
  \end{center}
\end{figure*}

\subsection{$C$ dependence of the mass spectrum}
\label{sec:CDependence}

The mass spectrum of the spin-$1$ hadrons presented in Fig.~\ref{fig:Spin1_G10_C12} is just a typical example.  In this case,
the value of $C$ is fixed such that the slight reduction of the $\rho$ meson mass in the superfluid phase suggested by lattice simulations is successfully reproduced. Next, we change the value of $C$ while keeping $g_\Phi=10$. When we take $C$ to be larger, e.g., $C=16$, the $\mu_q$ dependence of the spin-$1$ hadron masses is obtained as depicted in Fig.~\ref{fig:Spin1_G10_C16}. From this figure one can find that the mass spectrum in the hadronic phase is identical to the one with $(g_\Phi,C)=(10,12)$, Fig.~\ref{fig:Spin1_G10_C12}, as long as all the other parameters are the same. In the superfluid phase, the $\rho$ meson mass slightly increases with $\mu_q$, which is clearly different from the lattice result. This suggests that the relevant value of the parameter $C$ that controls the magnitude of the spin-$0$ and spin-$1$ mixing effect could not be so large. We note that, since the renormalization factor $Z_\pi$ (or $Z_\eta$), Eq.~(\ref{ZDef}), must be real, the inputs presented
in Sec.\ \ref{sec:Inputs} already constrains $C$ as $C\lesssim18.3$.

When we take $C$ to be smaller, e.g., $C=8$, the mass spectrum is evaluated as in Fig.~\ref{fig:Spin1_G10_C8}. The right panel indicates that, in this parameter choice, the mass of the lowest-lying $a_1$ - $B_{S}$ - $\bar{B}_{S}$ mixed state reaches zero at $\mu_q\approx 0.88 m_\pi^{(\rm H)}$, as the chemical potential increases in the superfluid phase. Since this mode includes an iso-triplet axialvector component, it is reasonable to argue that above this chemical potential, an axialvector condensed phase where $SU(2)_I$ isospin symmetry is broken emerges on top of the baryon superfluidity~\cite{Lenaghan:2001sd}, as exhibited by the shaded area in Fig.~\ref{fig:Spin1_G10_C8}. Thus, the true mass spectrum in this phase is obscure although we have still plotted the numerical result. For a self-consistent analysis, however, it would be necessary to include another mean field that is responsible for the axialvector condensed phase. 

Above the critical chemical potential $\mu_q\approx 0.88 m_\pi^{(\rm H)}$, the mass of the lowest-lying $\omega$ - $B_{AS}$ - $\bar{B}_{AS}$ mixed state also converges to zero at $\mu_q\approx 0.95m_\pi^{(\rm H)}$ as indicated in the left panel of Fig.~\ref{fig:Spin1_G10_C8}. Besides, the $\rho$ meson mass also becomes zero at $\mu_q\approx1.1m_\pi^{(\rm H)}$. These critical chemical potentials lie in the axialvector condensed phase, so that more precise determination of their values would require extension of the present exploratory analysis. From those findings, however, at least one could expect the existence of the vector condensed phase. We note that the axialvector condensation occurs prior to the vector condensation, reflecting the fact that the $B_S$ mass is invariably lighter than the $B_{AS}$ one at $\mu_q=\mu_q^{\rm cr}$, since $B_S$ is an $S$-wave state.

From the above analysis, one can infer that when $C$ is small enough, the appearance of the axialvector (and vector) condensate in the superfluid phase is favored. To see this tendency more clearly, in Fig.~\ref{fig:AVCritical}, we plot the critical chemical potential for the appearance of the axialvector condensed phase as a function of the mixing strength $C$.  In this figure, the pure baryon superfluid phase lies below the curve, while in the above shaded area the axialvector condensate emerges in the superfluid phase. The figure indeed indicates that the smaller value of $C$ triggers the axialvector condensation at lower $\mu_q$. In other words, the spin-$0$ and spin-$1$ mixing controlled by $C$ acts as a stabilizer to avoid emergence of the axialvector condensate, i.e., onset of the Bose-Einstein condensation of parity-even spin-$1$ hadrons, in the low density regime. We note that, even when we take a value of $C$ as large as possible, we find the critical chemical potential for the axialvector condensation at a certain value, which could be too high for the present model to be valid.

\begin{figure}[t]
\centering
\hspace*{-0.3cm} 
\includegraphics*[scale=0.07]{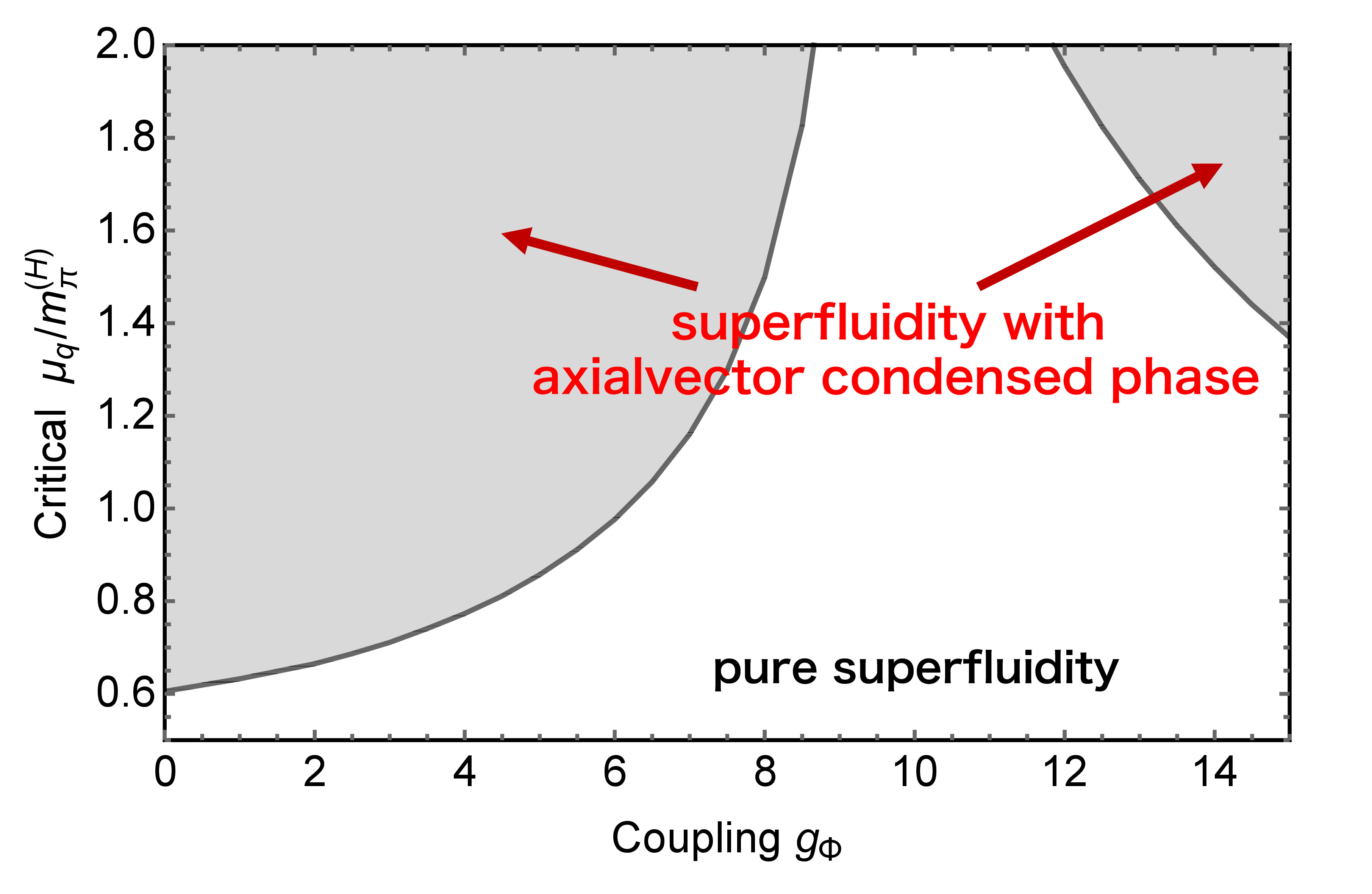}
\caption{Critical chemical potential for the axialvector condensation as a function of the coupling $ g_\Phi$. We take $C=12$.}
\label{fig:AVCritical2}
\end{figure}

\subsection{$g_\Phi$ dependence of the mass spectrum}
\label{sec:gPhiDependence}

Thus far we have only varied the value of $C$, while keeping $g_\Phi=10$, to focus on the spin-$0$ and spin-$1$ mixing effects on the hadron mass spectrum. Let us now examine effects of the coupling $g_\Phi$. When we employ $(g_\Phi,C)=(5,12)$, the mass spectrum is obtained as in Fig.~\ref{fig:Spin1_G5_C12}. This figure exhibits appearance of the axial vector condensate and possibly of the vector condensate similarly to Fig.~\ref{fig:Spin1_G10_C8}. One obvious distinction is the qualitative behavior of the $\rho$ meson mass: The $\mu_q$ dependence of the $\rho$ meson mass in Fig.~\ref{fig:Spin1_G5_C12} does not change from that in Fig.~\ref{fig:Spin1_G10_C12} in the sense that both show the same gradual decrease with $\mu_q$, whereas the $\rho$ meson mass in Fig.~\ref{fig:Spin1_G10_C8} decreases rapidly to zero.
This characteristic behavior can be understood by the fact that the $\rho$ meson mass has no dependence on $g_\Phi$, as shown in Eq.~(\ref{MRhoS}). Thus, $g_\Phi$ plays a role in changing the $\mu_q$ dependence of the $\omega$ - $B_{AS}$ - $\bar{B}_{AS}$ mixed states and the $a_1$ - $B_{S}$ - $\bar{B}_{S}$ mixed states, particularly the lowest one for each. Detailed consideration of th $\rho$ meson mass in the superfluid phase will be done in Sec.~\ref{sec:RhoMassRed}.

Depicted in Fig.~\ref{fig:AVCritical2} is the $g_\Phi$ dependence of the critical chemical potential for the axialvector condensation. In this figure the axialvector condensed phase is indicated by the shaded area. Figure~\ref{fig:AVCritical2} implies that the smaller value of $g_\Phi$ leads to the appearance of the axialvector condensate at lower $\mu_q$. Moreover, one can see that reentrant axialvector condensation occurs in a regime of large $g_\Phi$ in such a way that the intervening pure superfluid region shrinks with increasing $\mu_q$.

\subsection{Signs of $C$ and $g_\Phi$}
\label{sec:Sign}

We conclude this section by giving comments on the signs of $C$ and $g_\Phi$. Throughout the above numerical analysis, we have assumed $C>0$ and $g_\Phi>0$. When the signs are taken to be $C<0$ and $g_\Phi<0$, we can obtain qualitatively similar results for the mass spectrum although the detailed numerical values are slightly changed. For instance, a negatively larger value of $C$ acts to prevent the axialvector condensation and possible vector condensation from occurring in the superfluid phase. 
On the other hand, when we take $C>0$ and $g_\Phi<0$ or $C<0$ and $g_\Phi>0$, 
the resultant mass spectrum always exhibits both types of condensation in the range of $m_\pi^{\rm (H)}/2<\mu_q\lesssim 2m_\pi^{(\rm H)}$.

\section{Chiral partner structure}
\label{sec:ChiralPartner}

From the numerical analysis in Sec.~\ref{sec:MassSpectrum}, we have succeeded in gaining insights into roles of the mixing strength $C$ and the coupling $g_\Phi$ in determining the spin-$1$ hadron masses at finite $\mu_q$. In this section, by focusing on a high $\mu_q$ regime where chiral symmetry is sufficiently restored, we demonstrate the so-called {\it chiral partner structure} of the spin-$1$ hadrons by identifying the pairs of $1^+$ and $1^-$ hadrons that are degenerate in mass.

\begin{figure}[H]
\centering
\hspace*{-0.55cm} 
\includegraphics*[scale=0.098]{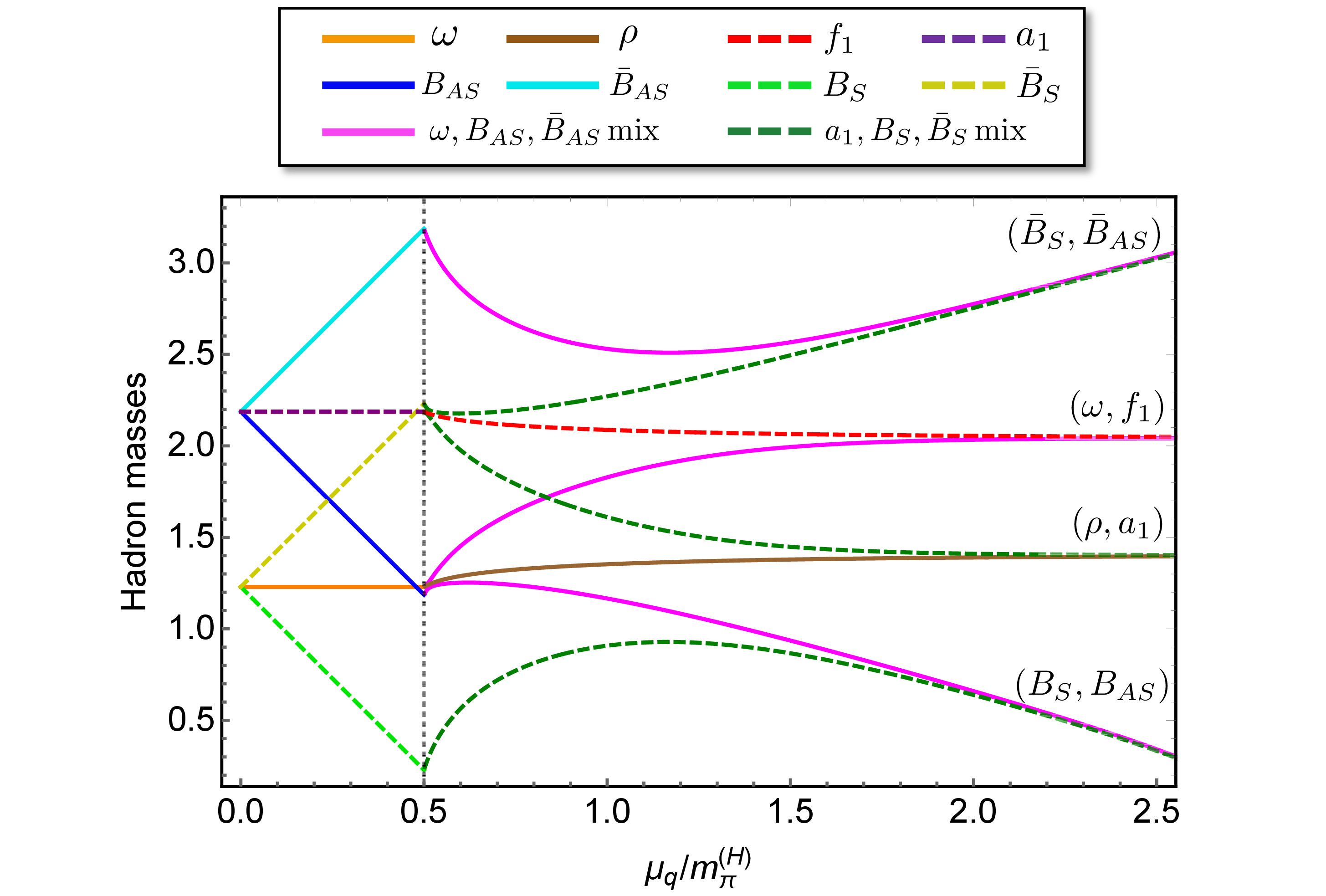}
\caption{$\mu_q$ dependence of all the $1^+$ and $1^-$ hadron masses for $(g_\Phi,C)=(10,16)$. The chiral partner structures are clearly shown by the degeneracy.}
\label{fig:ChiralPartner}
\end{figure}

At sufficiently high $\mu_q$, the four mean fields would asymptotically behave as $\sigma_0\to\sigma_\infty\mu_q^{-2}$, $\Delta\to \Delta_\infty$, $\bar{\omega}\to\bar{\omega}_\infty\mu_q$, and $\bar{V} \to \bar{V}_\infty\mu_q^{-1}$, as expected from the numerical results in Fig.~\ref{fig:MeanFields}, where $\sigma_\infty$, $\Delta_\infty$, $\bar{\omega}_\infty$, and $\bar{V}_\infty$ are constants. Using these asymptotic behaviors and the mass formulas derived in Appendix~\ref{sec:MassFormulas}, first, one can easily show $m_\omega = m_{f_1}$ and $ m_\rho = m_{a_1}$ at $\mu_q\to\infty$, where $\omega$ and $a_1$ mesons are decoupled from the respective mixings with the (anti)diquark baryons.\footnote{Here, the superscript ``$s$'' for the hadron masses representing the spatial component is omitted for simplicity. The same abbreviation applies to Eq.~(\ref{MRhoDiscuss}) to refer to the $\rho$ meson mass.}  That is, ($\omega, f_1$) and ($\rho,a_1$) can be regarded as the chiral partners even in the presence of the mean fields $\Delta$ and $\bar{\omega}$ in a high density region. Next, as for the remaining (anti)diquark baryons, $B_S$, $\bar{B}_S$, $B_{AS}$, and $\bar{B}_{AS}$, the asymptotic behaviors of the mean fields tell that the mixing structures for the $V_9$ - $V_{10}$ system and the $V'_{4}$ - $V_5'$ system become identical while the remaining bare masses satisfy $m_{V_9}=m_{V_4'}$ and $m_{V_{10}}=m_{V_5'}$ at $\mu_q\to\infty$, as can be seen from Appendix~\ref{sec:MassFormulas}. Thus, $(B_S,B_{AS})$ and $(\bar{B}_S,\bar{B}_{AS})$ are also regarded as the chiral partners.

The above analytic consideration of the chiral partner structure is, indeed, numerically confirmed as shown in Fig.~\ref{fig:ChiralPartner}. In this figure we have taken $(g_\Phi,C)=(10,16)$ to see the mass degeneracy clearly. Figure~\ref{fig:ChiralPartner} indicates that the mass degeneracy occurs between the following $1^+$ and $1^-$ hadrons,  $(B_S,B_{AS})$, $(\rho,a_1)$, $(\omega,f_1)$, and $(\bar{B}_S,\bar{B}_{AS})$ from below, at high $\mu_q$. We note that the chiral partner structure for the spin-$0$ hadrons in the absence of the $U(1)_A$ anomaly was examined in Ref.~\cite{Suenaga:2022uqn}, where mass degeneracy was demonstrated for $(B,B')$, $(\sigma,\pi)$, $(a_0,\eta)$, and $(\bar{B},\bar{B}')$.

\section{Discussions}
\label{sec:Discussions}

\subsection{Analytic derivation of $\mu_q^{\rm cr} = m_\pi^{(\rm H)}/2$ in the eLSM}
\label{sec:PhaseTransition}

In Sec.~\ref{sec:NumericalMF}, we have numerically seen that the phase transition from the hadronic phase to the baryon superfluid phase takes place just when $\mu_q$ coincides with $m_\pi^{(\rm H)}/2$, similarly to other chiral models ignoring spin-$1$ hadrons~\cite{Kogut:1999iv,Kogut:2000ek,Ratti:2004ra,Suenaga:2022uqn}. In this section we analytically prove such a universal property within our present eLSM based on a reasonable assumption.

The $\mu_q$ dependence of the mean fields, $\sigma_0$, $\Delta$, $\bar{\omega}$, and $\bar{V}$, has been determined by stationary conditions in Sec.\ \ref{sec:NumericalMF}. These conditions can be derived from Eq.~(\ref{ELSMRed}) as
\begin{widetext}
\begin{eqnarray}
\sigma_0:\frac{2\sqrt{2}h_q}{\sigma_0}-\frac{C}{\sqrt{2}\sigma_0}\mu_q\Delta\bar{V} -\frac{C_3}{4\sigma_0}\Delta\bar{V}\bar{\omega}+\frac{C_3}{8}(\bar{V}^2-\bar{\omega}^2)+\frac{C}{8}(\bar{V}^2+\bar{\omega}^2)  -m_0^2-\frac{\lambda_2}{4}(\sigma_0^2+\Delta^2)=0 \ , \label{GapSigmaELSM}
\end{eqnarray}
\begin{eqnarray}
\Delta: -\frac{C}{\sqrt{2}\Delta}\mu_q\bar{V}\sigma_0-\frac{C_3}{4\Delta}\bar{V}\sigma_0\bar{\omega}+4\mu_q^2+\sqrt{2}C\mu_q\bar{\omega}-\frac{C_3}{8}(\bar{V}^2-\bar{\omega}^2) + \frac{C}{8}(\bar{V}^2+\bar{\omega}^2)-m_0^2-\frac{\lambda_2}{4}(\sigma_0^2+\Delta^2)=0  \ , 
\label{GapDeltaELSM}
\end{eqnarray}
\begin{eqnarray}
\bar{\omega}:\frac{C}{\sqrt{2}\bar{\omega}}\mu_q\Delta^2-\frac{C_3}{4\bar{\omega}}\Delta\bar{V}\sigma_0-\frac{C_3}{8}(\sigma_0^2-\Delta^2) +\frac{C}{8}(\sigma_0^2+\Delta^2)  +m_1^2= 0 \ , \label{GapOmegaELSM}
\end{eqnarray}
and
\begin{eqnarray}
\bar{V}: -\frac{C}{\sqrt{2}\bar{V}}\mu_q\Delta\sigma_0-\frac{C_3}{4\bar{V}}\Delta\sigma_0\bar{\omega}+\frac{C_3}{8}(\sigma_0^2-\Delta^2)+ \frac{C}{8}(\sigma_0^2+\Delta^2) + m_1^2 = 0\ .  \label{GapV4ELSM}
\end{eqnarray}
\end{widetext}
The numerical solutions imply that emergence of $\bar{\omega}$ and $\bar{V}$ is accompanied by nonzero $\Delta$, i.e., by the onset of the baryon superfluidity and that the phase transition is of second order. Let us now suppose that the critical exponents of $\Delta$ and $\bar{V}$ are $+1/2$ while that of $\bar{\omega}$ is $+1$. Indeed, these exponents are suggested by the numerical result. Then, $\Delta$ and $\bar{V}$ take the form of
\begin{eqnarray}
&&\Delta \sim \Delta^{\rm cr}(\mu_q-\mu_q^{\rm cr})^{1/2} \ , \nonumber\\
&&\bar{V} \sim \bar{V}^{\rm cr}(\mu_q-\mu_q^{\rm cr})^{1/2} \ ,  \nonumber\\
&&\bar{\omega} \sim\bar{\omega}^{\rm cr} (\mu_q-\mu_q^{\rm cr})^{1} \ , \label{CriticalEx}
\end{eqnarray}
in the vicinity of the phase transition, where $\Delta^{\rm cr}>0$, $\bar{V}^{\rm cr}>0$, and $\bar{\omega}^{\rm cr}<0$. From these exponents, one finds that in the limit of $\mu_q\to\mu_q^{\rm cr}$, $\Delta^2/\bar{\omega}$, $\Delta\bar{V}/\bar{\omega}$, and $\Delta/\bar{V}$ approach nonzero values, while $\Delta\bar{\omega}/\bar{V}\to0$. Thus, the first and second terms of Eq.~(\ref{GapOmegaELSM}) and the first term in Eq.~(\ref{GapV4ELSM}), which remain finite at the critical chemical potential, act as catalyzers to yield nonzero $\bar{\omega}$ and $\bar{V}$ above $\mu_q=\mu_q^{\rm cr}$, respectively.

Given the critical behavior~(\ref{CriticalEx}), at $\mu_q=\mu_q^{\rm cr}$, the stationary conditions, Eqs.~(\ref{GapSigmaELSM})~-~(\ref{GapV4ELSM}), are reduced to
\begin{eqnarray}
\frac{2\sqrt{2}h_q}{\sigma_0^{\rm (H)}} - Z_\pi^{-2}\big(m_\pi^{(\rm H)}\big)^2=0\ ,\label{SigmaCr}
\end{eqnarray}
\begin{eqnarray}
 -\frac{C\mu^{\rm cr}_q\sigma_0^{\rm (H)}}{\sqrt{2}x}+4(\mu_q^{\rm cr})^2-Z_\pi^{-2}\big(m_\pi^{(\rm H)}\big)^2=0\ , \label{DeltaCr}
\end{eqnarray}
\begin{eqnarray}
\frac{C\mu_q^{\rm cr}y}{\sqrt{2}}-\frac{C_3\sigma_0^{\rm (H)}y}{4x} + \big(m_\rho^{(\rm H)}\big)^2 = 0\ , \label{OmegaCr}
\end{eqnarray}
and
\begin{eqnarray}
-\frac{C\mu^{\rm cr}_q\sigma_0^{\rm (H)} x}{\sqrt{2}}+(m_{f_1}^{(\rm H)})^2= 0\ ,  \label{VbarCr}
\end{eqnarray}
respectively, where we have used the hadron mass formulas in Appendix~\ref{sec:MassHadronic} and defined 
\begin{eqnarray}
x \equiv \frac{\Delta^{\rm cr}}{\bar{V}^{\rm cr}}>0\ ,  \ \ y \equiv \frac{\big(\Delta^{\rm cr}\big)^2}{\bar{\omega}^{\rm cr}}<0\ .
\end{eqnarray}
From Eqs.~(\ref{DeltaCr}) and~(\ref{VbarCr}) as well as the renormalization factor~(\ref{ZDef}), therefore, one can analytically prove
\begin{eqnarray}
\mu_q^{\rm cr} = \frac{m_\pi^{(\rm H)}}{2}\ ,
\end{eqnarray}
which was numerically confirmed in Fig.~\ref{fig:MeanFields}. We note that $C$ cannot be zero from the stationary condition for $\bar{V}$ at $\mu_q=\mu_q^{\rm cr}$, Eq.~(\ref{VbarCr}), as long as Eq.~(\ref{CriticalEx}) holds. In other words, $\bar{V}$ always vanishes when $C=0$, and similarly,  from Eq.~(\ref{OmegaCr}) one can see $\bar{\omega}=0$ at any $\mu_q$ when $C=0$.

\subsection{Comments on the $\rho$ mass reduction}
\label{sec:RhoMassRed}

Here we provide comments on the $\rho$ meson mass reduction in the superfluid phase.

As derived in Eq.~(\ref{MRhoS}), the $\rho$ meson mass is evaluated as
\begin{eqnarray}
m_\rho^2 = m_1^2+\frac{C-C_3}{8} (\sigma_0^2+\Delta^2) \ , \label{MRhoDiscuss}
\end{eqnarray}
in both the hadronic and superfluid phases. This universal structure stems from the fact that the $\rho$ meson is not contaminated by any  mixing with other hadrons, even in the presence of the superfluidity.  Then, if the combination $\sigma_0^2+\Delta^2$ is enhanced in the superfluid phase while $C<C_3$, the $m_\rho$ reduction observed on the lattice can be reproduced within the present eLSM. From Fig.~\ref{fig:MixMF}, however, one can see that the enhancement of the combination $\sigma_0^2+\Delta^2$ in the superfluid phase gets mild as the spin-$0$ and spin-$1$ mixing effect becomes prominent. That is why $m_\rho$ is relatively hard to change for a larger value of $C$.

Since $\mu_q$ dependence of the combination $\sigma_0^2+\Delta^2$ in the superfluid phase is rather monotonic as can be inferred from Fig.~\ref{fig:MixMF}, the resultant $m_\rho$ also changes almost linearly as a function of $\mu_q$. On the other hand, the lattice data would imply a rather abrupt reduction of $m_\rho$ just above $\mu_q=\mu_q^{\rm cr}$, although there remain error bars; $\mu_q$ dependence of $m_\rho$ measured by the lattice simulation would look convex downward~\cite{Murakami:2022lmq,Murakami2022}. One promising mechanism to yield such a downward-convex behavior of $m_\rho$ could be additional mixing with other states that are associated with the superfluidity but have yet to be considered in the present analysis. The $\rho$ meson, which is an iso-triplet state carrying $J^P=1^-$, may strongly mix with an iso-triplet and $J^P=1^-$ diquark $\tilde{B}^i\sim \epsilon^{ijk}[\psi^TC\Sigma^{jk}\tau_c^2\psi]_{\rm sym.}$, where $\Sigma^{\mu\nu} = \frac{i}{2}[\gamma^\mu,\gamma^\nu]$ is the antisymmetric tensor and the subscript ``sym.'' means the flavor symmetric structure. Inclusion of this new diquark state, however, requires us to introduce another quark bilinear operator $\tilde{\Phi}_{ij}^{\mu\nu} \sim\Psi_j^T\sigma^2\bar{\sigma}^{\mu\nu}\tau_c^2\Psi_i$, where $\bar{\sigma}^{\mu\nu} = \frac{i}{2}(\bar{\sigma}^\mu\sigma^\nu-\bar{\sigma}^\nu\sigma^\mu)$ with $\bar{\sigma}^\mu=({\bm 1}, -\sigma^i)$, which is beyond the scope of the present study. Thus, we leave detailed examination of the $\tilde{B}^i$ - $\rho$ mixing in the superfluid phase for future study.

\section{Conclusions}
\label{sec:Conclusions}

In summary, for dense QC$_2$D at zero temperature, we have constructed the extended linear sigma model, eLSM, in such a way as to respect the Pauli-G\"{u}rsey $SU(4)$ symmetry and to describe both the spin-$0$ and spin-$1$ hadrons.  Then, based on the eLSM, richness of the mass spectrum of the spin-$1$ hadrons in dense QC$_2$D has been explored.

In the baryon superfluid phase where the diquark condensate emerges, we have found that not only the scalar meson and scalar diquark baryon but also the time components of the vector meson and vector diquark baryon possess their mean field values, in the presence of spin-$0$ and spin-$1$ mixing. These mean fields are induced by violation of the Lorentz invariance as well as of $U(1)$ baryon-number conservation. Besides, we have analytically shown that the onset condition of superfluidity corresponds to $\mu_q=m_\pi^{(\rm H)}/2$ ($m_\pi^{(\rm H)}$ is the pion mass in the hadronic phase) and that the pion mass in the superfluid phase reads $m_\pi = 2\mu_q$. Moreover, the appearance of the NG boson associated with $U(1)_B$ violation has been confirmed. Those self-consistent properties, which were numerically indicated by lattice simulations,} are derived only when the above four types of mean fields are included.

Inclusion of the vector-meson and vector-diquark mean fields has led to suppression of the otherwise substantial increase of the scalar-diquark mean field in the superfluid phase via the spin-$0$ and spin-$1$ mixing. Simultaneously, it has been found that the slight reduction of the $\rho$ meson mass in the superfluid phase suggested by the lattice data is successfully reproduced in the presence of the significant spin-$0$ and spin-$1$ mixing. Furthermore, by varying the magnitude of this mixing and of coupling among the spin-$1$ hadrons, we have demonstrated the emergence of the axialvector condensate and the possible vector condensate. Those novel condensates are induced when the masses of the corresponding modes reach zero. Given such condensates, therefore, it is inevitable to investigate the in-medium masses of the spin-$1$ hadrons from the first-principles lattice calculation of dense QC$_2$D toward further delineation of the phase structures.

In addition, we have also discussed a possible existence of an iso-triplet $1^-$ diquark, motivated by the possible downward-convex behavior of the $\rho$ meson mass reduction in the superfluid phase as suggested by the lattice simulations~\cite{Murakami:2022lmq,Murakami2022}. This diquark is denoted by a tensor-type quark bilinear field, while no examination has been done so far. Therefore, it would be challenging to pursue properties of such a new diquark state from both effective theories and first-principles numerical studies. 

\section*{Acknowledgment}

The authors thank Masayasu Harada for useful comments on the HLS formalism for the spin-$1$ hadrons. D.S. was supported by the RIKEN special postdoctoral researcher program and by the Japan Society for the Promotion of Science (JSPS) KAKENHI Grants No.~23K03377. K.~M. is supported in part by JST SPRING with Grant Number JPMJSP2110, by Grants-in-Aid for JSPS Fellows (Nos.\ JP22J14889, JP22KJ1870), and by JSPS KAKENHI with Grant No.\ 22H04917.
The work of K.~I. is supported by JSPS KAKENHI with Grant Numbers 18H05406 and 23H01167.
The work of E.~I. is supported by JSPS KAKENHI with Grant Number 23H05439, 
JST PRESTO Grant Number JPMJPR2113, 
JSPS Grant-in-Aid for Transformative Research Areas (A) JP21H05190, 
JST Grant Number JPMJPF2221  
and also supported by Program for Promoting Researches on the Supercomputer ``Fugaku'' (Simulation for basic science: from fundamental laws of particles to creation of nuclei) and (Simulation for basic science: approaching the new quantum era), and Joint Institute for Computational Fundamental Science (JICFuS), Grant Number JPMXP1020230411.
The work of E.~I is supported also by Center for Gravitational Physics and Quantum
Information (CGPQI) at YITP. 
\appendix

\section{The Pauli-G\"{u}rsey $SU(4)$ symmetry in two-flavor QC$_2$D}
\label{sec:SU2Nf}

In this appendix, we briefly show emergence of the Pauli-G\"{u}rsey $SU(4)$ symmetry in QC$_2$D with two flavors~\cite{Kogut:1999iv,Kogut:2000ek}.

The QC$_2$D Lagrangian for massless $u$ and $d$ quarks is of the form
\begin{eqnarray}
{\cal L}_{\rm QC_2D} = \bar{\psi}i\Slash{D}\psi \ , \label{QC2DLagrangian}
\end{eqnarray}
where $\psi=(u,d)^T$ is the quark doublet and $D_\mu\psi = \partial_\mu\psi-ig_cA_\mu^aT_c^a\psi$ is the covariant derivative describing interactions between the quarks $\psi$ and gluons $A_\mu^a$. The $2\times2$ matrix $T_c^a=\tau_c^a/2$ is the $SU(2)_c$ generator ({\bf $\tau_c^a$} is the Pauli matrix for colors). Adopting the Weyl representation for the Dirac matrices, the Lagrangian~(\ref{QC2DLagrangian}) can be expressed in terms of left-handed and right-handed quarks as
\begin{eqnarray}
{\cal L}_{\rm QC_2D} &=& \psi^\dagger_R i\partial_\mu\sigma^\mu\psi_R -g_c\psi_R^\dagger A_\mu^aT_c^a\sigma^\mu \psi_R \nonumber\\
&+& \psi^\dagger_L i\partial_\mu\bar{\sigma}^\mu\psi_L -g_c\psi_L^\dagger A_\mu^aT_c^a\bar{\sigma}^\mu \psi_L . \label{QC2DLagrangian2}
\end{eqnarray}
In this Lagrangian, $u=(u_R,u_L)^T$ and $d=(d_R,d_L)^T$ in the Weyl representation, and the $2\times2$ matrices in the spinor space are defined by $\sigma^\mu=({\bm 1}, \sigma^i)$ and $\bar{\sigma}^\mu=({\bm 1},-\sigma^i)$ with the Pauli matrix $\sigma^i$. Here, we make use of the pseudoreal property of the Pauli matrix. Namely, using relations
\begin{eqnarray}
T_c^a = -\tau_c^2(T_c^a)^T\tau_c^2\ , \ \ \sigma^i = -\sigma^2(\sigma^i)^T\sigma^2\ ,
\end{eqnarray}
and accordingly introducing the ``conjugate quark fields'' 
\begin{eqnarray}
\tilde{\psi}_R \equiv \sigma^2{\bf \tau_c^2}\psi_R^* \ , \ \ \tilde{\psi}_L \equiv \sigma^2{\bf \tau_c^2}\psi_L^*\ ,
\end{eqnarray}
the Lagrangian~(\ref{QC2DLagrangian2}) can be expressed in a unified form as
\begin{eqnarray}
{\cal L}_{\rm QC_2D} = \Psi^\dagger i\partial_\mu\sigma^\mu\Psi-g\Psi^\dagger A_\mu^a\sigma^\mu\Psi\ . \label{QC2DFour} 
\end{eqnarray}
Here, we have described the quark fields by using
a four-component column vector defined as
\begin{eqnarray}
\Psi \equiv \left(
\begin{array}{c}
\psi_R \\
\tilde{\psi}_L \\
\end{array}
\right) =  \left(
\begin{array}{c}
u_R \\
d_R \\
\tilde{u}_L \\
\tilde{d}_L \\
\end{array}
\right) \ . \label{PsiFour}
\end{eqnarray}

The Lagrangian~(\ref{QC2DFour}) is obviously invariant under an $SU(4)$ transformation of 
\begin{eqnarray}
\Psi\to U\Psi\ \ \ {\rm with}\ \ \ U\in SU(4)\ ,
\end{eqnarray}
 rather than $SU(2)_L\times SU(2)_R$ chiral transformation. Such an extended symmetry is sometimes referred to as the Pauli-G\"{u}rsey $SU(4)$ symmetry~\cite{Kogut:1999iv,Kogut:2000ek}. As can be seen from Eq.~(\ref{PsiFour}), the Pauli-G\"{u}rsey $SU(4)$ symmetry is realized by treating $\psi$ and $\tilde{\psi}$ in a single multiplet, reflecting the fact that mesons and diquark baryons can be described in a unified way in two-flavor QC$_2$D. It should be noted that the $U(1)_B$ baryon-number transformation is generated by 
\begin{eqnarray}
\Psi \to {\rm e}^{-i\theta_q {J}} \Psi\  \ \ {\rm with} \ \  \ {J} \equiv \left(
\begin{array}{cc}
{\bm 1} & 0 \\
0 & -{\bm 1} \\
\end{array}
\right) \ , \label{BaryonPsi}
\end{eqnarray}
where ${\rm e}^{-i\theta_q{\bm J}}$ belongs to a subgroup of the Pauli-G\"{u}rsey $SU(4)$ group.

\section{Generators of $U(4)$ Lie algebra}
\label{sec:Generators}

In this appendix, we list the generators of $U(4)$ Lie algebra. 

The number of the $U(4)$ generators is $4\times4=16$. It is convenient to separate these $16$ generators into two sets $S^i$ ($i=1$ - $10$) and $X^a$ ($a=0$ - $5$) that satisfy
\begin{eqnarray}
 E(S^i)^T = -S^iE\ ,\ \ E(X^a)^T = X^aE\ ,
\end{eqnarray}
with the symplectic matrix
\begin{eqnarray}
E = \left(
\begin{array}{cc}
0 & {\bm 1}_f  \\
-{\bm 1}_f & 0 \\
\end{array}
\right)\ .
\end{eqnarray}
That is, the elements generated by $S^i$, $h={\rm e}^{-i\theta_S^iS^i}$, exhibit the following relation:
\begin{eqnarray}
hEh^T = E\ .
\end{eqnarray}
This relation means that $h$ belongs to the $Sp(4)$ group, which is the subgroup of the original $U(4)$ group.

More concretely, the generators $S^i$ belonging to the Lie algebra of $Sp(4)$ read
\begin{eqnarray}
&& S^{i=1-4} = \frac{1}{2\sqrt{2}}\left(
\begin{array}{cc}
\tau_f^i & 0 \\
0 & -(\tau_f^i)^T \\
\end{array}
\right)\ ,  \nonumber\\
&& S^{i=5-10} = \frac{1}{2\sqrt{2}}\left(
\begin{array}{cc}
0& B^i \\
(B^i)^\dagger &0 \\
\end{array}
\right)\ , \label{SDef}
\end{eqnarray}
with $\tau_f^4 = {\bm 1}_f$, $B^5={\bm 1}_f$, $B^6=i{\bm 1}_f$, $B^7=\tau_f^3$, $B^8=i\tau_f^3$, $B^9=\tau_f^1$, and $B^{10}=i\tau_f^1$. Meanwhile, the remaining generators belonging to the algebras of $U(1)$ and $SU(4)/Sp(4)$ are given by
\begin{eqnarray}
&& X^{a=0-3} = \frac{1}{2\sqrt{2}}\left(
\begin{array}{cc}
\tau_f^a & 0 \\
0 &(\tau_f^a)^T \\
\end{array}
\right)\ , \nonumber\\
&& X^{a=4,5} = \frac{1}{2\sqrt{2}}\left(
\begin{array}{cc}
0& D^a \\
(D^a)^\dagger & 0 \\
\end{array}
\right) \ , \label{XDef}
\end{eqnarray}
where $\tau_f^{a=0}={\bm 1}_f$ is the $2\times2$ unit matrix and $\tau_f^{a=1-3}$ are the Pauli matrices in the flavor space. Besides, $D^4=\tau_f^2$ and $D^5=i\tau_f^2$.

\section{Mass formulas}
\label{sec:MassFormulas}

In this appendix, we derive hadron mass formulas from the reduced eLSM Lagrangian~(\ref{ELSMRed}).

The mass formulas are derived by picking up quadratic terms of the hadron fields in the Lagrangian~(\ref{ELSMRed}) on top of the mean fields $\sigma_0$, $\Delta$, $\bar{\omega}$, and $\bar{V}$ defined by Eq.~(\ref{MeanFields}). The eLSM describes the mass spectrum of 16 hadrons in total, namely, eight spin-$0$ hadrons, $\eta$, $\pi$, $B$, $\bar{B}$, $\sigma$, $a_0$, $B'$, and $\bar{B}'$, and eight spin-$1$ hadrons, $\omega$, $\rho$, $B_S$, $\bar{B}_S$, $f_1$, $a_1$, $B_{AS}$, and $\bar{B}_{AS}$. As explained in Sec.~\ref{sec:Analytic}, those 16 hadrons are separated into the following four systems due to different mixing patterns:
\begin{enumerate}
\item $a_0$ - $\rho$ system,
\item $\eta$ - $B'$ - $\bar{B}'$- $f_1$ system,
\item $\pi$ - $a_1$ - $B_S$ - $\bar{B}_{S}$ system,
\item $\sigma$ - $B$ - $\bar{B}$ - $\omega$ - $B_{AS}$ - $\bar{B}_{AS}$ system.
\end{enumerate}
Then, in what follows we show the mass formulas for these four systems separately.

\subsection{$a_0$ - $\rho$ system}
\label{sec:Mass1}

As for the $a_0$ - $\rho$ system, the relevant Lagrangian is obtained as
\begin{eqnarray}
{\cal L}_{(1)} = {\cal L}_{a_0} + {\cal L}_{\rho} + {\cal L}_{a_0\rho}  \ ,
\end{eqnarray}
where each term reads
\begin{eqnarray}
{\cal L}_{a_0} = \frac{1}{2}\partial_\mu a_0\partial^\mu a_0-\frac{m_{a_0}^2}{2}a_0^2\ ,
\end{eqnarray}
\begin{eqnarray}
{\cal L}_{\rho} &=& -\frac{1}{4}( \partial_\mu \rho_{\nu}-\partial_\nu \rho_{\mu})^2+ \frac{(m^t_{\rho})^2}{2}(\rho^0)^2 + \frac{(m_{\rho}^s)^2}{2}\rho_{i}\rho^i \ , \nonumber\\
\end{eqnarray}
and
\begin{eqnarray}
{\cal L}_{a_0\rho} = \frac{C-C_3}{4}(\Delta\bar{V}_4+\sigma_0\bar{\omega})a_0\rho^0\ . \nonumber\\ \label{LA0Rho}
\end{eqnarray}
In these terms, we have suppressed the isospin indices for simplicity, and have defined the mass parameters by
\begin{eqnarray}
m_{a_0}^2  =  m_0^2+\frac{3\lambda_2}{4}(\sigma_0^2+\Delta^2) - \frac{C-C_3}{8}(\bar{V}^2+\bar{\omega}^2) \ ,  \label{MA02}
\end{eqnarray}
and
\begin{eqnarray}
(m_\rho^t)^2 = (m_\rho^s)^2 = m_1^2+\frac{C-C_3}{8} (\sigma_0^2+\Delta^2) \ , 
 \label{MRhoS}
\end{eqnarray}
where the superscripts ``{\it t}'' and ``{\it s}'' are attached to distinguish between time-component (unphysical) and spatial-component (physical) masses.

Thus, from Eq.~(\ref{LA0Rho}) one can see that the $a_0$ meson mixes with the time component of the $\rho$ meson due to the violation of the Lorentz invariance. For this reason, the mass of the $a_0$ meson does not coincide with $m_{a_0}$ provided by Eq.~(\ref{MA02}) but is determined by a pole of the propagator matrix for the $a_0$ - $\rho^0$ system at vanishing momentum ${\bm p}={\bm 0}$. The inverse of the propagator matrix $i{\cal D}_{(1)}^{-1}(p_0,{\bm 0})$ can be derived as
\begin{eqnarray}
i{\cal D}_{(1)}^{-1} (p_0,{\bm 0})= \left(
\begin{array}{cc}
{\cal M}_{a_0a_0} & {\cal M}_{a_0\rho} \\
{\cal M}_{\rho a_0} & {\cal M}_{\rho\rho} \\
\end{array}
\right)\ ,
\end{eqnarray}
with
\begin{eqnarray}
{\cal M}_{a_0a_0} = p_0^2-m_{a_0}^2 \ ,
\end{eqnarray}
\begin{eqnarray}
{\cal M}_{a_0\rho} = {\cal M}_{\rho a_0}= \frac{C-C_3}{4}(\Delta\bar{V}+\sigma_0\bar{\omega})\ , 
\end{eqnarray}
and
\begin{eqnarray}
{\cal M}_{\rho \rho} = (m_\rho^t)^2\ .
\end{eqnarray}
Therefore, the mass of $a_0$ at arbitrary $\mu_q$ is evaluated by numerically solving ${\rm det}\left[i{\cal D}_{(1)}^{-1} (p_0,{\bm 0})\right]=0$. Meanwhile, the spatial components of the $\rho$ meson, i.e., the physical states of $\rho$ do not join any mixing, and hence, the $\rho$ meson mass is identical to $m_\rho^s$, Eq.~(\ref{MRhoS}).

\subsection{$\eta$ - $B'$ - $\bar{B}'$ - $f_1$ system}
\label{sec:Mass2}

Employing a similar procedure demonstrated in Appendix~\ref{sec:Mass1}, the mass spectrum of the $\eta$ - $B'$ - $\bar{B}'$ - $f_1$ system can be evaluated. In this case, $\eta$, $B'$, $\bar{B}'$ (or $B'^4$, $B'^5$) and the time component of $f_1$ can mix. The inverse propagator of these four states at vanishing momentum ${\bm p}={\bm 0}$: $i{\cal D}_{(2)}^{-1}(p_0,{\bm 0})$, is given by
\begin{eqnarray}
i{\cal D}_{(2)}^{-1}(p_0,{\bm 0}) = \left(
\begin{array}{cccc}
{\cal M}_{\eta\eta} & {\cal M}_{\eta B_4'} & {\cal M}_{\eta B_5'} & {\cal M}_{\eta f_1} \\
{\cal M}_{B_4'\eta} & {\cal M}_{B_4' B_4'} & {\cal M}_{B_4' B_5'} & {\cal M}_{B_4' f_1} \\
{\cal M}_{B_5'\eta} & {\cal M}_{B_5' B_4'} & {\cal M}_{B_5' B_5'} & {\cal M}_{B_5' f_1} \\
{\cal M}_{f_1\eta} & {\cal M}_{f_1 B_4'} & {\cal M}_{f_1 B_5'} & {\cal M}_{f_1 f_1} \\
\end{array}
\right)\ , \nonumber\\
\end{eqnarray}
where each matrix element reads
\begin{eqnarray}
{\cal M}_{\eta\eta} = p_0^2-m_\eta^2 + \frac{C^2}{8}\bar{V}^2\ ,
\end{eqnarray}
\begin{eqnarray}
{\cal M}_{\eta B_4'} = -{\cal M}_{ B_4'\eta}=-i\frac{C}{\sqrt{2}}\bar{V}p_0\ ,
\end{eqnarray}
\begin{eqnarray}
{\cal M}_{\eta B_5'} = {\cal M}_{ B_5'\eta}  = -\frac{2\lambda_2}{4}\sigma_0\Delta  +\frac{C}{\sqrt{2}}\bar{V}\mu_q  + \frac{C_3}{4}\bar{V}\bar{\omega}\ , \nonumber\\
\end{eqnarray}
\begin{eqnarray}
{\cal M}_{\eta f_1} = -{\cal M}_{ f_1\eta}  = -i\frac{C}{2\sqrt{2}}\sigma_0p_0\ , \label{MEtaF1}
\end{eqnarray}
\begin{eqnarray}
{\cal M}_{B_4' B_4'} = p_0^2-m_{B_4'}^2+\frac{C^2}{8}\bar{V}^2 +\left(2\mu_q+\frac{C}{2\sqrt{2}}\bar{\omega}\right)^2\ ,  \nonumber\\
\end{eqnarray}
\begin{eqnarray}
{\cal M}_{B_4' B_5'} = -{\cal M}_{B_5' B_4'} = 2i\left(2\mu_q+\frac{C}{2\sqrt{2}}\bar{\omega}\right)p_0\ , 
\end{eqnarray}
\begin{eqnarray}
{\cal M}_{B_4' f_1} = {\cal M}_{f_1B_4' } =  \frac{C+C_3}{4} (\sigma_0\bar{V}-\Delta\bar{\omega}) -\sqrt{2}C\Delta\mu_q\ , \nonumber\\
\end{eqnarray}
\begin{eqnarray}
{\cal M}_{B_5' B_5'} = p_0^2-m_{B_5'}^2 + \left(2\mu_q+\frac{C}{2\sqrt{2}}\bar{\omega}\right)^2 \ ,
\end{eqnarray}
\begin{eqnarray}
{\cal M}_{B_5' f_1} = -{\cal M}_{f_1 B_5'} = i \frac{C}{2\sqrt{2}}\Delta p_0\ ,
\end{eqnarray}
and
\begin{eqnarray}
{\cal M}_{f_1 f_1} = (m_{f_1}^t)^2\ ,
\end{eqnarray}
with the mass parameters
\begin{eqnarray}
m_{B_4'}^2 &=& m_0^2+\frac{3\lambda_2}{4}(\sigma_0^2+\Delta^2) \nonumber\\
&& -\frac{C-C^2+C_3}{8}(\bar{V}^2+\bar{\omega}^2)\ , 
\end{eqnarray}
\begin{eqnarray}
m_{B_5'}^2 &=& m_0^2+\frac{\lambda_2}{4}(3\sigma_0^2+\Delta^2)\nonumber\\
&&-\frac{C-C^2+C_3}{8}\bar{\omega}^2 -\frac{C-C_3}{8}\bar{V}^2\ ,
\end{eqnarray}
\begin{eqnarray}
m_\eta^2 &=& m_0^2+\frac{\lambda_2}{4}(\sigma_0^2+3\Delta^2) \nonumber\\
&&-\frac{C-C_3}{8}\bar{\omega}^2 - \frac{C-C^2+C_3}{8}\bar{V}^2\ ,
\end{eqnarray}
and
\begin{eqnarray}
(m_{f_1}^t)^2 &=& m_1^2+\frac{C}{4}(\sigma_0^2+\Delta^2) \ .
\end{eqnarray}
In the presence of the mixing, the mass spectrum of the $\eta$ - $B'$- $\bar{B}'$ system is evaluated by numerically solving ${\rm det}\left[i{\cal D}_{(2)}^{-1}(p_0,{\bm 0})\right]=0$. Meanwhile, the space components of $f_1$ are decoupled, which allows us to simply obtain the physical $f_1$ meson mass as
\begin{eqnarray}
(m_{f_1}^s)^2 &=& m_1^2+\frac{C+C_3}{8}(\sigma_0^2+\Delta^2) \ .
\end{eqnarray}

\subsection{$\pi$ - $a_1$ - $B_S$ - $\bar{B}_{S}$ system}
\label{sec:Mass3}

Here, we derive the mass formulas for the $\pi$ - $a_1$ - $B_S$ - $\bar{B}_{S}$ system. In this case, $\pi$ and the time components of $a_1$, $B_S$, $\bar{B}_S$ (or $V^9$, $V^{10}$) can mix. In addition, the space components of $a_1$, $B_S$, $\bar{B}_S$ (or $V^9$, $V^{10}$) also mix with each other.

First, we consider the mixing among $\pi$ and the time components of $a_1$, $B_S$, and $\bar{B}_S$. The inverse propagator of these four states at vanishing momentum ${\bm p}={\bm 0}$, $i{\cal D}_{(3)}^{-1}(p_0,{\bm 0})$, is given by
\begin{eqnarray}
i{\cal D}_{(3)}^{-1}(p_0,{\bm 0})  = \left(
\begin{array}{cccc}
{\cal M}_{\pi\pi} & {\cal M}_{\pi a_1} & {\cal M}_{\pi V_9} & {\cal M}_{\pi V_{10}} \\
{\cal M}_{ a_1\pi} & {\cal M}_{a_1 a_1} & 0 & {\cal M}_{a_1 V_{10}} \\
{\cal M}_{V_9\pi} & 0& {\cal M}_{V_9V_9} &0\\
{\cal M}_{V_{10}\pi} & {\cal M}_{V_{10} a_1} & 0& {\cal M}_{V_{10}V_{10}} \\
\end{array}
\right) , \nonumber\\
\end{eqnarray}
where each matrix element reads
\begin{eqnarray}
{\cal M}_{\pi\pi} = p_0^2-m_\pi^2 \ ,
\end{eqnarray}
\begin{eqnarray}
{\cal M}_{\pi a_1}= -{\cal M}_{ a_1\pi}  = -i\frac{C}{2\sqrt{2}}\sigma_0 p_0\ ,
\end{eqnarray}
\begin{eqnarray}
{\cal M}_{\pi V_9}  = {\cal M}_{ V_9\pi} = \frac{C_3}{4}\sigma_0\bar{V}- \frac{C}{\sqrt{2}}\Delta\mu_q - \frac{C_3}{4}\Delta\bar{\omega}\ , 
\end{eqnarray}
\begin{eqnarray}
{\cal M}_{\pi V_{10}} = -{\cal M}_{ V_{10}\pi}  = i\frac{C}{2\sqrt{2}}\Delta p_0\ ,
\end{eqnarray}
\begin{eqnarray}
{\cal M}_{ a_1a_1} = (m_{a_1}^t)^2\ ,
\end{eqnarray}
\begin{eqnarray}
{\cal M}_{ a_1V_{10}} = {\cal M}_{V_{10}a_1} = -\frac{C_3}{4}\sigma_0\Delta \ , 
\end{eqnarray}
\begin{eqnarray}
{\cal M}_{V_9V_9} =(m_{V_9}^t)^2\ ,
\end{eqnarray}
and
\begin{eqnarray}
{\cal M}_{V_{10}V_{10}} =  (m_{V_{10}}^t)^2\ ,
\end{eqnarray}
with the mass parameters
\begin{eqnarray}
m_\pi^2 = m_0^2+\frac{\lambda_2}{4}(\sigma_0^2+\Delta^2)  -\frac{C-C_3}{8}(\bar{V}^2+\bar{\omega}^2) \ ,
\end{eqnarray}
\begin{eqnarray}
(m^t_{a_1})^2 = m_1^2 + \frac{C}{8}(\sigma_0^2+\Delta^2) + \frac{C_3}{8} (\sigma_0^2-\Delta^2)\ ,
\end{eqnarray}
\begin{eqnarray}
(m_{V_9}^t)^2 = m_1^2+\frac{C-C_3}{8}(\sigma_0^2+\Delta^2) \ ,
\end{eqnarray}
and
\begin{eqnarray}
(m_{V_{10}}^t)^2 = m_1^2+\frac{C}{8}(\sigma_0^2+\Delta^2)-\frac{C_3}{8}(\sigma_0^2-\Delta^2) \ .
\end{eqnarray}
In the presence of the mixing, the pion mass is evaluated by numerically solving ${\rm det}\left[i{\cal D}_{(3)}^{-1}(p_0,{\bm 0})\right]=0$. 

Next, we consider the mixing among the spatial components of $a_1$, $B_S$, and $\bar{B}_S$. In this case, the corresponding inverse propagator at vanishing momentum ${\bm p}={\bm 0}$, $i\tilde{\cal D}_{(3)}^{-1}(p_0,{\bm 0})$, is given by
\begin{eqnarray}
 i\tilde{\cal D}_{(3)}^{-1}(p_0,{\bm 0})  = \left(
\begin{array}{ccc}
 \tilde{\cal M}_{a_1 a_1} & \tilde{\cal M}_{a_1 V_{9}}  & \tilde{\cal M}_{a_1 V_{10}} \\
\tilde{\cal M}_{V_9a_1} & \tilde{\cal M}_{V_9V_9} & \tilde{\cal M}_{V_9V_{10}} \\
\tilde{\cal M}_{V_{10} a_1} &  \tilde{\cal M}_{V_{10}V_9} & \tilde{\cal M}_{V_{10}V_{10}} \\
\end{array}
\right)\ , \nonumber\\
\end{eqnarray}
where each matrix element reads
\begin{eqnarray}
\tilde{\cal M}_{a_1a_1} = -p_0^2-\frac{g_\Phi^2}{2}\bar{V}^2+ (m_{a_1}^s)^2\ ,
\end{eqnarray}
\begin{eqnarray}
\tilde{\cal M}_{a_1V_9} =-\tilde{\cal M}_{V_9a_1} =  \sqrt{2}ig_\Phi\bar{V}p_0\ ,
\end{eqnarray}
\begin{eqnarray}
\tilde{\cal M}_{a_1V_{10}}=\tilde{\cal M}_{V_{10}a_1} = -\frac{C_3}{4}\sigma_0\Delta -\sqrt{2}g_\Phi\bar{V}\mu_q - \frac{g_\Phi^2}{2}\bar{V}\bar{\omega} , \nonumber\\
\end{eqnarray}
\begin{eqnarray}
\tilde{\cal M}_{V_9V_9} = -p_0^2-\frac{g_\Phi^2}{2}\bar{V}^2-\left(2\mu_q+\frac{g_\Phi}{\sqrt{2}}\bar{\omega}\right)^2+ (m_{V_9}^s)^2 , \nonumber\\
\end{eqnarray}
\begin{eqnarray}
\tilde{\cal M}_{V_9V_{10}} =-\tilde{\cal M}_{V_{10}V_9}  = -2i\left(2\mu_q+\frac{g_\Phi}{\sqrt{2}}\bar{\omega}\right)p_0 \ ,
\end{eqnarray}
and
\begin{eqnarray}
\tilde{\cal M}_{V_{10}V_{10}} = -p_0^2-\left(2\mu_q+\frac{g_\Phi}{\sqrt{2}}\bar{\omega}\right)^2+ (m_{V_{10}}^s)^2\ ,
\end{eqnarray}
with the mass parameters 
\begin{eqnarray}
(m^s_{a_1})^2 = m_1^2+ \frac{C}{8}(\sigma_0^2+\Delta^2) + \frac{C_3}{8}(\sigma_0^2-\Delta^2)\ ,
\end{eqnarray}
\begin{eqnarray}
(m_{V_9}^s)^2 = m_1^2 + \frac{C-C_3}{8}(\sigma_0^2+\Delta^2) \ ,
\end{eqnarray}
and
\begin{eqnarray}
(m_{V_{10}}^s)^2 = m_1^2+ \frac{C}{8}(\sigma_0^2+\Delta^2)-\frac{C_3}{8}(\sigma_0^2-\Delta^2) \ .
\end{eqnarray}
In the presence of the mixing, the mass spectrum for the $a_1$ - $B_S$ - $\bar{B}_S$ system is evaluated by numerically solving ${\rm det}\left[i\tilde{\cal D}_{(3)}^{-1}(p_0,{\bm 0})\right]=0$.

\subsection{$\sigma$ - $B$ - $\bar{B}$ - $\omega$ - $B_{AS}$ - $\bar{B}_{AS}$  system}
\label{sec:Mass4}

Here, we derive the mass formulas for the $\sigma$ - $B$ - $\bar{B}$ - $\omega$ - $B_{AS}$ - $\bar{B}_{AS}$ system. In this case, $\sigma$, $B$, $\bar{B}$ (or $B^4$, $B^5$) and the time components of $\omega$, $B_{AS}$, $\bar{B}_{AS}$ (or $V'^4$, $V'^5$) can mix. In addition, the space components of $\omega$, $B_{AS}$, $\bar{B}_{AS}$ (or $V'^4$, $V'^5$) also mix with each other.

First, we consider the mixing among $\sigma$, $B$, $\bar{B}$ (or $B^4$, $B^5$) and the time components of $\omega$, $B_{AS}$, $\bar{B}_{AS}$. The inverse propagator of these six states at vanishing momentum ${\bm p}={\bm 0}$, $i{\cal D}_{(4)}^{-1}(p_0,{\bm 0})$, is given by
\begin{widetext}
\begin{eqnarray}
i{\cal D}_{(4)}^{-1}(p_0,{\bm 0}) = \left(
\begin{array}{cccccc}
{\cal M}_{\sigma\sigma} & {\cal M}_{\sigma B_4} & {\cal M}_{\sigma B_5} & {\cal M}_{\sigma\omega} & {\cal M}_{\sigma V_4'} & {\cal M}_{\sigma V_5'}   \\
{\cal M}_{B_4\sigma} & {\cal M}_{B_4 B_4} & {\cal M}_{B_4 B_5} & {\cal M}_{B_4\omega} & {\cal M}_{B_4 V_4'} & {\cal M}_{B_4 V_5'}   \\
{\cal M}_{B_5\sigma} & {\cal M}_{B_5 B_4} & {\cal M}_{B_5 B_5} & {\cal M}_{B_5\omega} & {\cal M}_{B_5 V_4'} & {\cal M}_{B_5 V_5'}   \\
{\cal M}_{\omega\sigma} & {\cal M}_{\omega B_4} & {\cal M}_{\omega B_5} & {\cal M}_{\omega\omega} & {\cal M}_{\omega V_4'} & 0 \\
{\cal M}_{V_4'\sigma} & {\cal M}_{V_4' B_4} & {\cal M}_{V_4' B_5} & {\cal M}_{V_4'\omega} & {\cal M}_{V_4' V_4'} & 0  \\
{\cal M}_{V_5'\sigma} & {\cal M}_{V_5' B_4} & {\cal M}_{V_5' B_5} & 0& 0 & {\cal M}_{V_5' V_5'}   \\
\end{array}
\right)\ , \label{iDSigma}
\end{eqnarray}
\end{widetext}
where each matrix element reads
\begin{eqnarray}
{\cal M}_{\sigma\sigma} = p_0^2+\frac{C^2}{8}\bar{V}^2-m_\sigma^2 \ ,
\end{eqnarray}
\begin{eqnarray}
{\cal M}_{\sigma B_4} = - {\cal M}_{ B_4\sigma}= i\frac{C}{\sqrt{2}}\bar{V}p_0\ ,
\end{eqnarray}
\begin{eqnarray}
{\cal M}_{\sigma B_5} = {\cal M}_{ B_5\sigma} = -\frac{C}{\sqrt{2}}\bar{V}\mu_q -\frac{\lambda_2}{2}\sigma_0\Delta -\frac{C_3\bar{\omega}}{4}\bar{V}\ , \nonumber\\
\end{eqnarray}
\begin{eqnarray}
{\cal M}_{\sigma \omega} = {\cal M}_{ \omega\sigma}  =  \frac{C-C_3}{4}\bar{\omega}\sigma_0 -\frac{C_3}{4}\Delta\bar{V} \ ,
\end{eqnarray}
\begin{eqnarray}
{\cal M}_{\sigma V_4'} = {\cal M}_{ V_4'\sigma}  = \frac{C+C_3}{4}\sigma_0\bar{V} -\frac{C}{\sqrt{2}}\Delta\mu_q -\frac{C_3}{4}\Delta\bar{\omega} \ , \nonumber\\
\end{eqnarray}
\begin{eqnarray}
{\cal M}_{\sigma V_5'} =-{\cal M}_{ V_5'\sigma} = i\frac{C}{2\sqrt{2}}\Delta p_0\ ,
\end{eqnarray}
\begin{eqnarray}
{\cal M}_{ B_4B_4} = p_0^2 + \frac{C^2}{8}\bar{V}^2  + \left(2\mu_q+\frac{C}{2\sqrt{2}}\bar{\omega}\right)^2 -m_{B_4}^2\ , \nonumber\\
\end{eqnarray}
\begin{eqnarray}
{\cal M}_{ B_4B_5} =-{\cal M}_{ B_5B_4} = 2i\left(2\mu_q+\frac{C}{2\sqrt{2}}\bar{\omega}\right)p_0\ ,
\end{eqnarray}
\begin{eqnarray}
{\cal M}_{ B_4\omega} = -{\cal M}_{ \omega B_4}= i\frac{C}{2\sqrt{2}}\Delta p_0\ ,
\end{eqnarray}
\begin{eqnarray}
{\cal M}_{ B_4V_4'} = -{\cal M}_{V_4'B_4}  = -i\frac{C}{2\sqrt{2}}\sigma_0 p_0\ ,
\end{eqnarray}
\begin{eqnarray}
{\cal M}_{ B_4V_5'} ={\cal M}_{ V_5'B_4} =\frac{C_3}{4}(\Delta\bar{V}_4+\sigma_0\bar{\omega})+\frac{C}{\sqrt{2}}\sigma_0\mu_q \ , \nonumber\\
\end{eqnarray}
\begin{eqnarray}
{\cal M}_{B_5B_5} = p_0^2 + \left(2\mu_q + \frac{C}{2\sqrt{2}}\bar{\omega}\right)^2 -m_{B_5}^2\ , 
\end{eqnarray}
\begin{eqnarray}
{\cal M}_{B_5\omega} = {\cal M}_{\omega B_5} &=& \sqrt{2}C\mu_q\Delta + \frac{C+C_3}{4}\Delta\bar{\omega} \nonumber\\
&& -\frac{C_3}{4}\sigma_0\bar{V}\ , 
\end{eqnarray}
\begin{eqnarray}
{\cal M}_{B_5V_4'} = {\cal M}_{V_4'B_5}  &=& \frac{C-C_3}{4}\Delta\bar{V} -\frac{C_3}{4}\sigma_0\bar{\omega} \nonumber\\
&& -\frac{C}{\sqrt{2}} \sigma_0\mu_q \ ,  
\end{eqnarray}
\begin{eqnarray}
{\cal M}_{B_5V_5'} = -{\cal M}_{V_5'B_5}  = -i\frac{C}{2\sqrt{2}}\sigma_0p_0\ ,
\end{eqnarray}
\begin{eqnarray}
{\cal M}_{\omega \omega} = (m_{\omega}^t)^2\ ,
\end{eqnarray}
\begin{eqnarray}
{\cal M}_{\omega V_4'} = {\cal M}_{ V_4' \omega}  = -\frac{C_3}{4}\sigma_0\Delta\ ,
\end{eqnarray}
\begin{eqnarray}
{\cal M}_{ V_4' V_4'} = (m_{V_4'}^t)^2\ ,
\end{eqnarray}
and
\begin{eqnarray}
{\cal M}_{V_5'V_5'} =  (m_{V_5'}^t)^2\ ,
\end{eqnarray}
with the mass parameters
\begin{eqnarray}
m_\sigma^2 &=& m_0^2+\frac{\lambda_2}{4}(3\sigma_0^2+\Delta^2) \nonumber\\
&& - \frac{C-C_3}{8}\bar{\omega}^2  -\frac{C-C^2+C_3}{8}\bar{V}^2\ ,
\end{eqnarray}
\begin{eqnarray}
m_{B_4}^2 &=& m_0^2+\frac{\lambda_2}{4}(\sigma_0^2+\Delta^2) \nonumber\\
&& -\frac{C-C^2+C_3}{8}(\bar{V}^2+\bar{\omega}^2 )\ ,
\end{eqnarray}
\begin{eqnarray}
m_{B_5}^2 &=& m_0^2+\frac{\lambda_2}{4}(\sigma_0^2+3\Delta^2) \nonumber\\
&& -\frac{C-C^2+C_3}{8}\bar{\omega}^2 -\frac{C-C_3}{8}\bar{V}^2 \ ,
\end{eqnarray} 
\begin{eqnarray}
(m_{\omega}^t)^2 = m_1^2+\frac{C}{8}(\sigma_0^2+\Delta^2)-\frac{C_3}{8}(\sigma_0^2-\Delta^2) \ ,
\end{eqnarray}
\begin{eqnarray}
(m_{V'_{4}}^t)^2 = m_1^2+\frac{C}{8}(\sigma_0^2+\Delta^2)+\frac{C_3}{8}(\sigma_0^2-\Delta^2) \ ,
\end{eqnarray}
and
\begin{eqnarray}
(m_{V'_{5}}^t)^2 = m_1^2+\frac{C+C_3}{8}(\sigma_0^2+\Delta^2) \ .
\end{eqnarray}
In the presence of the mixing, the mass spectrum for the $\sigma$ - $B$ - $\bar{B}$ system is evaluated by numerically solving $i{\cal D}_{(4)}^{-1}(p_0,{\bm 0})=0$.

Next, we consider the mixing among the spatial components of $\omega$, $B_{AS}$ and $\bar{B}_{AS}$. In this case, the corresponding inverse propagator at vanishing momentum ${\bm p}={\bm 0}$, $i\tilde{\cal D}_{(4)}^{-1}(p_0,{\bm 0})$, is given by
\begin{eqnarray}
i\tilde{\cal D}_{(4)}^{-1}(p_0,{\bm 0}) = \left(
\begin{array}{ccc}
 \tilde{\cal M}_{\omega\omega} &  \tilde{\cal M}_{\omega V_4'}  & \tilde{\cal M}_{\omega V_5'} \\
\tilde{\cal M}_{V_4'\omega} & \tilde{\cal M}_{V_4'V_4'} & \tilde{\cal M}_{V_4'V_5'} \\
\tilde{\cal M}_{V_5'\omega} &  \tilde{\cal M}_{V_5'V_4'} & \tilde{\cal M}_{V_5'V_5'} \\
\end{array}
\right)\ , \nonumber\\
\end{eqnarray}
where each matrix element reads
\begin{eqnarray}
\tilde{\cal M}_{\omega\omega} = -p_0^2-\frac{g_\Phi^2}{2}\bar{V}^2 + (m_{\omega}^s)^2\ ,
\end{eqnarray}
\begin{eqnarray}
\tilde{\cal M}_{\omega V_4'}  = \tilde{\cal M}_{ V_4'\omega} = -\frac{C_3}{4}\sigma_0\Delta+\sqrt{2}g_\Phi\bar{V}\mu_q+\frac{g_\Phi^2}{2}\bar{V}\bar{\omega} \ , \nonumber\\
\end{eqnarray}
\begin{eqnarray}
\tilde{\cal M}_{\omega V_5'} = -\tilde{\cal M}_{ V_5'\omega} = \sqrt{2}ig_\Phi\bar{V}p_0\ ,
\end{eqnarray}
\begin{eqnarray}
\tilde{\cal M}_{V_4'V_4'} = -p_0^2-\left(2\mu_q+\frac{g_\Phi}{\sqrt{2}}\bar{\omega}\right)^2 + (m_{V_4'}^s)^2\ ,
\end{eqnarray}
\begin{eqnarray}
\tilde{\cal M}_{V_4'V_5'} = -\tilde{\cal M}_{V_5'V_4'} = -2i\left(2\mu_q+\frac{g_\Phi}{\sqrt{2}}\bar{\omega}\right)p_0\ ,
\end{eqnarray}
and
\begin{eqnarray}
\tilde{\cal M}_{V_5'V_5'} = -p_0^2-\frac{g_\Phi^2}{2}\bar{V}^2-\left(2\mu_q+\frac{g_\Phi}{\sqrt{2}}\bar{\omega}\right)^2 + (m_{V_5'}^s)^2\ ,\nonumber\\
\end{eqnarray}
with the mass parameters
\begin{eqnarray}
(m_{\omega}^s)^2 = m_1^2+\frac{C}{8}(\sigma_0^2+\Delta^2)-\frac{C_3}{8}(\sigma_0^2-\Delta^2)\ , 
\end{eqnarray}
\begin{eqnarray}
(m_{V'_{4}}^s)^2 = m_1^2+\frac{C}{8}(\sigma_0^2+\Delta^2)+\frac{C_3}{8}(\sigma_0^2-\Delta^2) \ ,
\end{eqnarray}
and
\begin{eqnarray}
(m_{V'_{5}}^s)^2 = m_1^2+ \frac{C+C_3}{8}(\sigma_0^2+\Delta^2)  \ .
\end{eqnarray}
In the presence of the mixing, the mass spectrum for the $\omega$ - $B_{AS}$ - $\bar{B}_{AS}$ system is evaluated by numerically solving ${\rm det}\left[i\tilde{\cal D}_{(4)}^{-1}(p_0,{\bm 0})\right]=0$. 

\section{Masses in the hadronic phase}
\label{sec:MassHadronic}

As derived in Appendix~\ref{sec:MassFormulas}, in general, the hadron masses are evaluated by pole positions of the appropriate propagator matrices, which would be obtained numerically. When we focus on the hadronic phase, the mass formulas can be evaluated analytically. In this appendix, we exhibit the resultant mass formulas for all hadrons. 

Such formulas can simply be obtained by taking $\Delta=0$ and accordingly $\bar{\omega}=\bar{V}=0$, while keeping $\sigma_0\neq0$, in the mass formulas in Appendix~\ref{sec:MassFormulas}. In this limit, one can find
\begin{eqnarray}
 m_\omega^{(\rm H)} = m_{\rho}^{(\rm H)} = \sqrt{m_1^2 + \frac{C-C_3}{8}\big(\sigma_0^{(\rm H)}\big)^2}\ ,
 \end{eqnarray}
\begin{eqnarray}
m_{f_1}^{(\rm H)} = m_{a_1}^{(\rm H)} = \sqrt{m_1^2 + \frac{C+C_3}{8}\big(\sigma_0^{(\rm H)}\big)^2}\ , \label{MF1Had}
\end{eqnarray}
\begin{eqnarray}
m_{B_S}^{(\rm H)} &=&  m_{\omega}^{(\rm H)}-2\mu_q\ , \nonumber\\
m_{\bar{B}_S}^{(\rm H)} &=& m_{\omega}^{(\rm H)} + 2\mu_q \ , \label{MassBSHad}
\end{eqnarray}
\begin{eqnarray}
m_{B_{\rm AS}}^{(\rm H)} &=& m_{f_1}^{(\rm H)} -2\mu_q\ ,  \nonumber\\
m_{\bar{B}_{\rm AS}}^{(\rm H)} &=& m_{f_1}^{(\rm H)} + 2\mu_q \ , \label{MassBASHad}
\end{eqnarray}
for the spin-$1$ hadrons, while
\begin{eqnarray}
m_\pi^{(\rm H)} = Z_\pi \sqrt{m_0^2+\frac{\lambda_2}{4}\big(\sigma_0^{(\rm H)}\big)^2}\ ,  \label{MPiHad}
\end{eqnarray}
\begin{eqnarray}
m_\eta^{(\rm H)} = Z_\eta \sqrt{m_0^2+\frac{\lambda_2}{4}\big(\sigma_0^{(\rm H)}\big)^2}\ , \label{MEtaHad}
\end{eqnarray}
\begin{eqnarray}
m_{a_0}^{(\rm H)} = \sqrt{m_0^2+\frac{3\lambda_2}{4}\big(\sigma_0^{(\rm H)}\big)^2}\ , 
\end{eqnarray}
\begin{eqnarray}
m_\sigma^{(\rm H)} = \sqrt{m_0^2+3\frac{\lambda_2}{4}\big(\sigma_0^{(\rm H)}\big)^2} \ ,
\end{eqnarray}
\begin{eqnarray}
m_{B}^{(\rm H)} &=& m_\pi^{(\rm H)}-2\mu_q\ , \nonumber\\
m_{\bar{B}}^{(\rm H)} &=& m_\pi^{(\rm H)}+2\mu_q \ ,
\end{eqnarray}
\begin{eqnarray}
m_{B'}^{(\rm H)} &=&m_{a_0}^{(\rm H)} -2\mu_q\ , \nonumber\\
m_{\bar{B}'}^{(\rm H)} &=&m_{a_0}^{(\rm H)}  + 2\mu_q \ , 
\end{eqnarray}
for the spin-$0$ hadrons. In Eqs.~(\ref{MEtaHad}) and~(\ref{MPiHad}), the renormalization factors $Z_\pi$ and $Z_\eta$ are defined by
\begin{eqnarray}
Z_\pi &=& \left(1-\frac{C^2\big(\sigma_0^{(\rm H)}\big)^2}{8\big(m_{a_1}^{(\rm H)}\big)^2}\right)^{-1/2}\ , \nonumber\\
Z_\eta &=& \left(1-\frac{C^2\big(\sigma_0^{(\rm H)}\big)^2}{8\big(m_{f_1}^{(\rm H)}\big)^2}\right)^{-1/2}\ , \label{ZDef}
\end{eqnarray}
which stems from the $\pi$ - $a_1$ mixing and the $\eta$ - $f_1$ mixing, respectively. It should be noted that $Z_\pi = Z_\eta$ follows from $m_{f_1}^{(\rm H)} = m_{a_1}^{(\rm H)}$. These types of mixing originate from the spontaneous breakdown of chiral symmetry since $Z_\pi=Z_\eta=1$ when $(\sigma_0^{(\rm H)})=0$, as in the three-color eLSM~\cite{Parganlija:2012fy}.

\bibliography{reference}

\begin{thebibliography}{74}%
\makeatletter
\providecommand \@ifxundefined [1]{%
 \@ifx{#1\undefined}
}%
\providecommand \@ifnum [1]{%
 \ifnum #1\expandafter \@firstoftwo
 \else \expandafter \@secondoftwo
 \fi
}%
\providecommand \@ifx [1]{%
 \ifx #1\expandafter \@firstoftwo
 \else \expandafter \@secondoftwo
 \fi
}%
\providecommand \natexlab [1]{#1}%
\providecommand \enquote  [1]{``#1''}%
\providecommand \bibnamefont  [1]{#1}%
\providecommand \bibfnamefont [1]{#1}%
\providecommand \citenamefont [1]{#1}%
\providecommand \href@noop [0]{\@secondoftwo}%
\providecommand \href [0]{\begingroup \@sanitize@url \@href}%
\providecommand \@href[1]{\@@startlink{#1}\@@href}%
\providecommand \@@href[1]{\endgroup#1\@@endlink}%
\providecommand \@sanitize@url [0]{\catcode `\\12\catcode `\$12\catcode
  `\&12\catcode `\#12\catcode `\^12\catcode `\_12\catcode `\%12\relax}%
\providecommand \@@startlink[1]{}%
\providecommand \@@endlink[0]{}%
\providecommand \url  [0]{\begingroup\@sanitize@url \@url }%
\providecommand \@url [1]{\endgroup\@href {#1}{\urlprefix }}%
\providecommand \urlprefix  [0]{URL }%
\providecommand \Eprint [0]{\href }%
\providecommand \doibase [0]{http://dx.doi.org/}%
\providecommand \selectlanguage [0]{\@gobble}%
\providecommand \bibinfo  [0]{\@secondoftwo}%
\providecommand \bibfield  [0]{\@secondoftwo}%
\providecommand \translation [1]{[#1]}%
\providecommand \BibitemOpen [0]{}%
\providecommand \bibitemStop [0]{}%
\providecommand \bibitemNoStop [0]{.\EOS\space}%
\providecommand \EOS [0]{\spacefactor3000\relax}%
\providecommand \BibitemShut  [1]{\csname bibitem#1\endcsname}%
\let\auto@bib@innerbib\@empty
\bibitem [{\citenamefont {Holt}\ \emph {et~al.}(2016)\citenamefont {Holt},
  \citenamefont {Rho},\ and\ \citenamefont {Weise}}]{Holt:2014hma}%
  \BibitemOpen
  \bibfield  {author} {\bibinfo {author} {\bibfnamefont {Jeremy~W.}\
  \bibnamefont {Holt}}, \bibinfo {author} {\bibfnamefont {Mannque}\
  \bibnamefont {Rho}}, \ and\ \bibinfo {author} {\bibfnamefont {Wolfram}\
  \bibnamefont {Weise}},\ }\bibfield  {title} {\enquote {\bibinfo {title}
  {{Chiral symmetry and effective field theories for hadronic, nuclear and
  stellar matter}},}\ }\href {\doibase 10.1016/j.physrep.2015.10.011}
  {\bibfield  {journal} {\bibinfo  {journal} {Phys. Rept.}\ }\textbf {\bibinfo
  {volume} {621}},\ \bibinfo {pages} {2--75} (\bibinfo {year} {2016})},\
  \Eprint {http://arxiv.org/abs/1411.6681} {arXiv:1411.6681 [nucl-th]}
  \BibitemShut {NoStop}%
\bibitem [{\citenamefont {Metag}\ \emph {et~al.}(2017)\citenamefont {Metag},
  \citenamefont {Nanova},\ and\ \citenamefont {Paryev}}]{Metag:2017yuh}%
  \BibitemOpen
  \bibfield  {author} {\bibinfo {author} {\bibfnamefont {V.}~\bibnamefont
  {Metag}}, \bibinfo {author} {\bibfnamefont {M.}~\bibnamefont {Nanova}}, \
  and\ \bibinfo {author} {\bibfnamefont {E.~Ya.}\ \bibnamefont {Paryev}},\
  }\bibfield  {title} {\enquote {\bibinfo {title} {{Meson\textendash{}nucleus
  potentials and the search for meson\textendash{}nucleus bound states}},}\
  }\href {\doibase 10.1016/j.ppnp.2017.08.002} {\bibfield  {journal} {\bibinfo
  {journal} {Prog. Part. Nucl. Phys.}\ }\textbf {\bibinfo {volume} {97}},\
  \bibinfo {pages} {199--260} (\bibinfo {year} {2017})},\ \Eprint
  {http://arxiv.org/abs/1706.09654} {arXiv:1706.09654 [nucl-ex]} \BibitemShut
  {NoStop}%
\bibitem [{\citenamefont {Aarts}(2016)}]{Aarts:2015tyj}%
  \BibitemOpen
  \bibfield  {author} {\bibinfo {author} {\bibfnamefont {Gert}\ \bibnamefont
  {Aarts}},\ }\bibfield  {title} {\enquote {\bibinfo {title} {{Introductory
  lectures on lattice QCD at nonzero baryon number}},}\ }\href {\doibase
  10.1088/1742-6596/706/2/022004} {\bibfield  {journal} {\bibinfo  {journal}
  {J. Phys. Conf. Ser.}\ }\textbf {\bibinfo {volume} {706}},\ \bibinfo {pages}
  {022004} (\bibinfo {year} {2016})},\ \Eprint
  {http://arxiv.org/abs/1512.05145} {arXiv:1512.05145 [hep-lat]} \BibitemShut
  {NoStop}%
\bibitem [{\citenamefont {Nagata}(2022)}]{Nagata:2021ugx}%
  \BibitemOpen
  \bibfield  {author} {\bibinfo {author} {\bibfnamefont {Keitaro}\ \bibnamefont
  {Nagata}},\ }\bibfield  {title} {\enquote {\bibinfo {title} {{Finite-density
  lattice QCD and sign problem: Current status and open problems}},}\ }\href
  {\doibase 10.1016/j.ppnp.2022.103991} {\bibfield  {journal} {\bibinfo
  {journal} {Prog. Part. Nucl. Phys.}\ }\textbf {\bibinfo {volume} {127}},\
  \bibinfo {pages} {103991} (\bibinfo {year} {2022})},\ \Eprint
  {http://arxiv.org/abs/2108.12423} {arXiv:2108.12423 [hep-lat]} \BibitemShut
  {NoStop}%
\bibitem [{\citenamefont {Muroya}\ \emph
  {et~al.}(2003{\natexlab{a}})\citenamefont {Muroya}, \citenamefont {Nakamura},
  \citenamefont {Nonaka},\ and\ \citenamefont {Takaishi}}]{Muroya:2003qs}%
  \BibitemOpen
  \bibfield  {author} {\bibinfo {author} {\bibfnamefont {Shin}\ \bibnamefont
  {Muroya}}, \bibinfo {author} {\bibfnamefont {Atsushi}\ \bibnamefont
  {Nakamura}}, \bibinfo {author} {\bibfnamefont {Chiho}\ \bibnamefont
  {Nonaka}}, \ and\ \bibinfo {author} {\bibfnamefont {Tetsuya}\ \bibnamefont
  {Takaishi}},\ }\bibfield  {title} {\enquote {\bibinfo {title} {{Lattice QCD
  at finite density: An Introductory review}},}\ }\href {\doibase
  10.1143/PTP.110.615} {\bibfield  {journal} {\bibinfo  {journal} {Prog. Theor.
  Phys.}\ }\textbf {\bibinfo {volume} {110}},\ \bibinfo {pages} {615--668}
  (\bibinfo {year} {2003}{\natexlab{a}})},\ \Eprint
  {http://arxiv.org/abs/hep-lat/0306031} {arXiv:hep-lat/0306031} \BibitemShut
  {NoStop}%
\bibitem [{\citenamefont {Hands}\ \emph {et~al.}(1999)\citenamefont {Hands},
  \citenamefont {Kogut}, \citenamefont {Lombardo},\ and\ \citenamefont
  {Morrison}}]{Hands:1999md}%
  \BibitemOpen
  \bibfield  {author} {\bibinfo {author} {\bibfnamefont {Simon}\ \bibnamefont
  {Hands}}, \bibinfo {author} {\bibfnamefont {John~B.}\ \bibnamefont {Kogut}},
  \bibinfo {author} {\bibfnamefont {Maria-Paola}\ \bibnamefont {Lombardo}}, \
  and\ \bibinfo {author} {\bibfnamefont {Susan~E.}\ \bibnamefont {Morrison}},\
  }\bibfield  {title} {\enquote {\bibinfo {title} {{Symmetries and spectrum of
  SU(2) lattice gauge theory at finite chemical potential}},}\ }\href {\doibase
  10.1016/S0550-3213(99)00364-8} {\bibfield  {journal} {\bibinfo  {journal}
  {Nucl. Phys. B}\ }\textbf {\bibinfo {volume} {558}},\ \bibinfo {pages}
  {327--346} (\bibinfo {year} {1999})},\ \Eprint
  {http://arxiv.org/abs/hep-lat/9902034} {arXiv:hep-lat/9902034} \BibitemShut
  {NoStop}%
\bibitem [{\citenamefont {Kogut}\ \emph {et~al.}(2001)\citenamefont {Kogut},
  \citenamefont {Sinclair}, \citenamefont {Hands},\ and\ \citenamefont
  {Morrison}}]{Kogut:2001na}%
  \BibitemOpen
  \bibfield  {author} {\bibinfo {author} {\bibfnamefont {J.~B.}\ \bibnamefont
  {Kogut}}, \bibinfo {author} {\bibfnamefont {D.~K.}\ \bibnamefont {Sinclair}},
  \bibinfo {author} {\bibfnamefont {S.~J.}\ \bibnamefont {Hands}}, \ and\
  \bibinfo {author} {\bibfnamefont {S.~E.}\ \bibnamefont {Morrison}},\
  }\bibfield  {title} {\enquote {\bibinfo {title} {{Two color QCD at nonzero
  quark number density}},}\ }\href {\doibase 10.1103/PhysRevD.64.094505}
  {\bibfield  {journal} {\bibinfo  {journal} {Phys. Rev. D}\ }\textbf {\bibinfo
  {volume} {64}},\ \bibinfo {pages} {094505} (\bibinfo {year} {2001})},\
  \Eprint {http://arxiv.org/abs/hep-lat/0105026} {arXiv:hep-lat/0105026}
  \BibitemShut {NoStop}%
\bibitem [{\citenamefont {Hands}\ \emph {et~al.}(2001)\citenamefont {Hands},
  \citenamefont {Montvay}, \citenamefont {Scorzato},\ and\ \citenamefont
  {Skullerud}}]{Hands:2001ee}%
  \BibitemOpen
  \bibfield  {author} {\bibinfo {author} {\bibfnamefont {Simon}\ \bibnamefont
  {Hands}}, \bibinfo {author} {\bibfnamefont {Istvan}\ \bibnamefont {Montvay}},
  \bibinfo {author} {\bibfnamefont {Luigi}\ \bibnamefont {Scorzato}}, \ and\
  \bibinfo {author} {\bibfnamefont {Jonivar}\ \bibnamefont {Skullerud}},\
  }\bibfield  {title} {\enquote {\bibinfo {title} {{Diquark condensation in
  dense adjoint matter}},}\ }\href {\doibase 10.1007/s100520100836} {\bibfield
  {journal} {\bibinfo  {journal} {Eur. Phys. J. C}\ }\textbf {\bibinfo {volume}
  {22}},\ \bibinfo {pages} {451--461} (\bibinfo {year} {2001})},\ \Eprint
  {http://arxiv.org/abs/hep-lat/0109029} {arXiv:hep-lat/0109029} \BibitemShut
  {NoStop}%
\bibitem [{\citenamefont {Muroya}\ \emph
  {et~al.}(2003{\natexlab{b}})\citenamefont {Muroya}, \citenamefont
  {Nakamura},\ and\ \citenamefont {Nonaka}}]{Muroya:2002ry}%
  \BibitemOpen
  \bibfield  {author} {\bibinfo {author} {\bibfnamefont {Shin}\ \bibnamefont
  {Muroya}}, \bibinfo {author} {\bibfnamefont {Atsushi}\ \bibnamefont
  {Nakamura}}, \ and\ \bibinfo {author} {\bibfnamefont {Chiho}\ \bibnamefont
  {Nonaka}},\ }\bibfield  {title} {\enquote {\bibinfo {title} {{Behavior of
  hadrons at finite density: Lattice study of color SU(2) QCD}},}\ }\href
  {\doibase 10.1016/S0370-2693(02)03065-4} {\bibfield  {journal} {\bibinfo
  {journal} {Phys. Lett. B}\ }\textbf {\bibinfo {volume} {551}},\ \bibinfo
  {pages} {305--310} (\bibinfo {year} {2003}{\natexlab{b}})},\ \Eprint
  {http://arxiv.org/abs/hep-lat/0211010} {arXiv:hep-lat/0211010} \BibitemShut
  {NoStop}%
\bibitem [{\citenamefont {Chandrasekharan}\ and\ \citenamefont
  {Jiang}(2006)}]{Chandrasekharan:2006tz}%
  \BibitemOpen
  \bibfield  {author} {\bibinfo {author} {\bibfnamefont {Shailesh}\
  \bibnamefont {Chandrasekharan}}\ and\ \bibinfo {author} {\bibfnamefont
  {Fu-Jiun}\ \bibnamefont {Jiang}},\ }\bibfield  {title} {\enquote {\bibinfo
  {title} {{Phase-diagram of two-color lattice QCD in the chiral limit}},}\
  }\href {\doibase 10.1103/PhysRevD.74.014506} {\bibfield  {journal} {\bibinfo
  {journal} {Phys. Rev. D}\ }\textbf {\bibinfo {volume} {74}},\ \bibinfo
  {pages} {014506} (\bibinfo {year} {2006})},\ \Eprint
  {http://arxiv.org/abs/hep-lat/0602031} {arXiv:hep-lat/0602031} \BibitemShut
  {NoStop}%
\bibitem [{\citenamefont {Hands}\ \emph {et~al.}(2006)\citenamefont {Hands},
  \citenamefont {Kim},\ and\ \citenamefont {Skullerud}}]{Hands:2006ve}%
  \BibitemOpen
  \bibfield  {author} {\bibinfo {author} {\bibfnamefont {Simon}\ \bibnamefont
  {Hands}}, \bibinfo {author} {\bibfnamefont {Seyong}\ \bibnamefont {Kim}}, \
  and\ \bibinfo {author} {\bibfnamefont {Jon-Ivar}\ \bibnamefont {Skullerud}},\
  }\bibfield  {title} {\enquote {\bibinfo {title} {{Deconfinement in dense
  2-color QCD}},}\ }\href {\doibase 10.1140/epjc/s2006-02621-8} {\bibfield
  {journal} {\bibinfo  {journal} {Eur. Phys. J. C}\ }\textbf {\bibinfo {volume}
  {48}},\ \bibinfo {pages} {193} (\bibinfo {year} {2006})},\ \Eprint
  {http://arxiv.org/abs/hep-lat/0604004} {arXiv:hep-lat/0604004} \BibitemShut
  {NoStop}%
\bibitem [{\citenamefont {Hands}\ \emph {et~al.}(2008)\citenamefont {Hands},
  \citenamefont {Sitch},\ and\ \citenamefont {Skullerud}}]{Hands:2007uc}%
  \BibitemOpen
  \bibfield  {author} {\bibinfo {author} {\bibfnamefont {Simon}\ \bibnamefont
  {Hands}}, \bibinfo {author} {\bibfnamefont {Peter}\ \bibnamefont {Sitch}}, \
  and\ \bibinfo {author} {\bibfnamefont {Jon-Ivar}\ \bibnamefont {Skullerud}},\
  }\bibfield  {title} {\enquote {\bibinfo {title} {{Hadron Spectrum in a
  Two-Colour Baryon-Rich Medium}},}\ }\href {\doibase
  10.1016/j.physletb.2008.01.078} {\bibfield  {journal} {\bibinfo  {journal}
  {Phys. Lett. B}\ }\textbf {\bibinfo {volume} {662}},\ \bibinfo {pages}
  {405--412} (\bibinfo {year} {2008})},\ \Eprint
  {http://arxiv.org/abs/0710.1966} {arXiv:0710.1966 [hep-lat]} \BibitemShut
  {NoStop}%
\bibitem [{\citenamefont {Hands}\ \emph {et~al.}(2010)\citenamefont {Hands},
  \citenamefont {Kim},\ and\ \citenamefont {Skullerud}}]{Hands:2010gd}%
  \BibitemOpen
  \bibfield  {author} {\bibinfo {author} {\bibfnamefont {Simon}\ \bibnamefont
  {Hands}}, \bibinfo {author} {\bibfnamefont {Seyong}\ \bibnamefont {Kim}}, \
  and\ \bibinfo {author} {\bibfnamefont {Jon-Ivar}\ \bibnamefont {Skullerud}},\
  }\bibfield  {title} {\enquote {\bibinfo {title} {{A Quarkyonic Phase in Dense
  Two Color Matter?}}}\ }\href {\doibase 10.1103/PhysRevD.81.091502} {\bibfield
   {journal} {\bibinfo  {journal} {Phys. Rev. D}\ }\textbf {\bibinfo {volume}
  {81}},\ \bibinfo {pages} {091502} (\bibinfo {year} {2010})},\ \Eprint
  {http://arxiv.org/abs/1001.1682} {arXiv:1001.1682 [hep-lat]} \BibitemShut
  {NoStop}%
\bibitem [{\citenamefont {Cotter}\ \emph {et~al.}(2013)\citenamefont {Cotter},
  \citenamefont {Giudice}, \citenamefont {Hands},\ and\ \citenamefont
  {Skullerud}}]{Cotter:2012mb}%
  \BibitemOpen
  \bibfield  {author} {\bibinfo {author} {\bibfnamefont {Seamus}\ \bibnamefont
  {Cotter}}, \bibinfo {author} {\bibfnamefont {Pietro}\ \bibnamefont
  {Giudice}}, \bibinfo {author} {\bibfnamefont {Simon}\ \bibnamefont {Hands}},
  \ and\ \bibinfo {author} {\bibfnamefont {Jon-Ivar}\ \bibnamefont
  {Skullerud}},\ }\bibfield  {title} {\enquote {\bibinfo {title} {{Towards the
  phase diagram of dense two-color matter}},}\ }\href {\doibase
  10.1103/PhysRevD.87.034507} {\bibfield  {journal} {\bibinfo  {journal} {Phys.
  Rev. D}\ }\textbf {\bibinfo {volume} {87}},\ \bibinfo {pages} {034507}
  (\bibinfo {year} {2013})},\ \Eprint {http://arxiv.org/abs/1210.4496}
  {arXiv:1210.4496 [hep-lat]} \BibitemShut {NoStop}%
\bibitem [{\citenamefont {Hands}\ \emph {et~al.}(2012)\citenamefont {Hands},
  \citenamefont {Kim},\ and\ \citenamefont {Skullerud}}]{Hands:2012yy}%
  \BibitemOpen
  \bibfield  {author} {\bibinfo {author} {\bibfnamefont {Simon}\ \bibnamefont
  {Hands}}, \bibinfo {author} {\bibfnamefont {Seyong}\ \bibnamefont {Kim}}, \
  and\ \bibinfo {author} {\bibfnamefont {Jon-Ivar}\ \bibnamefont {Skullerud}},\
  }\bibfield  {title} {\enquote {\bibinfo {title} {{Non-relativistic spectrum
  of two-color QCD at non-zero baryon density}},}\ }\href {\doibase
  10.1016/j.physletb.2012.04.002} {\bibfield  {journal} {\bibinfo  {journal}
  {Phys. Lett. B}\ }\textbf {\bibinfo {volume} {711}},\ \bibinfo {pages}
  {199--204} (\bibinfo {year} {2012})},\ \Eprint
  {http://arxiv.org/abs/1202.4353} {arXiv:1202.4353 [hep-lat]} \BibitemShut
  {NoStop}%
\bibitem [{\citenamefont {Boz}\ \emph {et~al.}(2013)\citenamefont {Boz},
  \citenamefont {Cotter}, \citenamefont {Fister}, \citenamefont {Mehta},\ and\
  \citenamefont {Skullerud}}]{Boz:2013rca}%
  \BibitemOpen
  \bibfield  {author} {\bibinfo {author} {\bibfnamefont {Tamer}\ \bibnamefont
  {Boz}}, \bibinfo {author} {\bibfnamefont {Seamus}\ \bibnamefont {Cotter}},
  \bibinfo {author} {\bibfnamefont {Leonard}\ \bibnamefont {Fister}}, \bibinfo
  {author} {\bibfnamefont {Dhagash}\ \bibnamefont {Mehta}}, \ and\ \bibinfo
  {author} {\bibfnamefont {Jon-Ivar}\ \bibnamefont {Skullerud}},\ }\bibfield
  {title} {\enquote {\bibinfo {title} {{Phase transitions and gluodynamics in
  2-colour matter at high density}},}\ }\href {\doibase
  10.1140/epja/i2013-13087-6} {\bibfield  {journal} {\bibinfo  {journal} {Eur.
  Phys. J. A}\ }\textbf {\bibinfo {volume} {49}},\ \bibinfo {pages} {87}
  (\bibinfo {year} {2013})},\ \Eprint {http://arxiv.org/abs/1303.3223}
  {arXiv:1303.3223 [hep-lat]} \BibitemShut {NoStop}%
\bibitem [{\citenamefont {Braguta}\ \emph {et~al.}(2016)\citenamefont
  {Braguta}, \citenamefont {Ilgenfritz}, \citenamefont {Kotov}, \citenamefont
  {Molochkov},\ and\ \citenamefont {Nikolaev}}]{Braguta:2016cpw}%
  \BibitemOpen
  \bibfield  {author} {\bibinfo {author} {\bibfnamefont {V.~V.}\ \bibnamefont
  {Braguta}}, \bibinfo {author} {\bibfnamefont {E.~M.}\ \bibnamefont
  {Ilgenfritz}}, \bibinfo {author} {\bibfnamefont {A.~Yu.}\ \bibnamefont
  {Kotov}}, \bibinfo {author} {\bibfnamefont {A.~V.}\ \bibnamefont
  {Molochkov}}, \ and\ \bibinfo {author} {\bibfnamefont {A.~A.}\ \bibnamefont
  {Nikolaev}},\ }\bibfield  {title} {\enquote {\bibinfo {title} {{Study of the
  phase diagram of dense two-color QCD within lattice simulation}},}\ }\href
  {\doibase 10.1103/PhysRevD.94.114510} {\bibfield  {journal} {\bibinfo
  {journal} {Phys. Rev. D}\ }\textbf {\bibinfo {volume} {94}},\ \bibinfo
  {pages} {114510} (\bibinfo {year} {2016})},\ \Eprint
  {http://arxiv.org/abs/1605.04090} {arXiv:1605.04090 [hep-lat]} \BibitemShut
  {NoStop}%
\bibitem [{\citenamefont {Puhr}\ and\ \citenamefont
  {Buividovich}(2017)}]{Puhr:2016kzp}%
  \BibitemOpen
  \bibfield  {author} {\bibinfo {author} {\bibfnamefont {M.}~\bibnamefont
  {Puhr}}\ and\ \bibinfo {author} {\bibfnamefont {P.~V.}\ \bibnamefont
  {Buividovich}},\ }\bibfield  {title} {\enquote {\bibinfo {title} {{Numerical
  Study of Nonperturbative Corrections to the Chiral Separation Effect in
  Quenched Finite-Density QCD}},}\ }\href {\doibase
  10.1103/PhysRevLett.118.192003} {\bibfield  {journal} {\bibinfo  {journal}
  {Phys. Rev. Lett.}\ }\textbf {\bibinfo {volume} {118}},\ \bibinfo {pages}
  {192003} (\bibinfo {year} {2017})},\ \Eprint
  {http://arxiv.org/abs/1611.07263} {arXiv:1611.07263 [hep-lat]} \BibitemShut
  {NoStop}%
\bibitem [{\citenamefont {Boz}\ \emph {et~al.}(2019)\citenamefont {Boz},
  \citenamefont {Hajizadeh}, \citenamefont {Maas},\ and\ \citenamefont
  {Skullerud}}]{Boz:2018crd}%
  \BibitemOpen
  \bibfield  {author} {\bibinfo {author} {\bibfnamefont {Tamer}\ \bibnamefont
  {Boz}}, \bibinfo {author} {\bibfnamefont {Ouraman}\ \bibnamefont
  {Hajizadeh}}, \bibinfo {author} {\bibfnamefont {Axel}\ \bibnamefont {Maas}},
  \ and\ \bibinfo {author} {\bibfnamefont {Jon-Ivar}\ \bibnamefont
  {Skullerud}},\ }\bibfield  {title} {\enquote {\bibinfo {title}
  {{Finite-density gauge correlation functions in QC2D}},}\ }\href {\doibase
  10.1103/PhysRevD.99.074514} {\bibfield  {journal} {\bibinfo  {journal} {Phys.
  Rev. D}\ }\textbf {\bibinfo {volume} {99}},\ \bibinfo {pages} {074514}
  (\bibinfo {year} {2019})},\ \Eprint {http://arxiv.org/abs/1812.08517}
  {arXiv:1812.08517 [hep-lat]} \BibitemShut {NoStop}%
\bibitem [{\citenamefont {Astrakhantsev}\ \emph {et~al.}(2019)\citenamefont
  {Astrakhantsev}, \citenamefont {Bornyakov}, \citenamefont {Braguta},
  \citenamefont {Ilgenfritz}, \citenamefont {Kotov}, \citenamefont {Nikolaev},\
  and\ \citenamefont {Rothkopf}}]{Astrakhantsev:2018uzd}%
  \BibitemOpen
  \bibfield  {author} {\bibinfo {author} {\bibfnamefont {N.~Yu.}\ \bibnamefont
  {Astrakhantsev}}, \bibinfo {author} {\bibfnamefont {V.~G.}\ \bibnamefont
  {Bornyakov}}, \bibinfo {author} {\bibfnamefont {V.~V.}\ \bibnamefont
  {Braguta}}, \bibinfo {author} {\bibfnamefont {E.~M.}\ \bibnamefont
  {Ilgenfritz}}, \bibinfo {author} {\bibfnamefont {A.~Yu.}\ \bibnamefont
  {Kotov}}, \bibinfo {author} {\bibfnamefont {A.~A.}\ \bibnamefont {Nikolaev}},
  \ and\ \bibinfo {author} {\bibfnamefont {A.}~\bibnamefont {Rothkopf}},\
  }\bibfield  {title} {\enquote {\bibinfo {title} {{Lattice study of static
  quark-antiquark interactions in dense quark matter}},}\ }\href {\doibase
  10.1007/JHEP05(2019)171} {\bibfield  {journal} {\bibinfo  {journal} {JHEP}\
  }\textbf {\bibinfo {volume} {05}},\ \bibinfo {pages} {171} (\bibinfo {year}
  {2019})},\ \Eprint {http://arxiv.org/abs/1808.06466} {arXiv:1808.06466
  [hep-lat]} \BibitemShut {NoStop}%
\bibitem [{\citenamefont {Iida}\ \emph {et~al.}(2020)\citenamefont {Iida},
  \citenamefont {Itou},\ and\ \citenamefont {Lee}}]{Iida:2019rah}%
  \BibitemOpen
  \bibfield  {author} {\bibinfo {author} {\bibfnamefont {Kei}\ \bibnamefont
  {Iida}}, \bibinfo {author} {\bibfnamefont {Etsuko}\ \bibnamefont {Itou}}, \
  and\ \bibinfo {author} {\bibfnamefont {Tong-Gyu}\ \bibnamefont {Lee}},\
  }\bibfield  {title} {\enquote {\bibinfo {title} {{Two-colour QCD phases and
  the topology at low temperature and high density}},}\ }\href {\doibase
  10.1007/JHEP01(2020)181} {\bibfield  {journal} {\bibinfo  {journal} {JHEP}\
  }\textbf {\bibinfo {volume} {01}},\ \bibinfo {pages} {181} (\bibinfo {year}
  {2020})},\ \Eprint {http://arxiv.org/abs/1910.07872} {arXiv:1910.07872
  [hep-lat]} \BibitemShut {NoStop}%
\bibitem [{\citenamefont {Wilhelm}\ \emph {et~al.}(2019)\citenamefont
  {Wilhelm}, \citenamefont {Holicki}, \citenamefont {Smith}, \citenamefont
  {Wellegehausen},\ and\ \citenamefont {von Smekal}}]{Wilhelm:2019fvp}%
  \BibitemOpen
  \bibfield  {author} {\bibinfo {author} {\bibfnamefont {Jonas}\ \bibnamefont
  {Wilhelm}}, \bibinfo {author} {\bibfnamefont {Lukas}\ \bibnamefont
  {Holicki}}, \bibinfo {author} {\bibfnamefont {Dominik}\ \bibnamefont
  {Smith}}, \bibinfo {author} {\bibfnamefont {Bj\"orn}\ \bibnamefont
  {Wellegehausen}}, \ and\ \bibinfo {author} {\bibfnamefont {Lorenz}\
  \bibnamefont {von Smekal}},\ }\bibfield  {title} {\enquote {\bibinfo {title}
  {{Continuum Goldstone spectrum of two-color QCD at finite density with
  staggered quarks}},}\ }\href {\doibase 10.1103/PhysRevD.100.114507}
  {\bibfield  {journal} {\bibinfo  {journal} {Phys. Rev. D}\ }\textbf {\bibinfo
  {volume} {100}},\ \bibinfo {pages} {114507} (\bibinfo {year} {2019})},\
  \Eprint {http://arxiv.org/abs/1910.04495} {arXiv:1910.04495 [hep-lat]}
  \BibitemShut {NoStop}%
\bibitem [{\citenamefont {Buividovich}\ \emph
  {et~al.}(2021{\natexlab{a}})\citenamefont {Buividovich}, \citenamefont
  {Smith},\ and\ \citenamefont {von Smekal}}]{Buividovich:2020gnl}%
  \BibitemOpen
  \bibfield  {author} {\bibinfo {author} {\bibfnamefont {P.~V.}\ \bibnamefont
  {Buividovich}}, \bibinfo {author} {\bibfnamefont {D.}~\bibnamefont {Smith}},
  \ and\ \bibinfo {author} {\bibfnamefont {L.}~\bibnamefont {von Smekal}},\
  }\bibfield  {title} {\enquote {\bibinfo {title} {{Numerical study of the
  chiral separation effect in two-color QCD at finite density}},}\ }\href
  {\doibase 10.1103/PhysRevD.104.014511} {\bibfield  {journal} {\bibinfo
  {journal} {Phys. Rev. D}\ }\textbf {\bibinfo {volume} {104}},\ \bibinfo
  {pages} {014511} (\bibinfo {year} {2021}{\natexlab{a}})},\ \Eprint
  {http://arxiv.org/abs/2012.05184} {arXiv:2012.05184 [hep-lat]} \BibitemShut
  {NoStop}%
\bibitem [{\citenamefont {Iida}\ \emph {et~al.}(2021)\citenamefont {Iida},
  \citenamefont {Itou},\ and\ \citenamefont {Lee}}]{Iida:2020emi}%
  \BibitemOpen
  \bibfield  {author} {\bibinfo {author} {\bibfnamefont {Kei}\ \bibnamefont
  {Iida}}, \bibinfo {author} {\bibfnamefont {Etsuko}\ \bibnamefont {Itou}}, \
  and\ \bibinfo {author} {\bibfnamefont {Tong-Gyu}\ \bibnamefont {Lee}},\
  }\bibfield  {title} {\enquote {\bibinfo {title} {{Relative scale setting for
  two-color QCD with $N_f$=2 Wilson fermions}},}\ }\href {\doibase
  10.1093/ptep/ptaa170} {\bibfield  {journal} {\bibinfo  {journal} {PTEP}\
  }\textbf {\bibinfo {volume} {2021}},\ \bibinfo {pages} {013B05} (\bibinfo
  {year} {2021})},\ \Eprint {http://arxiv.org/abs/2008.06322} {arXiv:2008.06322
  [hep-lat]} \BibitemShut {NoStop}%
\bibitem [{\citenamefont {Astrakhantsev}\ \emph {et~al.}(2020)\citenamefont
  {Astrakhantsev}, \citenamefont {Braguta}, \citenamefont {Ilgenfritz},
  \citenamefont {Kotov},\ and\ \citenamefont
  {Nikolaev}}]{Astrakhantsev:2020tdl}%
  \BibitemOpen
  \bibfield  {author} {\bibinfo {author} {\bibfnamefont {N.}~\bibnamefont
  {Astrakhantsev}}, \bibinfo {author} {\bibfnamefont {V.~V.}\ \bibnamefont
  {Braguta}}, \bibinfo {author} {\bibfnamefont {E.~M.}\ \bibnamefont
  {Ilgenfritz}}, \bibinfo {author} {\bibfnamefont {A.~Yu.}\ \bibnamefont
  {Kotov}}, \ and\ \bibinfo {author} {\bibfnamefont {A.~A.}\ \bibnamefont
  {Nikolaev}},\ }\bibfield  {title} {\enquote {\bibinfo {title} {{Lattice study
  of thermodynamic properties of dense QC$_2$D}},}\ }\href {\doibase
  10.1103/PhysRevD.102.074507} {\bibfield  {journal} {\bibinfo  {journal}
  {Phys. Rev. D}\ }\textbf {\bibinfo {volume} {102}},\ \bibinfo {pages}
  {074507} (\bibinfo {year} {2020})},\ \Eprint
  {http://arxiv.org/abs/2007.07640} {arXiv:2007.07640 [hep-lat]} \BibitemShut
  {NoStop}%
\bibitem [{\citenamefont {Bornyakov}\ \emph {et~al.}(2020)\citenamefont
  {Bornyakov}, \citenamefont {Braguta}, \citenamefont {Nikolaev},\ and\
  \citenamefont {Rogalyov}}]{Bornyakov:2020kyz}%
  \BibitemOpen
  \bibfield  {author} {\bibinfo {author} {\bibfnamefont {V.~G.}\ \bibnamefont
  {Bornyakov}}, \bibinfo {author} {\bibfnamefont {V.~V.}\ \bibnamefont
  {Braguta}}, \bibinfo {author} {\bibfnamefont {A.~A.}\ \bibnamefont
  {Nikolaev}}, \ and\ \bibinfo {author} {\bibfnamefont {R.~N.}\ \bibnamefont
  {Rogalyov}},\ }\bibfield  {title} {\enquote {\bibinfo {title} {{Effects of
  Dense Quark Matter on Gluon Propagators in Lattice QC$_2$D}},}\ }\href
  {\doibase 10.1103/PhysRevD.102.114511} {\bibfield  {journal} {\bibinfo
  {journal} {Phys. Rev. D}\ }\textbf {\bibinfo {volume} {102}},\ \bibinfo
  {pages} {114511} (\bibinfo {year} {2020})},\ \Eprint
  {http://arxiv.org/abs/2003.00232} {arXiv:2003.00232 [hep-lat]} \BibitemShut
  {NoStop}%
\bibitem [{\citenamefont {Buividovich}\ \emph {et~al.}(2020)\citenamefont
  {Buividovich}, \citenamefont {Smith},\ and\ \citenamefont {von
  Smekal}}]{Buividovich:2020dks}%
  \BibitemOpen
  \bibfield  {author} {\bibinfo {author} {\bibfnamefont {P.~V.}\ \bibnamefont
  {Buividovich}}, \bibinfo {author} {\bibfnamefont {D.}~\bibnamefont {Smith}},
  \ and\ \bibinfo {author} {\bibfnamefont {L.}~\bibnamefont {von Smekal}},\
  }\bibfield  {title} {\enquote {\bibinfo {title} {{Electric conductivity in
  finite-density $SU(2)$ lattice gauge theory with dynamical fermions}},}\
  }\href {\doibase 10.1103/PhysRevD.102.094510} {\bibfield  {journal} {\bibinfo
   {journal} {Phys. Rev. D}\ }\textbf {\bibinfo {volume} {102}},\ \bibinfo
  {pages} {094510} (\bibinfo {year} {2020})},\ \Eprint
  {http://arxiv.org/abs/2007.05639} {arXiv:2007.05639 [hep-lat]} \BibitemShut
  {NoStop}%
\bibitem [{\citenamefont {Buividovich}\ \emph
  {et~al.}(2021{\natexlab{b}})\citenamefont {Buividovich}, \citenamefont
  {Smith},\ and\ \citenamefont {von Smekal}}]{Buividovich:2021fsa}%
  \BibitemOpen
  \bibfield  {author} {\bibinfo {author} {\bibfnamefont {P.~V.}\ \bibnamefont
  {Buividovich}}, \bibinfo {author} {\bibfnamefont {D.}~\bibnamefont {Smith}},
  \ and\ \bibinfo {author} {\bibfnamefont {L.}~\bibnamefont {von Smekal}},\
  }\bibfield  {title} {\enquote {\bibinfo {title} {{Static magnetic
  susceptibility in finite-density $SU\left( 2\right) $ lattice gauge
  theory}},}\ }\href {\doibase 10.1140/epja/s10050-021-00604-7} {\bibfield
  {journal} {\bibinfo  {journal} {Eur. Phys. J. A}\ }\textbf {\bibinfo {volume}
  {57}},\ \bibinfo {pages} {293} (\bibinfo {year} {2021}{\natexlab{b}})},\
  \Eprint {http://arxiv.org/abs/2104.10012} {arXiv:2104.10012 [hep-lat]}
  \BibitemShut {NoStop}%
\bibitem [{\citenamefont {Iida}\ and\ \citenamefont
  {Itou}(2022)}]{Iida:2022hyy}%
  \BibitemOpen
  \bibfield  {author} {\bibinfo {author} {\bibfnamefont {Kei}\ \bibnamefont
  {Iida}}\ and\ \bibinfo {author} {\bibfnamefont {Etsuko}\ \bibnamefont
  {Itou}},\ }\bibfield  {title} {\enquote {\bibinfo {title} {{Velocity of Sound
  beyond the High-Density Relativistic Limit from Lattice Simulation of Dense
  Two-Color QCD}},}\ }\href@noop {} {\  (\bibinfo {year} {2022})},\ \Eprint
  {http://arxiv.org/abs/2207.01253} {arXiv:2207.01253 [hep-ph]} \BibitemShut
  {NoStop}%
\bibitem [{\citenamefont {Murakami}\ \emph
  {et~al.}(2023{\natexlab{a}})\citenamefont {Murakami}, \citenamefont {Itou},\
  and\ \citenamefont {Iida}}]{Murakami:2023ejc}%
  \BibitemOpen
  \bibfield  {author} {\bibinfo {author} {\bibfnamefont {Kotaro}\ \bibnamefont
  {Murakami}}, \bibinfo {author} {\bibfnamefont {Etsuko}\ \bibnamefont {Itou}},
  \ and\ \bibinfo {author} {\bibfnamefont {Kei}\ \bibnamefont {Iida}},\
  }\bibfield  {title} {\enquote {\bibinfo {title} {{Chemical potential
  (in)dependence of hadron scatterings in the hadronic phase of QCD-like
  theories and its applications}},}\ }\href@noop {} {\  (\bibinfo {year}
  {2023}{\natexlab{a}})},\ \Eprint {http://arxiv.org/abs/2309.08143}
  {arXiv:2309.08143 [hep-lat]} \BibitemShut {NoStop}%
\bibitem [{\citenamefont {Braguta}(2023)}]{Braguta:2023yhd}%
  \BibitemOpen
  \bibfield  {author} {\bibinfo {author} {\bibfnamefont {Victor~V.}\
  \bibnamefont {Braguta}},\ }\bibfield  {title} {\enquote {\bibinfo {title}
  {{Phase Diagram of Dense Two-Color QCD at Low Temperatures}},}\ }\href
  {\doibase 10.3390/sym15071466} {\bibfield  {journal} {\bibinfo  {journal}
  {Symmetry}\ }\textbf {\bibinfo {volume} {15}},\ \bibinfo {pages} {1466}
  (\bibinfo {year} {2023})}\BibitemShut {NoStop}%
\bibitem [{\citenamefont {Kogut}\ \emph {et~al.}(1999)\citenamefont {Kogut},
  \citenamefont {Stephanov},\ and\ \citenamefont {Toublan}}]{Kogut:1999iv}%
  \BibitemOpen
  \bibfield  {author} {\bibinfo {author} {\bibfnamefont {J.~B.}\ \bibnamefont
  {Kogut}}, \bibinfo {author} {\bibfnamefont {Misha~A.}\ \bibnamefont
  {Stephanov}}, \ and\ \bibinfo {author} {\bibfnamefont {D.}~\bibnamefont
  {Toublan}},\ }\bibfield  {title} {\enquote {\bibinfo {title} {{On two color
  QCD with baryon chemical potential}},}\ }\href {\doibase
  10.1016/S0370-2693(99)00971-5} {\bibfield  {journal} {\bibinfo  {journal}
  {Phys. Lett. B}\ }\textbf {\bibinfo {volume} {464}},\ \bibinfo {pages}
  {183--191} (\bibinfo {year} {1999})},\ \Eprint
  {http://arxiv.org/abs/hep-ph/9906346} {arXiv:hep-ph/9906346} \BibitemShut
  {NoStop}%
\bibitem [{\citenamefont {Kogut}\ \emph {et~al.}(2000)\citenamefont {Kogut},
  \citenamefont {Stephanov}, \citenamefont {Toublan}, \citenamefont
  {Verbaarschot},\ and\ \citenamefont {Zhitnitsky}}]{Kogut:2000ek}%
  \BibitemOpen
  \bibfield  {author} {\bibinfo {author} {\bibfnamefont {J.~B.}\ \bibnamefont
  {Kogut}}, \bibinfo {author} {\bibfnamefont {Misha~A.}\ \bibnamefont
  {Stephanov}}, \bibinfo {author} {\bibfnamefont {D.}~\bibnamefont {Toublan}},
  \bibinfo {author} {\bibfnamefont {J.~J.~M.}\ \bibnamefont {Verbaarschot}}, \
  and\ \bibinfo {author} {\bibfnamefont {A.}~\bibnamefont {Zhitnitsky}},\
  }\bibfield  {title} {\enquote {\bibinfo {title} {{QCD - like theories at
  finite baryon density}},}\ }\href {\doibase 10.1016/S0550-3213(00)00242-X}
  {\bibfield  {journal} {\bibinfo  {journal} {Nucl. Phys. B}\ }\textbf
  {\bibinfo {volume} {582}},\ \bibinfo {pages} {477--513} (\bibinfo {year}
  {2000})},\ \Eprint {http://arxiv.org/abs/hep-ph/0001171}
  {arXiv:hep-ph/0001171} \BibitemShut {NoStop}%
\bibitem [{\citenamefont {Lenaghan}\ \emph {et~al.}(2002)\citenamefont
  {Lenaghan}, \citenamefont {Sannino},\ and\ \citenamefont
  {Splittorff}}]{Lenaghan:2001sd}%
  \BibitemOpen
  \bibfield  {author} {\bibinfo {author} {\bibfnamefont {J.~T.}\ \bibnamefont
  {Lenaghan}}, \bibinfo {author} {\bibfnamefont {F.}~\bibnamefont {Sannino}}, \
  and\ \bibinfo {author} {\bibfnamefont {K.}~\bibnamefont {Splittorff}},\
  }\bibfield  {title} {\enquote {\bibinfo {title} {{The Superfluid and
  conformal phase transitions of two color QCD}},}\ }\href {\doibase
  10.1103/PhysRevD.65.054002} {\bibfield  {journal} {\bibinfo  {journal} {Phys.
  Rev. D}\ }\textbf {\bibinfo {volume} {65}},\ \bibinfo {pages} {054002}
  (\bibinfo {year} {2002})},\ \Eprint {http://arxiv.org/abs/hep-ph/0107099}
  {arXiv:hep-ph/0107099} \BibitemShut {NoStop}%
\bibitem [{\citenamefont {Splittorff}\ \emph {et~al.}(2002)\citenamefont
  {Splittorff}, \citenamefont {Toublan},\ and\ \citenamefont
  {Verbaarschot}}]{Splittorff:2001fy}%
  \BibitemOpen
  \bibfield  {author} {\bibinfo {author} {\bibfnamefont {K.}~\bibnamefont
  {Splittorff}}, \bibinfo {author} {\bibfnamefont {D.}~\bibnamefont {Toublan}},
  \ and\ \bibinfo {author} {\bibfnamefont {J.~J.~M.}\ \bibnamefont
  {Verbaarschot}},\ }\bibfield  {title} {\enquote {\bibinfo {title} {{Diquark
  condensate in QCD with two colors at next-to-leading order}},}\ }\href
  {\doibase 10.1016/S0550-3213(01)00536-3} {\bibfield  {journal} {\bibinfo
  {journal} {Nucl. Phys. B}\ }\textbf {\bibinfo {volume} {620}},\ \bibinfo
  {pages} {290--314} (\bibinfo {year} {2002})},\ \Eprint
  {http://arxiv.org/abs/hep-ph/0108040} {arXiv:hep-ph/0108040} \BibitemShut
  {NoStop}%
\bibitem [{\citenamefont {Ratti}\ and\ \citenamefont
  {Weise}(2004)}]{Ratti:2004ra}%
  \BibitemOpen
  \bibfield  {author} {\bibinfo {author} {\bibfnamefont {Claudia}\ \bibnamefont
  {Ratti}}\ and\ \bibinfo {author} {\bibfnamefont {Wolfram}\ \bibnamefont
  {Weise}},\ }\bibfield  {title} {\enquote {\bibinfo {title} {{Thermodynamics
  of two-colour QCD and the Nambu Jona-Lasinio model}},}\ }\href {\doibase
  10.1103/PhysRevD.70.054013} {\bibfield  {journal} {\bibinfo  {journal} {Phys.
  Rev. D}\ }\textbf {\bibinfo {volume} {70}},\ \bibinfo {pages} {054013}
  (\bibinfo {year} {2004})},\ \Eprint {http://arxiv.org/abs/hep-ph/0406159}
  {arXiv:hep-ph/0406159} \BibitemShut {NoStop}%
\bibitem [{\citenamefont {Sun}\ \emph {et~al.}(2007)\citenamefont {Sun},
  \citenamefont {He},\ and\ \citenamefont {Zhuang}}]{Sun:2007fc}%
  \BibitemOpen
  \bibfield  {author} {\bibinfo {author} {\bibfnamefont {Gao-feng}\
  \bibnamefont {Sun}}, \bibinfo {author} {\bibfnamefont {Lianyi}\ \bibnamefont
  {He}}, \ and\ \bibinfo {author} {\bibfnamefont {Pengfei}\ \bibnamefont
  {Zhuang}},\ }\bibfield  {title} {\enquote {\bibinfo {title} {{BEC-BCS
  crossover in the Nambu-Jona-Lasinio model of QCD}},}\ }\href {\doibase
  10.1103/PhysRevD.75.096004} {\bibfield  {journal} {\bibinfo  {journal} {Phys.
  Rev. D}\ }\textbf {\bibinfo {volume} {75}},\ \bibinfo {pages} {096004}
  (\bibinfo {year} {2007})},\ \Eprint {http://arxiv.org/abs/hep-ph/0703159}
  {arXiv:hep-ph/0703159} \BibitemShut {NoStop}%
\bibitem [{\citenamefont {Fukushima}\ and\ \citenamefont
  {Iida}(2007)}]{Fukushima:2007bj}%
  \BibitemOpen
  \bibfield  {author} {\bibinfo {author} {\bibfnamefont {Kenji}\ \bibnamefont
  {Fukushima}}\ and\ \bibinfo {author} {\bibfnamefont {Kei}\ \bibnamefont
  {Iida}},\ }\bibfield  {title} {\enquote {\bibinfo {title}
  {{Larkin-Ovchinnikov-Fulde-Ferrell state in two-color quark matter}},}\
  }\href {\doibase 10.1103/PhysRevD.76.054004} {\bibfield  {journal} {\bibinfo
  {journal} {Phys. Rev. D}\ }\textbf {\bibinfo {volume} {76}},\ \bibinfo
  {pages} {054004} (\bibinfo {year} {2007})},\ \Eprint
  {http://arxiv.org/abs/0705.0792} {arXiv:0705.0792 [hep-ph]} \BibitemShut
  {NoStop}%
\bibitem [{\citenamefont {Brauner}\ \emph {et~al.}(2009)\citenamefont
  {Brauner}, \citenamefont {Fukushima},\ and\ \citenamefont
  {Hidaka}}]{Brauner:2009gu}%
  \BibitemOpen
  \bibfield  {author} {\bibinfo {author} {\bibfnamefont {Tomas}\ \bibnamefont
  {Brauner}}, \bibinfo {author} {\bibfnamefont {Kenji}\ \bibnamefont
  {Fukushima}}, \ and\ \bibinfo {author} {\bibfnamefont {Yoshimasa}\
  \bibnamefont {Hidaka}},\ }\bibfield  {title} {\enquote {\bibinfo {title}
  {{Two-color quark matter: U(1)(A) restoration, superfluidity, and quarkyonic
  phase}},}\ }\href {\doibase 10.1103/PhysRevD.81.119904} {\bibfield  {journal}
  {\bibinfo  {journal} {Phys. Rev. D}\ }\textbf {\bibinfo {volume} {80}},\
  \bibinfo {pages} {074035} (\bibinfo {year} {2009})},\ \bibinfo {note}
  {[Erratum: Phys.Rev.D 81, 119904 (2010)]},\ \Eprint
  {http://arxiv.org/abs/0907.4905} {arXiv:0907.4905 [hep-ph]} \BibitemShut
  {NoStop}%
\bibitem [{\citenamefont {Kanazawa}\ \emph {et~al.}(2009)\citenamefont
  {Kanazawa}, \citenamefont {Wettig},\ and\ \citenamefont
  {Yamamoto}}]{Kanazawa:2009ks}%
  \BibitemOpen
  \bibfield  {author} {\bibinfo {author} {\bibfnamefont {Takuya}\ \bibnamefont
  {Kanazawa}}, \bibinfo {author} {\bibfnamefont {Tilo}\ \bibnamefont {Wettig}},
  \ and\ \bibinfo {author} {\bibfnamefont {Naoki}\ \bibnamefont {Yamamoto}},\
  }\bibfield  {title} {\enquote {\bibinfo {title} {{Chiral Lagrangian and
  spectral sum rules for dense two-color QCD}},}\ }\href {\doibase
  10.1088/1126-6708/2009/08/003} {\bibfield  {journal} {\bibinfo  {journal}
  {JHEP}\ }\textbf {\bibinfo {volume} {08}},\ \bibinfo {pages} {003} (\bibinfo
  {year} {2009})},\ \Eprint {http://arxiv.org/abs/0906.3579} {arXiv:0906.3579
  [hep-ph]} \BibitemShut {NoStop}%
\bibitem [{\citenamefont {Harada}\ \emph {et~al.}(2010)\citenamefont {Harada},
  \citenamefont {Nonaka},\ and\ \citenamefont {Yamaoka}}]{Harada:2010vy}%
  \BibitemOpen
  \bibfield  {author} {\bibinfo {author} {\bibfnamefont {Masayasu}\
  \bibnamefont {Harada}}, \bibinfo {author} {\bibfnamefont {Chiho}\
  \bibnamefont {Nonaka}}, \ and\ \bibinfo {author} {\bibfnamefont {Tetsuro}\
  \bibnamefont {Yamaoka}},\ }\bibfield  {title} {\enquote {\bibinfo {title}
  {{Masses of vector bosons in two-color dense QCD based on the hidden local
  symmetry}},}\ }\href {\doibase 10.1103/PhysRevD.81.096003} {\bibfield
  {journal} {\bibinfo  {journal} {Phys. Rev. D}\ }\textbf {\bibinfo {volume}
  {81}},\ \bibinfo {pages} {096003} (\bibinfo {year} {2010})},\ \Eprint
  {http://arxiv.org/abs/1002.4705} {arXiv:1002.4705 [hep-ph]} \BibitemShut
  {NoStop}%
\bibitem [{\citenamefont {Andersen}\ and\ \citenamefont
  {Brauner}(2010)}]{Andersen:2010vu}%
  \BibitemOpen
  \bibfield  {author} {\bibinfo {author} {\bibfnamefont {Jens~O.}\ \bibnamefont
  {Andersen}}\ and\ \bibinfo {author} {\bibfnamefont {Tomas}\ \bibnamefont
  {Brauner}},\ }\bibfield  {title} {\enquote {\bibinfo {title} {{Phase diagram
  of two-color quark matter at nonzero baryon and isospin density}},}\ }\href
  {\doibase 10.1103/PhysRevD.81.096004} {\bibfield  {journal} {\bibinfo
  {journal} {Phys. Rev. D}\ }\textbf {\bibinfo {volume} {81}},\ \bibinfo
  {pages} {096004} (\bibinfo {year} {2010})},\ \Eprint
  {http://arxiv.org/abs/1001.5168} {arXiv:1001.5168 [hep-ph]} \BibitemShut
  {NoStop}%
\bibitem [{\citenamefont {Zhang}\ \emph {et~al.}(2010)\citenamefont {Zhang},
  \citenamefont {Brauner},\ and\ \citenamefont {Rischke}}]{Zhang:2010kn}%
  \BibitemOpen
  \bibfield  {author} {\bibinfo {author} {\bibfnamefont {Tian}\ \bibnamefont
  {Zhang}}, \bibinfo {author} {\bibfnamefont {Tomas}\ \bibnamefont {Brauner}},
  \ and\ \bibinfo {author} {\bibfnamefont {Dirk~H.}\ \bibnamefont {Rischke}},\
  }\bibfield  {title} {\enquote {\bibinfo {title} {{QCD-like theories at
  nonzero temperature and density}},}\ }\href {\doibase
  10.1007/JHEP06(2010)064} {\bibfield  {journal} {\bibinfo  {journal} {JHEP}\
  }\textbf {\bibinfo {volume} {06}},\ \bibinfo {pages} {064} (\bibinfo {year}
  {2010})},\ \Eprint {http://arxiv.org/abs/1005.2928} {arXiv:1005.2928
  [hep-ph]} \BibitemShut {NoStop}%
\bibitem [{\citenamefont {He}(2010)}]{He:2010nb}%
  \BibitemOpen
  \bibfield  {author} {\bibinfo {author} {\bibfnamefont {Lianyi}\ \bibnamefont
  {He}},\ }\bibfield  {title} {\enquote {\bibinfo {title} {{Nambu-Jona-Lasinio
  model description of weakly interacting Bose condensate and BEC-BCS crossover
  in dense QCD-like theories}},}\ }\href {\doibase 10.1103/PhysRevD.82.096003}
  {\bibfield  {journal} {\bibinfo  {journal} {Phys. Rev. D}\ }\textbf {\bibinfo
  {volume} {82}},\ \bibinfo {pages} {096003} (\bibinfo {year} {2010})},\
  \Eprint {http://arxiv.org/abs/1007.1920} {arXiv:1007.1920 [hep-ph]}
  \BibitemShut {NoStop}%
\bibitem [{\citenamefont {Strodthoff}\ \emph {et~al.}(2012)\citenamefont
  {Strodthoff}, \citenamefont {Schaefer},\ and\ \citenamefont {von
  Smekal}}]{Strodthoff:2011tz}%
  \BibitemOpen
  \bibfield  {author} {\bibinfo {author} {\bibfnamefont {Nils}\ \bibnamefont
  {Strodthoff}}, \bibinfo {author} {\bibfnamefont {Bernd-Jochen}\ \bibnamefont
  {Schaefer}}, \ and\ \bibinfo {author} {\bibfnamefont {Lorenz}\ \bibnamefont
  {von Smekal}},\ }\bibfield  {title} {\enquote {\bibinfo {title}
  {{Quark-meson-diquark model for two-color QCD}},}\ }\href {\doibase
  10.1103/PhysRevD.85.074007} {\bibfield  {journal} {\bibinfo  {journal} {Phys.
  Rev. D}\ }\textbf {\bibinfo {volume} {85}},\ \bibinfo {pages} {074007}
  (\bibinfo {year} {2012})},\ \Eprint {http://arxiv.org/abs/1112.5401}
  {arXiv:1112.5401 [hep-ph]} \BibitemShut {NoStop}%
\bibitem [{\citenamefont {Imai}\ \emph {et~al.}(2013)\citenamefont {Imai},
  \citenamefont {Toki},\ and\ \citenamefont {Weise}}]{Imai:2012hr}%
  \BibitemOpen
  \bibfield  {author} {\bibinfo {author} {\bibfnamefont {Shotaro}\ \bibnamefont
  {Imai}}, \bibinfo {author} {\bibfnamefont {Hiroshi}\ \bibnamefont {Toki}}, \
  and\ \bibinfo {author} {\bibfnamefont {Wolfram}\ \bibnamefont {Weise}},\
  }\bibfield  {title} {\enquote {\bibinfo {title} {{Quark-Hadron Matter at
  Finite Temperature and Density in a Two-Color PNJL model}},}\ }\href
  {\doibase 10.1016/j.nuclphysa.2013.06.001} {\bibfield  {journal} {\bibinfo
  {journal} {Nucl. Phys. A}\ }\textbf {\bibinfo {volume} {913}},\ \bibinfo
  {pages} {71--102} (\bibinfo {year} {2013})},\ \Eprint
  {http://arxiv.org/abs/1210.1307} {arXiv:1210.1307 [nucl-th]} \BibitemShut
  {NoStop}%
\bibitem [{\citenamefont {Strodthoff}\ and\ \citenamefont {von
  Smekal}(2014)}]{Strodthoff:2013cua}%
  \BibitemOpen
  \bibfield  {author} {\bibinfo {author} {\bibfnamefont {Nils}\ \bibnamefont
  {Strodthoff}}\ and\ \bibinfo {author} {\bibfnamefont {Lorenz}\ \bibnamefont
  {von Smekal}},\ }\bibfield  {title} {\enquote {\bibinfo {title}
  {{Polyakov-Quark-Meson-Diquark Model for two-color QCD}},}\ }\href {\doibase
  10.1016/j.physletb.2014.03.008} {\bibfield  {journal} {\bibinfo  {journal}
  {Phys. Lett. B}\ }\textbf {\bibinfo {volume} {731}},\ \bibinfo {pages}
  {350--357} (\bibinfo {year} {2014})},\ \Eprint
  {http://arxiv.org/abs/1306.2897} {arXiv:1306.2897 [hep-ph]} \BibitemShut
  {NoStop}%
\bibitem [{\citenamefont {Khan}\ \emph {et~al.}(2015)\citenamefont {Khan},
  \citenamefont {Pawlowski}, \citenamefont {Rennecke},\ and\ \citenamefont
  {Scherer}}]{Khan:2015puu}%
  \BibitemOpen
  \bibfield  {author} {\bibinfo {author} {\bibfnamefont {Naseemuddin}\
  \bibnamefont {Khan}}, \bibinfo {author} {\bibfnamefont {Jan~M.}\ \bibnamefont
  {Pawlowski}}, \bibinfo {author} {\bibfnamefont {Fabian}\ \bibnamefont
  {Rennecke}}, \ and\ \bibinfo {author} {\bibfnamefont {Michael~M.}\
  \bibnamefont {Scherer}},\ }\bibfield  {title} {\enquote {\bibinfo {title}
  {{The Phase Diagram of QC2D from Functional Methods}},}\ }\href@noop {} {\
  (\bibinfo {year} {2015})},\ \Eprint {http://arxiv.org/abs/1512.03673}
  {arXiv:1512.03673 [hep-ph]} \BibitemShut {NoStop}%
\bibitem [{\citenamefont {Duarte}\ \emph {et~al.}(2016)\citenamefont {Duarte},
  \citenamefont {Allen}, \citenamefont {Farias}, \citenamefont {Manso},
  \citenamefont {Ramos},\ and\ \citenamefont {Scoccola}}]{Duarte:2015ppa}%
  \BibitemOpen
  \bibfield  {author} {\bibinfo {author} {\bibfnamefont {Dyana~C.}\
  \bibnamefont {Duarte}}, \bibinfo {author} {\bibfnamefont {P.~G.}\
  \bibnamefont {Allen}}, \bibinfo {author} {\bibfnamefont {R.~L.~S.}\
  \bibnamefont {Farias}}, \bibinfo {author} {\bibfnamefont {Pedro H.~A.}\
  \bibnamefont {Manso}}, \bibinfo {author} {\bibfnamefont {Rudnei~O.}\
  \bibnamefont {Ramos}}, \ and\ \bibinfo {author} {\bibfnamefont {N.~N.}\
  \bibnamefont {Scoccola}},\ }\bibfield  {title} {\enquote {\bibinfo {title}
  {{BEC-BCS crossover in a cold and magnetized two color NJL model}},}\ }\href
  {\doibase 10.1103/PhysRevD.93.025017} {\bibfield  {journal} {\bibinfo
  {journal} {Phys. Rev. D}\ }\textbf {\bibinfo {volume} {93}},\ \bibinfo
  {pages} {025017} (\bibinfo {year} {2016})},\ \Eprint
  {http://arxiv.org/abs/1510.02756} {arXiv:1510.02756 [hep-ph]} \BibitemShut
  {NoStop}%
\bibitem [{\citenamefont {Chao}(2020)}]{Chao:2018czo}%
  \BibitemOpen
  \bibfield  {author} {\bibinfo {author} {\bibfnamefont {Jingyi}\ \bibnamefont
  {Chao}},\ }\bibfield  {title} {\enquote {\bibinfo {title} {{Phase diagram of
  two-color QCD matter at finite baryon and axial isospin densities}},}\ }\href
  {\doibase 10.1088/1674-1137/44/3/034108} {\bibfield  {journal} {\bibinfo
  {journal} {Chin. Phys. C}\ }\textbf {\bibinfo {volume} {44}},\ \bibinfo
  {pages} {034108} (\bibinfo {year} {2020})},\ \Eprint
  {http://arxiv.org/abs/1808.01928} {arXiv:1808.01928 [hep-ph]} \BibitemShut
  {NoStop}%
\bibitem [{\citenamefont {Adhikari}\ \emph {et~al.}(2018)\citenamefont
  {Adhikari}, \citenamefont {Beleznay},\ and\ \citenamefont
  {Mannarelli}}]{Adhikari:2018kzh}%
  \BibitemOpen
  \bibfield  {author} {\bibinfo {author} {\bibfnamefont {Prabal}\ \bibnamefont
  {Adhikari}}, \bibinfo {author} {\bibfnamefont {Soma~B.}\ \bibnamefont
  {Beleznay}}, \ and\ \bibinfo {author} {\bibfnamefont {Massimo}\ \bibnamefont
  {Mannarelli}},\ }\bibfield  {title} {\enquote {\bibinfo {title} {{Finite
  Density Two Color Chiral Perturbation Theory Revisited}},}\ }\href {\doibase
  10.1140/epjc/s10052-018-5934-6} {\bibfield  {journal} {\bibinfo  {journal}
  {Eur. Phys. J. C}\ }\textbf {\bibinfo {volume} {78}},\ \bibinfo {pages} {441}
  (\bibinfo {year} {2018})},\ \Eprint {http://arxiv.org/abs/1803.00490}
  {arXiv:1803.00490 [hep-th]} \BibitemShut {NoStop}%
\bibitem [{\citenamefont {Contant}\ and\ \citenamefont
  {Huber}(2020)}]{Contant:2019lwf}%
  \BibitemOpen
  \bibfield  {author} {\bibinfo {author} {\bibfnamefont {Romain}\ \bibnamefont
  {Contant}}\ and\ \bibinfo {author} {\bibfnamefont {Markus~Q.}\ \bibnamefont
  {Huber}},\ }\bibfield  {title} {\enquote {\bibinfo {title} {{Dense two-color
  QCD from Dyson-Schwinger equations}},}\ }\href {\doibase
  10.1103/PhysRevD.101.014016} {\bibfield  {journal} {\bibinfo  {journal}
  {Phys. Rev. D}\ }\textbf {\bibinfo {volume} {101}},\ \bibinfo {pages}
  {014016} (\bibinfo {year} {2020})},\ \Eprint
  {http://arxiv.org/abs/1909.12796} {arXiv:1909.12796 [hep-ph]} \BibitemShut
  {NoStop}%
\bibitem [{\citenamefont {Suenaga}\ and\ \citenamefont
  {Kojo}(2019)}]{Suenaga:2019jjv}%
  \BibitemOpen
  \bibfield  {author} {\bibinfo {author} {\bibfnamefont {Daiki}\ \bibnamefont
  {Suenaga}}\ and\ \bibinfo {author} {\bibfnamefont {Toru}\ \bibnamefont
  {Kojo}},\ }\bibfield  {title} {\enquote {\bibinfo {title} {{Gluon propagator
  in two-color dense QCD: Massive Yang-Mills approach at one-loop}},}\ }\href
  {\doibase 10.1103/PhysRevD.100.076017} {\bibfield  {journal} {\bibinfo
  {journal} {Phys. Rev. D}\ }\textbf {\bibinfo {volume} {100}},\ \bibinfo
  {pages} {076017} (\bibinfo {year} {2019})},\ \Eprint
  {http://arxiv.org/abs/1905.08751} {arXiv:1905.08751 [hep-ph]} \BibitemShut
  {NoStop}%
\bibitem [{\citenamefont {Khunjua}\ \emph {et~al.}(2020)\citenamefont
  {Khunjua}, \citenamefont {Klimenko},\ and\ \citenamefont
  {Zhokhov}}]{Khunjua:2020xws}%
  \BibitemOpen
  \bibfield  {author} {\bibinfo {author} {\bibfnamefont {T.~G.}\ \bibnamefont
  {Khunjua}}, \bibinfo {author} {\bibfnamefont {K.~G.}\ \bibnamefont
  {Klimenko}}, \ and\ \bibinfo {author} {\bibfnamefont {R.~N.}\ \bibnamefont
  {Zhokhov}},\ }\bibfield  {title} {\enquote {\bibinfo {title} {{The dual
  properties of chiral and isospin asymmetric dense quark matter formed of
  two-color quarks}},}\ }\href {\doibase 10.1007/JHEP06(2020)148} {\bibfield
  {journal} {\bibinfo  {journal} {JHEP}\ }\textbf {\bibinfo {volume} {06}},\
  \bibinfo {pages} {148} (\bibinfo {year} {2020})},\ \Eprint
  {http://arxiv.org/abs/2003.10562} {arXiv:2003.10562 [hep-ph]} \BibitemShut
  {NoStop}%
\bibitem [{\citenamefont {Khunjua}\ \emph {et~al.}(2022)\citenamefont
  {Khunjua}, \citenamefont {Klimenko},\ and\ \citenamefont
  {Zhokhov}}]{Khunjua:2021oxf}%
  \BibitemOpen
  \bibfield  {author} {\bibinfo {author} {\bibfnamefont {T.~G.}\ \bibnamefont
  {Khunjua}}, \bibinfo {author} {\bibfnamefont {K.~G.}\ \bibnamefont
  {Klimenko}}, \ and\ \bibinfo {author} {\bibfnamefont {R.~N.}\ \bibnamefont
  {Zhokhov}},\ }\bibfield  {title} {\enquote {\bibinfo {title} {{Influence of
  chiral chemical potential \ensuremath{\mu}5 on phase structure of the
  two-color quark matter}},}\ }\href {\doibase 10.1103/PhysRevD.106.045008}
  {\bibfield  {journal} {\bibinfo  {journal} {Phys. Rev. D}\ }\textbf {\bibinfo
  {volume} {106}},\ \bibinfo {pages} {045008} (\bibinfo {year} {2022})},\
  \Eprint {http://arxiv.org/abs/2105.04952} {arXiv:2105.04952 [hep-ph]}
  \BibitemShut {NoStop}%
\bibitem [{\citenamefont {Kojo}\ and\ \citenamefont
  {Suenaga}(2021)}]{Kojo:2021knn}%
  \BibitemOpen
  \bibfield  {author} {\bibinfo {author} {\bibfnamefont {Toru}\ \bibnamefont
  {Kojo}}\ and\ \bibinfo {author} {\bibfnamefont {Daiki}\ \bibnamefont
  {Suenaga}},\ }\bibfield  {title} {\enquote {\bibinfo {title} {{Thermal quarks
  and gluon propagators in two-color dense QCD}},}\ }\href {\doibase
  10.1103/PhysRevD.103.094008} {\bibfield  {journal} {\bibinfo  {journal}
  {Phys. Rev. D}\ }\textbf {\bibinfo {volume} {103}},\ \bibinfo {pages}
  {094008} (\bibinfo {year} {2021})},\ \Eprint
  {http://arxiv.org/abs/2102.07231} {arXiv:2102.07231 [hep-ph]} \BibitemShut
  {NoStop}%
\bibitem [{\citenamefont {Suenaga}\ and\ \citenamefont
  {Kojo}(2021)}]{Suenaga:2021bjz}%
  \BibitemOpen
  \bibfield  {author} {\bibinfo {author} {\bibfnamefont {Daiki}\ \bibnamefont
  {Suenaga}}\ and\ \bibinfo {author} {\bibfnamefont {Toru}\ \bibnamefont
  {Kojo}},\ }\bibfield  {title} {\enquote {\bibinfo {title} {{Delineating
  chiral separation effect in two-color dense QCD}},}\ }\href {\doibase
  10.1103/PhysRevD.104.034038} {\bibfield  {journal} {\bibinfo  {journal}
  {Phys. Rev. D}\ }\textbf {\bibinfo {volume} {104}},\ \bibinfo {pages}
  {034038} (\bibinfo {year} {2021})},\ \Eprint
  {http://arxiv.org/abs/2105.10538} {arXiv:2105.10538 [hep-ph]} \BibitemShut
  {NoStop}%
\bibitem [{\citenamefont {Kojo}\ and\ \citenamefont
  {Suenaga}(2022)}]{Kojo:2021hqh}%
  \BibitemOpen
  \bibfield  {author} {\bibinfo {author} {\bibfnamefont {Toru}\ \bibnamefont
  {Kojo}}\ and\ \bibinfo {author} {\bibfnamefont {Daiki}\ \bibnamefont
  {Suenaga}},\ }\bibfield  {title} {\enquote {\bibinfo {title} {{Peaks of sound
  velocity in two color dense QCD: Quark saturation effects and semishort range
  correlations}},}\ }\href {\doibase 10.1103/PhysRevD.105.076001} {\bibfield
  {journal} {\bibinfo  {journal} {Phys. Rev. D}\ }\textbf {\bibinfo {volume}
  {105}},\ \bibinfo {pages} {076001} (\bibinfo {year} {2022})},\ \Eprint
  {http://arxiv.org/abs/2110.02100} {arXiv:2110.02100 [hep-ph]} \BibitemShut
  {NoStop}%
\bibitem [{\citenamefont {Suenaga}\ \emph {et~al.}(2023)\citenamefont
  {Suenaga}, \citenamefont {Murakami}, \citenamefont {Itou},\ and\
  \citenamefont {Iida}}]{Suenaga:2022uqn}%
  \BibitemOpen
  \bibfield  {author} {\bibinfo {author} {\bibfnamefont {Daiki}\ \bibnamefont
  {Suenaga}}, \bibinfo {author} {\bibfnamefont {Kotaro}\ \bibnamefont
  {Murakami}}, \bibinfo {author} {\bibfnamefont {Etsuko}\ \bibnamefont {Itou}},
  \ and\ \bibinfo {author} {\bibfnamefont {Kei}\ \bibnamefont {Iida}},\
  }\bibfield  {title} {\enquote {\bibinfo {title} {{Probing the hadron mass
  spectrum in dense two-color QCD with the linear sigma model}},}\ }\href
  {\doibase 10.1103/PhysRevD.107.054001} {\bibfield  {journal} {\bibinfo
  {journal} {Phys. Rev. D}\ }\textbf {\bibinfo {volume} {107}},\ \bibinfo
  {pages} {054001} (\bibinfo {year} {2023})},\ \Eprint
  {http://arxiv.org/abs/2211.01789} {arXiv:2211.01789 [hep-ph]} \BibitemShut
  {NoStop}%
\bibitem [{\citenamefont {Kawaguchi}\ and\ \citenamefont
  {Suenaga}(2023)}]{Kawaguchi:2023olk}%
  \BibitemOpen
  \bibfield  {author} {\bibinfo {author} {\bibfnamefont {Mamiya}\ \bibnamefont
  {Kawaguchi}}\ and\ \bibinfo {author} {\bibfnamefont {Daiki}\ \bibnamefont
  {Suenaga}},\ }\bibfield  {title} {\enquote {\bibinfo {title} {{Fate of the
  topological susceptibility in two-color dense QCD}},}\ }\href {\doibase
  10.1007/JHEP08(2023)189} {\bibfield  {journal} {\bibinfo  {journal} {JHEP}\
  }\textbf {\bibinfo {volume} {08}},\ \bibinfo {pages} {189} (\bibinfo {year}
  {2023})},\ \Eprint {http://arxiv.org/abs/2305.18682} {arXiv:2305.18682
  [hep-ph]} \BibitemShut {NoStop}%
\bibitem [{\citenamefont {Murakami}\ \emph
  {et~al.}(2023{\natexlab{b}})\citenamefont {Murakami}, \citenamefont
  {Suenaga}, \citenamefont {Iida},\ and\ \citenamefont
  {Itou}}]{Murakami:2022lmq}%
  \BibitemOpen
  \bibfield  {author} {\bibinfo {author} {\bibfnamefont {Kotaro}\ \bibnamefont
  {Murakami}}, \bibinfo {author} {\bibfnamefont {Daiki}\ \bibnamefont
  {Suenaga}}, \bibinfo {author} {\bibfnamefont {Kei}\ \bibnamefont {Iida}}, \
  and\ \bibinfo {author} {\bibfnamefont {Etsuko}\ \bibnamefont {Itou}},\
  }\bibfield  {title} {\enquote {\bibinfo {title} {{Measurement of hadron
  masses in 2-color finite density QCD}},}\ }\href {\doibase
  10.22323/1.430.0154} {\bibfield  {journal} {\bibinfo  {journal} {PoS}\
  }\textbf {\bibinfo {volume} {LATTICE2022}},\ \bibinfo {pages} {154} (\bibinfo
  {year} {2023}{\natexlab{b}})},\ \Eprint {http://arxiv.org/abs/2211.13472}
  {arXiv:2211.13472 [hep-lat]} \BibitemShut {NoStop}%
\bibitem [{\citenamefont {Murakami}\ \emph {et~al.}()\citenamefont {Murakami},
  \citenamefont {Suenaga}, \citenamefont {Itou},\ and\ \citenamefont
  {Iida}}]{Murakami2022}%
  \BibitemOpen
  \bibfield  {author} {\bibinfo {author} {\bibfnamefont {Kotaro}\ \bibnamefont
  {Murakami}}, \bibinfo {author} {\bibfnamefont {Daiki}\ \bibnamefont
  {Suenaga}}, \bibinfo {author} {\bibfnamefont {Etsuko}\ \bibnamefont {Itou}},
  \ and\ \bibinfo {author} {\bibfnamefont {Kei}\ \bibnamefont {Iida}},\
  }\href@noop {} {}\bibinfo {note} {In preparation}\BibitemShut {NoStop}%
\bibitem [{\citenamefont {Kawakami}\ and\ \citenamefont
  {Harada}(2018)}]{Kawakami:2018olq}%
  \BibitemOpen
  \bibfield  {author} {\bibinfo {author} {\bibfnamefont {Yohei}\ \bibnamefont
  {Kawakami}}\ and\ \bibinfo {author} {\bibfnamefont {Masayasu}\ \bibnamefont
  {Harada}},\ }\bibfield  {title} {\enquote {\bibinfo {title} {{Analysis of
  $\Lambda_c(2595)$, $\Lambda_c(2625)$, $\Lambda_b(5912)$, $\Lambda_b(5920)$
  based on a chiral partner structure}},}\ }\href {\doibase
  10.1103/PhysRevD.97.114024} {\bibfield  {journal} {\bibinfo  {journal} {Phys.
  Rev. D}\ }\textbf {\bibinfo {volume} {97}},\ \bibinfo {pages} {114024}
  (\bibinfo {year} {2018})},\ \Eprint {http://arxiv.org/abs/1804.04872}
  {arXiv:1804.04872 [hep-ph]} \BibitemShut {NoStop}%
\bibitem [{\citenamefont {Kawakami}\ and\ \citenamefont
  {Harada}(2019)}]{Kawakami:2019hpp}%
  \BibitemOpen
  \bibfield  {author} {\bibinfo {author} {\bibfnamefont {Yohei}\ \bibnamefont
  {Kawakami}}\ and\ \bibinfo {author} {\bibfnamefont {Masayasu}\ \bibnamefont
  {Harada}},\ }\bibfield  {title} {\enquote {\bibinfo {title} {{Singly heavy
  baryons with chiral partner structure in a three-flavor chiral model}},}\
  }\href {\doibase 10.1103/PhysRevD.99.094016} {\bibfield  {journal} {\bibinfo
  {journal} {Phys. Rev. D}\ }\textbf {\bibinfo {volume} {99}},\ \bibinfo
  {pages} {094016} (\bibinfo {year} {2019})},\ \Eprint
  {http://arxiv.org/abs/1902.06774} {arXiv:1902.06774 [hep-ph]} \BibitemShut
  {NoStop}%
\bibitem [{\citenamefont {Harada}\ \emph {et~al.}(2020)\citenamefont {Harada},
  \citenamefont {Liu}, \citenamefont {Oka},\ and\ \citenamefont
  {Suzuki}}]{Harada:2019udr}%
  \BibitemOpen
  \bibfield  {author} {\bibinfo {author} {\bibfnamefont {Masayasu}\
  \bibnamefont {Harada}}, \bibinfo {author} {\bibfnamefont {Yan-Rui}\
  \bibnamefont {Liu}}, \bibinfo {author} {\bibfnamefont {Makoto}\ \bibnamefont
  {Oka}}, \ and\ \bibinfo {author} {\bibfnamefont {Kei}\ \bibnamefont
  {Suzuki}},\ }\bibfield  {title} {\enquote {\bibinfo {title} {{Chiral
  effective theory of diquarks and the $U_A(1)$ anomaly}},}\ }\href {\doibase
  10.1103/PhysRevD.101.054038} {\bibfield  {journal} {\bibinfo  {journal}
  {Phys. Rev. D}\ }\textbf {\bibinfo {volume} {101}},\ \bibinfo {pages}
  {054038} (\bibinfo {year} {2020})},\ \Eprint
  {http://arxiv.org/abs/1912.09659} {arXiv:1912.09659 [hep-ph]} \BibitemShut
  {NoStop}%
\bibitem [{\citenamefont {Kim}\ \emph {et~al.}(2020)\citenamefont {Kim},
  \citenamefont {Hiyama}, \citenamefont {Oka},\ and\ \citenamefont
  {Suzuki}}]{Kim:2020imk}%
  \BibitemOpen
  \bibfield  {author} {\bibinfo {author} {\bibfnamefont {Yonghee}\ \bibnamefont
  {Kim}}, \bibinfo {author} {\bibfnamefont {Emiko}\ \bibnamefont {Hiyama}},
  \bibinfo {author} {\bibfnamefont {Makoto}\ \bibnamefont {Oka}}, \ and\
  \bibinfo {author} {\bibfnamefont {Kei}\ \bibnamefont {Suzuki}},\ }\bibfield
  {title} {\enquote {\bibinfo {title} {{Spectrum of singly heavy baryons from a
  chiral effective theory of diquarks}},}\ }\href {\doibase
  10.1103/PhysRevD.102.014004} {\bibfield  {journal} {\bibinfo  {journal}
  {Phys. Rev. D}\ }\textbf {\bibinfo {volume} {102}},\ \bibinfo {pages}
  {014004} (\bibinfo {year} {2020})},\ \Eprint
  {http://arxiv.org/abs/2003.03525} {arXiv:2003.03525 [hep-ph]} \BibitemShut
  {NoStop}%
\bibitem [{\citenamefont {Kawakami}\ \emph {et~al.}(2020)\citenamefont
  {Kawakami}, \citenamefont {Harada}, \citenamefont {Oka},\ and\ \citenamefont
  {Suzuki}}]{Kawakami:2020sxd}%
  \BibitemOpen
  \bibfield  {author} {\bibinfo {author} {\bibfnamefont {Yohei}\ \bibnamefont
  {Kawakami}}, \bibinfo {author} {\bibfnamefont {Masayasu}\ \bibnamefont
  {Harada}}, \bibinfo {author} {\bibfnamefont {Makoto}\ \bibnamefont {Oka}}, \
  and\ \bibinfo {author} {\bibfnamefont {Kei}\ \bibnamefont {Suzuki}},\
  }\bibfield  {title} {\enquote {\bibinfo {title} {{Suppression of decay widths
  in singly heavy baryons induced by the $U_A (1)$ anomaly}},}\ }\href
  {\doibase 10.1103/PhysRevD.102.114004} {\bibfield  {journal} {\bibinfo
  {journal} {Phys. Rev. D}\ }\textbf {\bibinfo {volume} {102}},\ \bibinfo
  {pages} {114004} (\bibinfo {year} {2020})},\ \Eprint
  {http://arxiv.org/abs/2009.06243} {arXiv:2009.06243 [hep-ph]} \BibitemShut
  {NoStop}%
\bibitem [{\citenamefont {Suenaga}\ and\ \citenamefont
  {Hosaka}(2021)}]{Suenaga:2021qri}%
  \BibitemOpen
  \bibfield  {author} {\bibinfo {author} {\bibfnamefont {Daiki}\ \bibnamefont
  {Suenaga}}\ and\ \bibinfo {author} {\bibfnamefont {Atsushi}\ \bibnamefont
  {Hosaka}},\ }\bibfield  {title} {\enquote {\bibinfo {title} {{Novel
  pentaquark picture for singly heavy baryons from chiral symmetry}},}\ }\href
  {\doibase 10.1103/PhysRevD.104.034009} {\bibfield  {journal} {\bibinfo
  {journal} {Phys. Rev. D}\ }\textbf {\bibinfo {volume} {104}},\ \bibinfo
  {pages} {034009} (\bibinfo {year} {2021})},\ \Eprint
  {http://arxiv.org/abs/2101.09764} {arXiv:2101.09764 [hep-ph]} \BibitemShut
  {NoStop}%
\bibitem [{\citenamefont {Suenaga}\ and\ \citenamefont
  {Hosaka}(2022)}]{Suenaga:2022ajn}%
  \BibitemOpen
  \bibfield  {author} {\bibinfo {author} {\bibfnamefont {Daiki}\ \bibnamefont
  {Suenaga}}\ and\ \bibinfo {author} {\bibfnamefont {Atsushi}\ \bibnamefont
  {Hosaka}},\ }\bibfield  {title} {\enquote {\bibinfo {title} {{Decays of
  Roper-like singly heavy baryons in a chiral model}},}\ }\href {\doibase
  10.1103/PhysRevD.105.074036} {\bibfield  {journal} {\bibinfo  {journal}
  {Phys. Rev. D}\ }\textbf {\bibinfo {volume} {105}},\ \bibinfo {pages}
  {074036} (\bibinfo {year} {2022})},\ \Eprint
  {http://arxiv.org/abs/2202.07804} {arXiv:2202.07804 [hep-ph]} \BibitemShut
  {NoStop}%
\bibitem [{\citenamefont {Suenaga}\ and\ \citenamefont
  {Oka}(2023)}]{Suenaga:2023tcy}%
  \BibitemOpen
  \bibfield  {author} {\bibinfo {author} {\bibfnamefont {Daiki}\ \bibnamefont
  {Suenaga}}\ and\ \bibinfo {author} {\bibfnamefont {Makoto}\ \bibnamefont
  {Oka}},\ }\bibfield  {title} {\enquote {\bibinfo {title} {{Axial anomaly
  effect on the chiral-partner structure of diquarks at high temperature}},}\
  }\href {\doibase 10.1103/PhysRevD.108.014030} {\bibfield  {journal} {\bibinfo
   {journal} {Phys. Rev. D}\ }\textbf {\bibinfo {volume} {108}},\ \bibinfo
  {pages} {014030} (\bibinfo {year} {2023})},\ \Eprint
  {http://arxiv.org/abs/2305.09730} {arXiv:2305.09730 [hep-ph]} \BibitemShut
  {NoStop}%
\bibitem [{\citenamefont {Takada}\ \emph {et~al.}(2023)\citenamefont {Takada},
  \citenamefont {Suenaga}, \citenamefont {Harada}, \citenamefont {Hosaka},\
  and\ \citenamefont {Oka}}]{Takada:2023evq}%
  \BibitemOpen
  \bibfield  {author} {\bibinfo {author} {\bibfnamefont {Hiroto}\ \bibnamefont
  {Takada}}, \bibinfo {author} {\bibfnamefont {Daiki}\ \bibnamefont {Suenaga}},
  \bibinfo {author} {\bibfnamefont {Masayasu}\ \bibnamefont {Harada}}, \bibinfo
  {author} {\bibfnamefont {Atsushi}\ \bibnamefont {Hosaka}}, \ and\ \bibinfo
  {author} {\bibfnamefont {Makoto}\ \bibnamefont {Oka}},\ }\bibfield  {title}
  {\enquote {\bibinfo {title} {{Axial anomaly effect on three-quark and
  five-quark singly heavy baryons}},}\ }\href {\doibase
  10.1103/PhysRevD.108.054033} {\bibfield  {journal} {\bibinfo  {journal}
  {Phys. Rev. D}\ }\textbf {\bibinfo {volume} {108}},\ \bibinfo {pages}
  {054033} (\bibinfo {year} {2023})},\ \Eprint
  {http://arxiv.org/abs/2307.15304} {arXiv:2307.15304 [hep-ph]} \BibitemShut
  {NoStop}%
\bibitem [{\citenamefont {Parganlija}\ \emph {et~al.}(2013)\citenamefont
  {Parganlija}, \citenamefont {Kovacs}, \citenamefont {Wolf}, \citenamefont
  {Giacosa},\ and\ \citenamefont {Rischke}}]{Parganlija:2012fy}%
  \BibitemOpen
  \bibfield  {author} {\bibinfo {author} {\bibfnamefont {Denis}\ \bibnamefont
  {Parganlija}}, \bibinfo {author} {\bibfnamefont {Peter}\ \bibnamefont
  {Kovacs}}, \bibinfo {author} {\bibfnamefont {Gyorgy}\ \bibnamefont {Wolf}},
  \bibinfo {author} {\bibfnamefont {Francesco}\ \bibnamefont {Giacosa}}, \ and\
  \bibinfo {author} {\bibfnamefont {Dirk~H.}\ \bibnamefont {Rischke}},\
  }\bibfield  {title} {\enquote {\bibinfo {title} {{Meson vacuum phenomenology
  in a three-flavor linear sigma model with (axial-)vector mesons}},}\ }\href
  {\doibase 10.1103/PhysRevD.87.014011} {\bibfield  {journal} {\bibinfo
  {journal} {Phys. Rev. D}\ }\textbf {\bibinfo {volume} {87}},\ \bibinfo
  {pages} {014011} (\bibinfo {year} {2013})},\ \Eprint
  {http://arxiv.org/abs/1208.0585} {arXiv:1208.0585 [hep-ph]} \BibitemShut
  {NoStop}%
\bibitem [{\citenamefont {Suenaga}\ and\ \citenamefont
  {Lakaschus}(2020)}]{Suenaga:2019urn}%
  \BibitemOpen
  \bibfield  {author} {\bibinfo {author} {\bibfnamefont {Daiki}\ \bibnamefont
  {Suenaga}}\ and\ \bibinfo {author} {\bibfnamefont {Phillip}\ \bibnamefont
  {Lakaschus}},\ }\bibfield  {title} {\enquote {\bibinfo {title}
  {{Comprehensive study of mass modifications of light mesons in nuclear matter
  in the three-flavor extended linear $\sigma$ model}},}\ }\href {\doibase
  10.1103/PhysRevC.101.035209} {\bibfield  {journal} {\bibinfo  {journal}
  {Phys. Rev. C}\ }\textbf {\bibinfo {volume} {101}},\ \bibinfo {pages}
  {035209} (\bibinfo {year} {2020})},\ \Eprint
  {http://arxiv.org/abs/1908.10509} {arXiv:1908.10509 [nucl-th]} \BibitemShut
  {NoStop}%
\bibitem [{\citenamefont {Walecka}(1974)}]{Walecka:1974qa}%
  \BibitemOpen
  \bibfield  {author} {\bibinfo {author} {\bibfnamefont {J.~D.}\ \bibnamefont
  {Walecka}},\ }\bibfield  {title} {\enquote {\bibinfo {title} {{A Theory of
  highly condensed matter}},}\ }\href {\doibase 10.1016/0003-4916(74)90208-5}
  {\bibfield  {journal} {\bibinfo  {journal} {Annals Phys.}\ }\textbf {\bibinfo
  {volume} {83}},\ \bibinfo {pages} {491--529} (\bibinfo {year}
  {1974})}\BibitemShut {NoStop}%
\end{thebibliography}%

\end{document}